\documentclass[a4paper,11pt]{article}

\usepackage{jheppub} % for details on the use of the package, please
\usepackage[T1]{fontenc} % if needed
\usepackage{float}
\usepackage{amsmath}
\usepackage{amssymb}             
\usepackage{amsfonts}            
\usepackage{color,colortbl}
\usepackage{rotating}
\definecolor{myBlue}{rgb}{0.1,0.57,0.96}
\definecolor{myBlue2}{rgb}{0.25,0.55,0.79}
\definecolor{myorange}{rgb}{0.92,0.49,0.24}
\newcommand{\be}{\begin{equation}}
\newcommand{\ee}{\end{equation}}
\newcommand{\beq}{\begin{equation}}
\newcommand{\beql}[1]{\begin{equation}\label{#1}}
\newcommand{\eeq}{\end{equation}}
\newcommand{\ba}{\begin{array}}
\newcommand{\ea}{\end{array}}
\newcommand{\bea}{\begin{eqnarray}}
\newcommand{\beal}[1]{\begin{eqnarray}\label{#1}}
\newcommand{\eea}{\end{eqnarray}}
\newcommand{\ben}{\begin{enumerate}}
\newcommand{\een}{\end{enumerate}}
\newcommand{\bean}{\begin{eqnarray*}}
\newcommand{\eean}{\end{eqnarray*}}
\newcommand{\eref}[1]{(\ref{#1})}
\newcommand{\sref}[1]{\S\ref{#1}}

\newcommand{\fref}[1]{Figure \ref{#1}}
\newcommand{\btab}[1]{\begin{tabular}{#1}}
\newcommand{\etab}{\end{tabular}}

\def \Black {\pdfsetcolor{0 0 0 1}}

\newcommand{\comment}[1]{}

\newcommand{\qed}{\nobreak \ifvmode \relax \else
      \ifdim\lastskip<1.5em \hskip-\lastskip
      \hskip1.5em plus0em minus0.5em \fi \nobreak
      \vrule height0.75em width0.5em depth0.25em\fi}

\begin{document}

%===============================================================================
\title{\hspace{0.37cm} BFT$_2$: a General Class of $2d$  $\mathcal{N}=(0,2)$ Theories, \\
\hspace{2.5cm} 3-Manifolds and Toric Geometry}
%===============================================================================

\author[a,b,c]{Sebasti\'an Franco,} 
\author[d]{Xingyang Yu}

\affiliation[a]{
Physics Department, The City College of the CUNY \\
160 Convent Avenue, New York, NY 10031, USA}

\affiliation[b]{Physics Program and $^c$Initiative for the Theoretical Sciences \\
The Graduate School and University Center, The City University of New York  \\
365 Fifth Avenue, New York NY 10016, USA}

\affiliation[d]{Center for Cosmology and Particle Physics,\\
Department of Physics, New York University,\\
726 Broadway, New York, NY 10003, USA}

\emailAdd{sfranco@ccny.cuny.edu}
\emailAdd{xy1038@nyu.edu}

%=================================================================
\abstract{We introduce and initiate the study of a general class of $2d$ $\mathcal{N}=(0,2)$ quiver gauge theories, defined in terms of certain 2-dimensional CW complexes on oriented 3-manifolds. We refer to this class of theories as BFT$_2$’s. They are natural generalizations of Brane Brick Models, which capture the gauge theories on D1-branes probing toric Calabi-Yau 4-folds. The dynamics and triality of the gauge theories translate into simple transformations of the underlying CW complexes. We introduce various combinatorial tools for analyzing these theories and investigate their connections to toric Calabi-Yau manifolds, which arise as their master and moduli spaces. Invariance of the moduli space is indeed a powerful criterion for identifying theories in the same triality class. We also investigate the reducibility of these theories.}
%=================================================================

\maketitle
\flushbottom

%=================================================================
\section{Introduction}
%=================================================================

In this paper we will introduce and initiate the study of a general class of $2d$ $\mathcal{N}=(0,2)$ gauge theories, which are defined in terms of 2-dimensional CW complexes on oriented 3-manifolds (possibly with boundaries). Remarkably, many aspects of the dynamics of these theories translate into simple transformations of these objects, several questions enjoy a combinatorial description, and the theories have a close relation to toric geometry.

In short, the new theories are generalizations of {\it brane brick models} (BBMs) to arbitrary 3-manifolds. BBMs encode the $2d$ $\mathcal{N}=(0,2)$ quiver gauge theories on the worldvolume of D1-branes probing non-compact toric Calabi-Yau 4-fold (CY$_4$) singularities \cite{Franco:2015tna,Franco:2015tya,Franco:2016nwv,Franco:2016qxh,Franco:2016fxm,Franco:2017cjj,Franco:2018qsc}.  BBMs indeed significantly simplify the map between the gauge theory and the underlying CY$_4$ geometry. A BBM is a type IIA brane configuration consisting of D4-branes suspended from an NS5-brane that wraps a holomorphic surface $\Sigma$ determined by the probed CY$_4$. BBMs are connected to the D1-branes at the singularities via T-duality. The most important information carried by a BBM is encoded in a 2-dimensional CW complex on $\mathbb{T}^3$, the tropical limit of the coamoeba projection of $\Sigma$.

We will refer to the new class of theories as BFT$_2$’s. The motivation for this name is that these theories are natural generalizations of {\it bipartite field theories} (BFTs), a class of $4d$ $\mathcal{N}=1$ gauge theories defined by bipartite graphs on Riemann surfaces (with boundaries) \cite{Franco:2012mm,Franco:2012wv,Franco:2013pg,Franco:2013ana,Franco:2014nca,Franco:2018vqd} (see also \cite{Xie:2012mr,Heckman:2012jh} for closely related constructions). Having said that, we will see that the notion of bipartiteness does not seem to be relevant for this class of models. Therefore, the ``B’’ letter might be interpreted as ``bicolored”. We expect similar generalizations of $m$-dimers \cite{Franco:2019bmx}, which we will denote BFT$_m$, to exist for general $m$. 

BFT$_2$’s fall within the modern and fruitful approach of defining quantum field theories in terms of geometric or combinatorial objects. In such constructions, the dynamics of the quantum field theory is often geometrized or translated into simple modifications of the underlying object. Moreover, complicated theories can usually be constructed by assembling simple building blocks. Examples of this general strategy abound, see e.g. for an incomplete selection \cite{Franco:2005rj,Gaiotto:2009we,Benini:2009mz,Bah:2012dg}.

This paper is organized as follows. \sref{section_(0,2)_review} provides a brief review of $2d$ $\mathcal{N}=(0,2)$ gauge theories. \sref{section_BFT2s} introduces BFT$_2$’s and discusses their structure. \sref{section_combinatorial_tools} introduces various combinatorial tools, which are useful for the analysis of BFT$_2$’s. \sref{section_graph_modifications} explains the map between several transformations of the underlying CW complex and the dynamics of BFT$_2$. \sref{section_BFT2s_and_toric_geometry} discusses two toric CYs that are associated to every BFT$_2$, its master and moduli spaces. \sref{section_geometry_from_flows} presents an alternative procedure, based on flows, to map BFT$_2$’s to the corresponding toric CYs. Additional explicit examples are presented in \sref{section_additional_examples}, to illustrate the dependence of the master and moduli spaces on the geometry of the underlying 3-manifold. \sref{section_triality_and_moduli_space} discusses the invariance of the moduli space under triality, motivating its application to identify theories connected by the corresponding moves. \sref{section_reduction_CY} studies the interplay between triality, reduction and the moduli space. In \sref{section_conclusions}, we conclude and present directions for future research.

%=================================================================
\section{2d $\mathcal{N}=(0,2)$ Gauge Theories: A Brief Review}
%=================================================================

\label{section_(0,2)_review}

We now present a lightning review of $2d$ $\mathcal{N}=(0,2)$ gauge theories. More detailed presentations can be found in \cite{Witten:1993yc,GarciaCompean:1998kh,Gadde:2013lxa,Franco:2015tna}. These theories can be conveniently formulated in $2d$ $\mathcal{N}=(0,2)$ superspace: $(x^\alpha,\theta^+,\bar{\theta}^+)$, $\alpha=0,1$. The elementary building blocks are three types of superfields: 

\begin{itemize}

\item{\bf Vector.} It contains a gauge boson $v_\alpha$ $(\alpha =0,1)$, adjoint chiral fermions $\chi_-$, $\bar{\chi}_-$ (gaugini) and a real auxiliary field $D$. From now on, $\pm$ subindices indicate fermion chirality.

\item{\bf Chiral.} Its on-shell degrees of freedom are a complex scalar $\phi$ and a chiral fermion $\psi_+$.

\item{\bf Fermi.} Its only on-shell degree of freedom is a chiral fermion $\lambda_-$. In addition, it contains an auxiliary field $G$. Every Fermi field $\Lambda_a$ is associated to two holomorphic functions of chiral superfields, which we call $J^a$ and $E_a$. $\Lambda_a$ couples to $J^a$ through a gauge invariant term of the form
\beq
L_J = -\int d^2y \, d\theta^+  \left( \Lambda_a J^a |_{\bar{\theta}^+=0}\right)-h.c. \, ,
\eeq
while $E_a$ is introduced as a deformation of the chirality condition for $\Lambda_a$. For these reasons, $E_a$ has the same quantum numbers (e.g. gauge quantum numbers) of $\Lambda_a$, while $J^a$ has conjugate quantum numbers. Sometimes, it is practical to summarize the information regarding the $J^a$ and $E_a$ functions via a $2d$ $\mathcal{N}=(0,2)$ ``superpotential”, which schematically has the following form
\beq
W=\sum_a (\Lambda_a J^a + \overline{\Lambda}^a E_a) \, .
\eeq
The theories are symmetric under the exchange of $\Lambda_a \leftrightarrow \overline{\Lambda}^a$ for any $a$, which should be accompanied by the switch of $J^a \leftrightarrow E_a$.

\end{itemize}

Integrating out the auxiliary fields in the vector multiplets, we obtain a usual $D$-term potential. For abelian theories, it becomes
\beq
V_D = \sum_\alpha \left( \sum_i q_{\alpha i} |\phi_i|^2 - t_\alpha \right)^2 \,,
\label{V_D}
\eeq
where $\alpha$ runs over the different factors of the gauge group and the $t_\alpha$ are complexified Fayet-Iliopoulos parameters.

Integrating out the auxiliary fields $G_a$ in the Fermis $\Lambda_a$, we obtain 
an analog of an $F$-term potential, 
\beq
V_F = \sum_a \left( \mathrm{tr}|E_a(\phi)|^2 +  \mathrm{tr}|J^a(\phi)|^2 \right)\,,
\label{V_JE}
\eeq
where $\phi$ represents the scalar components of chiral fields, and Yukawa-like interactions between pairs of fermions and scalars.

Consistency of the theory requires that the $J$- and $E$-terms satisfy the following ``orthogonality” condition
\beq
\sum_a \mathrm{tr} \left[ E_a J^a \right]= 0 \, ,
\label{trace_condition}
\eeq
which is often referred to as the {\it trace condition}.

Finally, gauge anomalies need to vanish. In \sref{section_anomaly_cancellation} we will elaborate on what this cancellation entails for the class of theories considered in this paper.

%=================================================================
\section{BFT$_2$ Theories}
%=================================================================

\label{section_BFT2s}

In this section, we introduce BFT$_2$ theories, an infinite class of $2d$ $\mathcal{N}=(0,2)$ gauge theories defined by a 2-dimensional CW complex $G$ on a 3-manifold $M_3$, possibly with boundaries $\partial M_3$. These theories are natural generalizations of BBMs, which encode the gauge theories on D1-branes probing toric CY$_4$ singularities and correspond to 2-dimensional CW complexes on $\mathbb{T}^3$ \cite{Franco:2015tna,Franco:2015tya,Franco:2016nwv,Franco:2016qxh,Franco:2016fxm,Franco:2017cjj,Franco:2018qsc}. 

\begin{figure}[ht]
	\centering
	\includegraphics[width=9cm]{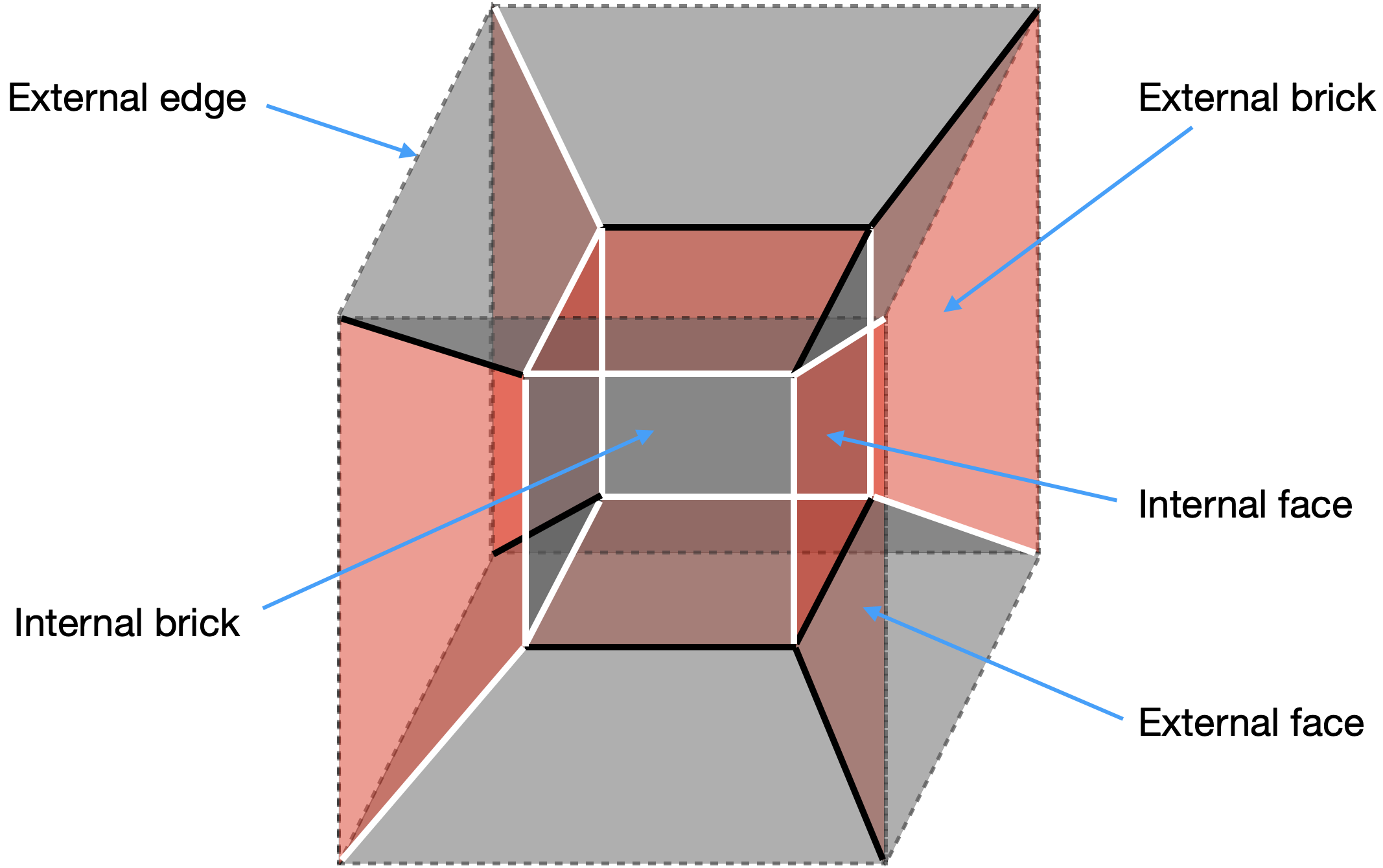}
	\caption{A BFT$_2$ defined on a 3-ball with $S^2$ boundary. We refer to this simple theory as the {\it cube model}.}
	\label{cube model}
\end{figure}

In \fref{cube model} we present a simple example of a BFT$_2$ defined on a 3-ball with $S^2$ boundary.\footnote{For brevity, we will often refer to $G$ and the corresponding gauge theory as  BFT$_2$ interchangeably.} We will call it the {\it cube model}.

%=================================================================
%=================================================================

We refer to the basic elements of $G$ as bricks, faces, edges and vertices, where we have sorted them according to decreasing dimension, from $3d$ to $0d$. There are two distinct types of faces, which we will identify by coloring them gray and red. It is natural to classify these elements as internal or external, depending on their location with respect to $\partial M_3$. Internal bricks are those surrounded entirely by faces in $G$, while all others are external. We refer to faces with at least one edge on $\partial M_3$ as external. Finally, external edges are those in $\partial M_3$. We represent them by dotted lines. We restrict external edges to be connected to a single face. To prevent clutter in the figures, we will leave $\partial M_3$ implicit. \fref{cube model} shows the distinction between internal and external elements. In the coming section we elaborate on their interpretation and additional properties.

%=================================================================
\subsection{The Dictionary}
%=================================================================

\label{section_dictionary}

Below, we present the map between the elements of $G$ in $M_3$ and the corresponding $2d$ $\mathcal{N}=(0,2)$ gauge theory.

\bigskip

\begin{itemize}
%=================================================================
	\item \textbf{Bricks.} Every brick is associated to a $U(N)$ group.\footnote{In fact, in general it is possible for different bricks to have different ranks while keeping the theory consistent, i.e. free of gauge anomalies. In the case of BBMs, this corresponds to populating the bricks with different numbers of D4-branes. Throughout this paper, we will equal ranks for all the bricks. In fact, we will shortly identify a condition such that the BFT$_2$ is anomaly free for equal ranks.}
	Internal and external bricks are mapped to gauge and global groups, respectively. This map is motivated by similar systems admitting an interpretation as brane configurations in string theory. In such cases, bricks are filled with stacks of D-branes suspended from heavier branes represented by $G$. Internal bricks are finite along all the directions transverse to the ones in which the gauge theory lives, therefore giving rise to gauge symmetries in $2d$. On the contrary, external bricks can have an infinite extension in the internal dimensions, so D-branes filling them lead to global symmetries in $2d$. As previously mentioned, BFT$_2$’s on $\mathbb{T}^3$ without boundaries correspond to BBMs, which are indeed configurations consisting of D4-branes suspended from an NS5-brane. Whether more general BFT$_2$’s admit a string theory realization is an interesting question, which we differ to future work.

\bigskip

%=================================================================
\item \textbf{Faces.} 
%=================================================================
Every face corresponds to a matter field transforming in the bifundamental representation of the two bricks it separates (or in the adjoint representation, if the brick on both sides is the same). There are two types of faces, gray or red, which represent the two types of matter superfields of $2d$ $\mathcal{N}=(0,2)$ SUSY: chiral (gray) and Fermi (red). Chiral faces can have an arbitrary even number of edges, while Fermi faces have four edges. Both constraints on the number of edges are explained below. Moreover, chiral faces are oriented, while Fermi faces are unoriented, reflecting the symmetry between Fermis and conjugate Fermis. Below we introduce a prescription for orienting chiral faces. 

Following the map between brick and gauge or global symmetry groups, faces correspond to: bifundamentals of the gauge group (when they separate two internal bricks), fundamental or antifundamental flavors (internal/external) or gauge singlets in a bifundamental representation of the global symmetry group (external/external).

%=================================================================
\paragraph{External faces: dynamical vs. non-dynamical}. 
%=================================================================
Let us consider the interpretation of fields associated to external faces more carefully. Thinking about brane configurations that could lead to gauge theories of this type, it is natural to contemplate the possibility of such external fields being dynamical or not. For the discussions in this paper, whether external chiral fields are regarded as dynamical or not is not so crucial. Such a distinction would only affect the interpretation of the geometries studied in \sref{section_BFT2s_and_toric_geometry}. On the contrary, the treatment of external Fermis is more important. In what follows, we will regard them as dynamical fields. More concretely, this choice means that these Fermis participate in the trace condition (see \sref{section_combinatorics_trace_condition}) and that we will require their $J$- and $E$-terms to vanish when computing moduli spaces (see \sref{section_BFT2s_and_toric_geometry}). Interestingly, in the context of D-branes at CY$_4$ singularities it is indeed possible to obtain dynamical Fermis between global symmetry groups, by e.g. considering stacks of flavors D5-branes with $2d$ intersections \cite{flavors_to_appear}.
	
\bigskip

%=================================================================	
\item \textbf{Edges.} 
%=================================================================
At every internal edge, a single Fermi face intersects with a number of chiral faces.\footnote{In fact, it is possible for Fermis to meet at an edge. Explicit examples of such instances, in the context of BBMs, can be found in \cite{Franco:2015tya,Franco:2016nwv}. In such configurations, the determination of the superpotential from the CW complex is more involved. However, all the ideas that we present in the paper, particularly the concept of perfect matchings and the correspondence between BFT$_2$'s and toric geometry, apply to those theories without changes. For simplicity, in what follows we will continue restricting to the class of BFT$_2$'s in which Fermis do not have common edges.} Such an edge translates into a monomial in the $2d$ $\mathcal{N}=(0,2)$ superpotential, i.e. a $J$- or $E$-term coupling.\footnote{The distinction between the two types of couplings is simply determined by the choice of orientation for the Fermi represented by the red face.} 
This monomial is the product of all the chiral fields and the single Fermi field (or its conjugate, as necessary by the orientation of the chiral fields) intersecting on it. We define the valence of an edge as the number of faces intersecting on it, which in turn determines the order of the corresponding superpotential term. Internal edges are colored white or black, depending on whether the corresponding monomial has a positive or negative sign, respectively. 

External edges do not have a superpotential interpretation since a single face terminates on them. We therefore do not assign any color to them.

A BFT$_2$ with edge coloring fully encodes the superpotential, i.e. it even determines the signs of its monomials. It is interesting to note that this is a new addition to this class of constructions. In particular, this information was left implicit in 
previous studies of BBMs. We will see that including this data reveals interesting new structures.

%=================================================================
\paragraph{Fermi faces and the toric condition of the superpotential.} 
%=================================================================
As we mentioned earlier, we restrict Fermi faces to have four edges. In addition, we will assign opposite signs to opposite internal edges of every Fermi, as illustrated in \fref{edge_coloring}.

%=================================================================
\begin{figure}[H]
	\centering
	\includegraphics[width=5cm]{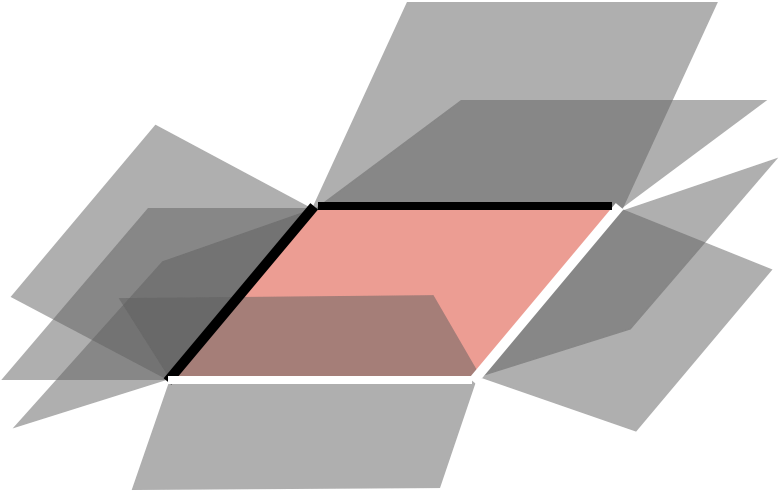}
	\caption{Opposite internal edges of a Fermi have opposite colors. This translates in the toric condition of the superpotential.} 
	\label{edge_coloring}
\end{figure}
%=================================================================

These conditions result in a special structure for the superpotential, often referred to as the {\it toric condition} \cite{Franco:2015tna}. For every Fermi face, each pair of opposite edges translates into two superpotential couplings of the same type (i.e. either both $J$- or both $E$-terms)\footnote{Which edges correspond to $J$- or $E$-terms depend on the orientation of chiral fields, as we explain later.} with opposite signs. Therefore, for every Fermi $\Lambda_a$ we have the following contributions to the superpotential
\beq
	\Lambda_a (J^a_+(X)-J^a_-(X))+\bar{\Lambda}^a(E_{a-} (X)-E_{a+}(X)) \, .
	\label{toric_superpotential}
\eeq
If a Fermi face is such that some of its edges is external, the corresponding term in \eref{toric_superpotential} will simply be missing. In \sref{section_J_and_E_flatness_PMs} we will discuss how the vanishing of $J$- and $E$-terms should be implemented in such cases.

The toric condition is crucial for the combinatorial description of BFT$_2$'s and their connection to toric geometry.

\bigskip	

%=================================================================
\item \textbf{Vertices.} 
%=================================================================
Interestingly, vertices have been overlooked in earlier studies of BBMs, despite being rather natural to ask for their interpretation. A deeper understanding of vertices will reveal several structures of BFT$_2$'s that were previously unnoticed.

We will define BFT$_2$'s such that two Fermis intersect at their corners at every vertex.\footnote{Multiple Fermis can meet at vertices in configurations with Fermis that are adjacent over edges, but we will not consider them in this paper. More generally, one could imagine configurations in which Fermis are not adjacent over edges, but more than two Fermis meet at a vertex. Such setups do satisfy our definition of BFT$_2$.} Therefore, four edges meet at every vertex. This structure plays an important role in the vanishing of the trace condition, as we will explain in \sref{section_combinatorics_trace_condition}.

\fref{vertex in G} represents the simplest vertex configuration, in which each of the four edges intersecting in it is shared by a Fermi and two chirals, therefore corresponding to a cubic term in the superpotential. More generally, any of the chirals in the figure can be replaced by multiple chirals terminating on the same edge.

%=================================================================
\begin{figure}[H]
	\centering
	\includegraphics[width=4cm]{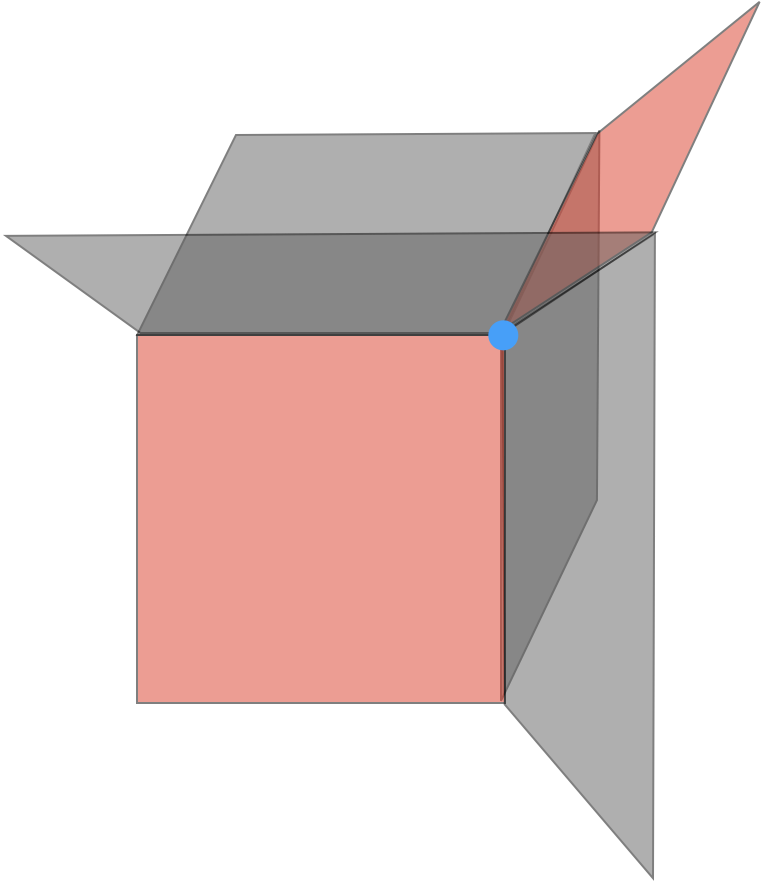}
	\caption{A simple vertex configuration in a BFT$_2$, where the vertex is highlighted in blue. It is shared by two Fermi faces and four edges. }
	\label{vertex in G}
\end{figure}
%=================================================================

\end{itemize}

%=================================================================
\subsection{The Dual Quiver}
%=================================================================

\begin{figure}[ht]
	\centering
	\includegraphics[width=12cm]{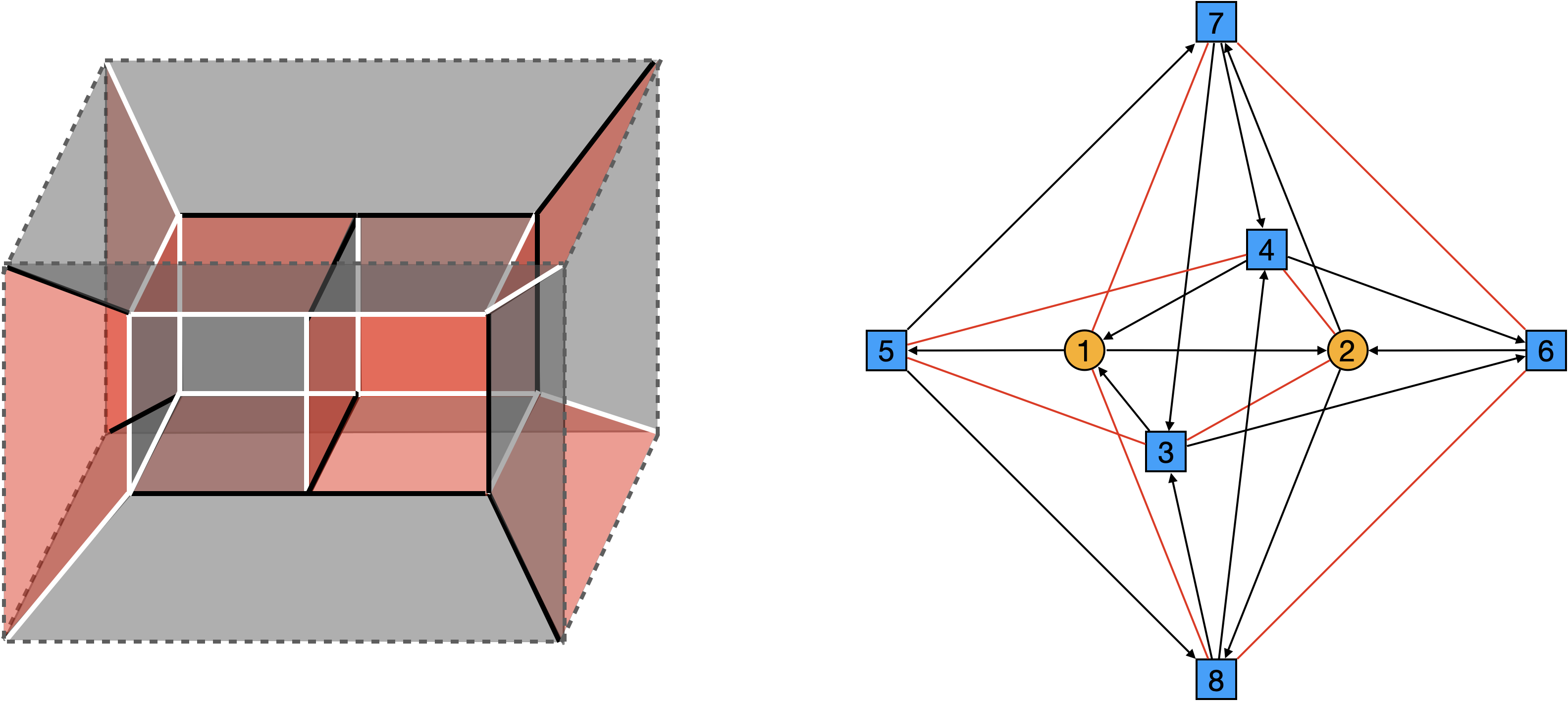}
	\caption{A BFT$_2$ on a 3-ball and its dual quiver. Yellow round and blue square nodes correspond to gauge and global symmetry groups, respectively. Every plaquette in the dual quiver corresponds to a monomial in the superpotential.}
	\label{dual_quiver}
\end{figure}

The complete information defining a BFT$_2$ is equivalently captured by a special quiver embedded in $M_3$, to which we will refer as the {\it dual quiver}. The dual quiver is obtained from $G$ via standard graphical dualization. It not only summarizes the symmetries (gauge and global) plus field content of the theory, like an ordinary quiver, but it also encodes its superpotential. Every {\it minimal plaquette} in the dual quiver, i.e. every plaquette dual to an edge in $G$, corresponds to a term in the superpotential.\footnote{In principle, we could color plaquettes in the dual quiver to capture information regarding the signs of the superpotential terms. For simplicity of the figures, we will avoid doing so.}  \fref{dual_quiver} shows an example of a dual quiver. In the case of BBMs, dual quivers correspond to the so-called periodic quivers \cite{Franco:2015tya,Franco:2016nwv,Franco:2016qxh,Franco:2016fxm,Franco:2018qsc}.

%=================================================================
%=================================================================

\begin{figure}[H]
	\centering
	\includegraphics[width=11cm]{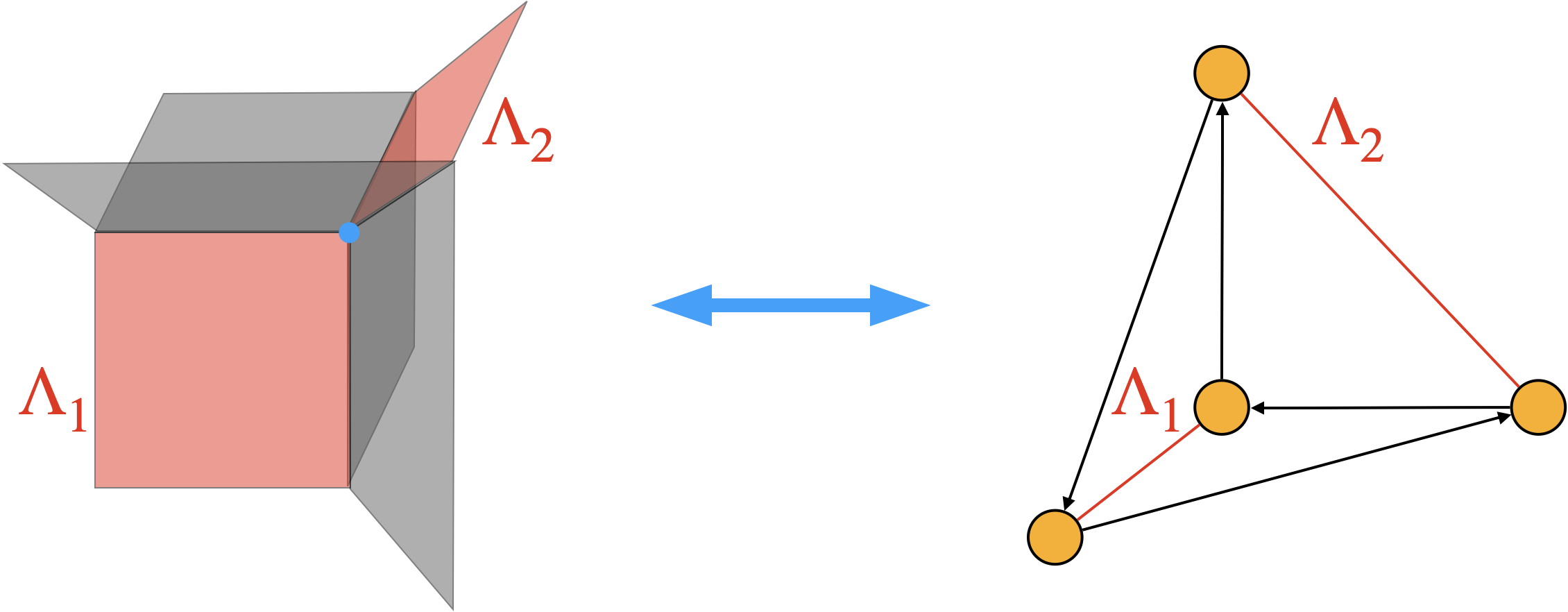}
	\caption{A vertex in a BFT$_2$ and the corresponding tetrahedron in the dual quiver.}
	\label{vertex-tetrahedron}
\end{figure}

%=================================================================
\paragraph{Chiral field orientation.}
%=================================================================
In order to determine the orientation of chiral fields in the dual quiver, it is sufficient to understand what happens at the vertices of $G$.\footnote{See \cite{Franco:2015tya} for an earlier discussion of the orientation of chiral fields in the dual quiver.} \fref{vertex-tetrahedron} shows a vertex in $G$ and the corresponding piece in the dual quiver, which takes the form of a tetrahedron. The six edges of the tetrahedron are divided into two Fermis and four chirals. Moreover, its four triangular faces are the plaquettes representing the four superpotential terms associated to the edges of $G$ intersecting at the vertex. As in our previous discussion of vertices, every chiral arrow may actually represent multiple consecutive chirals. In such case, the associated plaquette would represent a superpotential term that is of higher than cubic order. 
	
%=================================================================
%=================================================================

This perspective leads to a beautiful prescription for orienting the chiral fields. The orientation of any of the chiral arrows in this tetrahedron fixes the orientations of all other chiral fields. In particular, as shown in \fref{vertex-tetrahedron}, every corner of the tetrahedron is such that it contains exactly one field of each possible type: a Fermi, an incoming chiral and an outgoing chiral. The entire dual quiver is obtained by combining several tetrahedra. More precisely, as we mentioned before, these elementary building blocks may not be strictly tetrahedra if some of the arrows represent multiple chirals.

A $2d$ $\mathcal{N}=(0,2)$ quiver is equivalent to the quiver obtained by reversing all the chiral fields. This is the case because such transformation corresponds to reversing all fields in the quiver, i.e. exchanging the role of fundamental and antifundamental representations for all gauge and global groups, combined with the Fermi-conjugate Fermi symmetry. Therefore, once we pick the orientation of an arbitrary chiral field in the theory, the previous prescription exhausts all the freedom in the orientation of the chirals.

%=================================================================
\subsection{Additional Structural Constraints}
%=================================================================

We now discuss additional constraints on the structure of BFT$_2$’s.

%=================================================================
\subsubsection{Gauge Anomaly Cancellation}
%=================================================================

\label{section_anomaly_cancellation}

Since we are assuming that the BFT$_2$’s are consistent when the ranks of all bricks are equal, the cancellation of the $SU(N)^2$ non-abelian anomaly for every gauge group requires that each internal brick satisfies
\beq
	n_{i}^{\chi}-n_{i}^{F}=2 \, ,
\label{anomaly cancellation}
\eeq
where $n_{i}^{\chi}$ and $n_{i}^{F}$ are the numbers of chiral and Fermi faces for internal brick $i$, respectively. Given that external bricks correspond to global symmetry groups, they do not need to satisfy this condition, i.e. they generically can have non-vanishing 't Hooft anomalies.

On the other hand, the $U(1)$ factors of the bricks can generically have $U(1)_i^2$ and mixed $U(1)_i U(1)_j$ abelian anomalies \cite{Franco:2015tna}. If the BFT$_2$ has a string theory origin, abelian gauge anomalies hint at the existence of additional Fermi multiplets in appropriate determinant representations, located at specific places of the CW complex (see e.g \cite{Costello:2018fnz,Hanany:2018hlz, Gaiotto:2019jvo,Okazaki:2019bok} for related discussions). While the presence of these fields is important and interesting, we will not consider this issue any further. For the T-dual branes at singularities, these anomalies are generally cancelled by bulk fields in the closed string sector via the Green-Schwarz mechanism \cite{Mohri:1997ef}. The interplay or possible map between the two mechanisms described above remains an interesting open question.

%=================================================================
\subsubsection{The Combinatorics of the Trace Condition}
%=================================================================

\label{section_combinatorics_trace_condition}

The trace condition \eref{trace_condition} is a highly non-trivial constraint on the superpotential of $2d$ $\mathcal{N}=(0,2)$ theories. Remarkably, this condition admits a beautiful combinatorial solution within the class of BFT$_2$ theories. 

\begin{figure}[H]
	\centering
	\includegraphics[width=4.5cm]{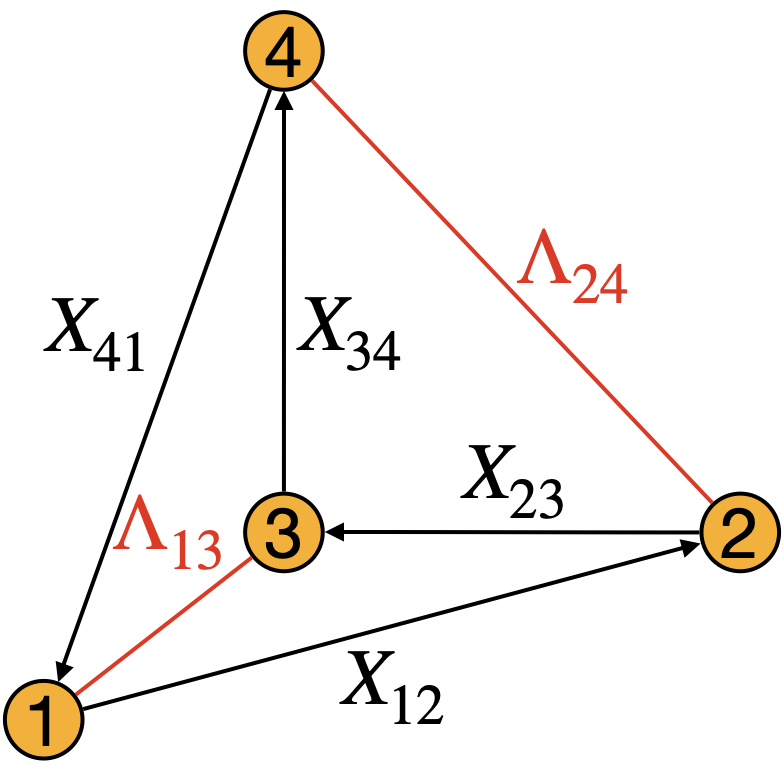}
	\caption{Tetrahedron in the dual quiver corresponding to a vertex, with fields labelled.}
	\label{tetrahedron labelled}
\end{figure}

\begin{figure}[H]
	\centering
	\includegraphics[width=9.7cm]{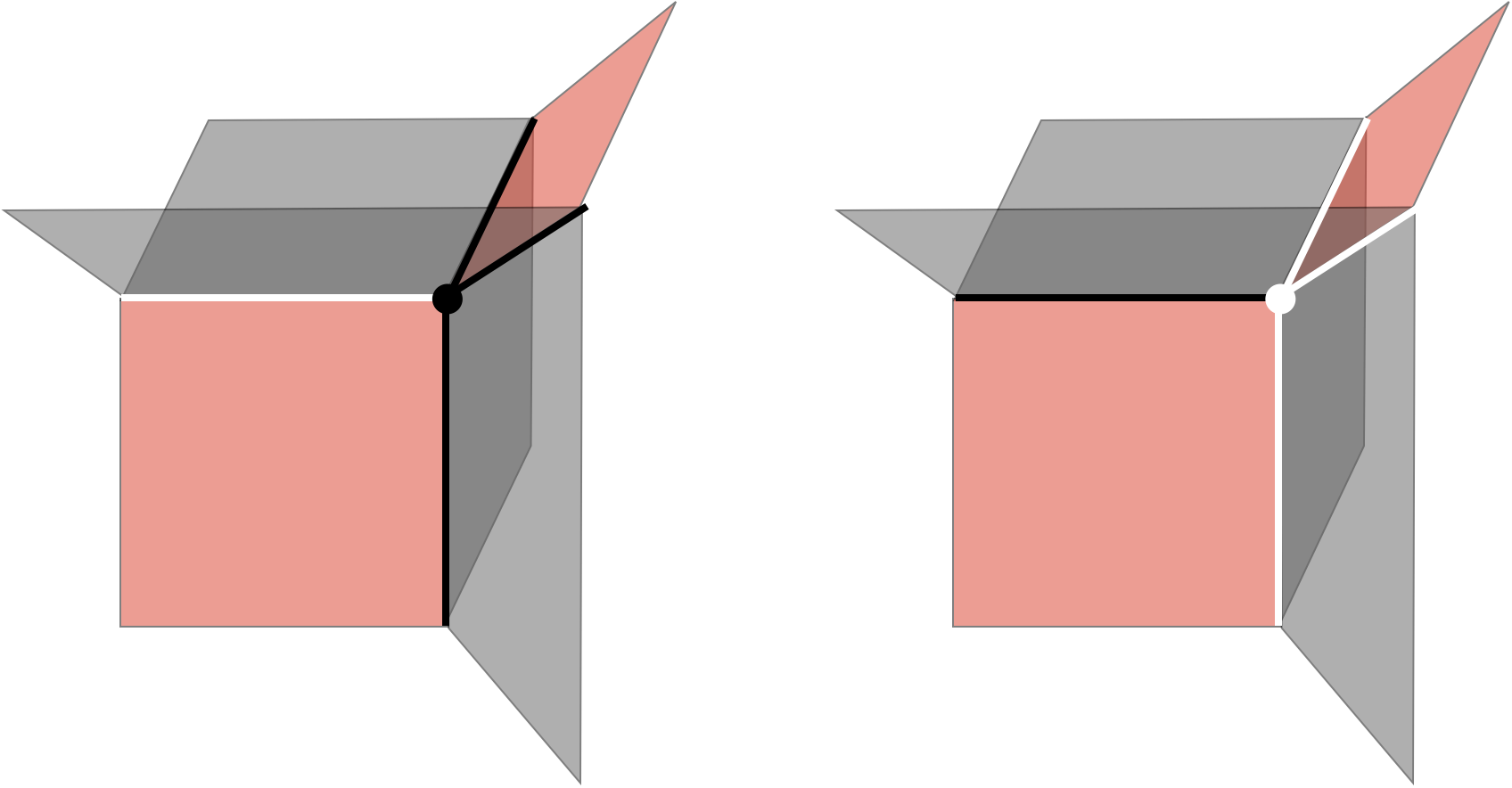}
	\caption{Two possible types of vertices with either three negative and one positive edges, or three positive and one negative edges.}
	\label{two types of vertices}
\end{figure}

Every internal vertex is associated to two holomorphic gauge invariant monomials in chiral fields constructed as follows. Consider a vertex $\alpha$ at which four edges $j_1$, $e_1$, $j_2$ and $e_2$ intersect, where $j_1$ and $e_1$ belong to the Fermi face $\Lambda_1$, while $j_2$ and $e_2$ belong to the Fermi face $\Lambda_2$. The four edges translate into two $J$- and two $E$-terms in the superpotential as follows: $j_1 \leftrightarrow \Lambda ^1J_1$, $e_1 \leftrightarrow \bar{\Lambda}_1E^1$, $j_2\leftrightarrow \Lambda^2 J_2$ and $e_2\leftrightarrow \bar{\Lambda}_2E^2$. Therefore, we can associate the vertex to the following sum
\beq
\label{vertex dictionary}
		\alpha \longleftrightarrow J_1 \cdot E^1+J_2 \cdot E^2 \, .
\eeq
The trace condition is the sum of such contributions over all the vertices of $G$. In fact, the two terms in \eref{vertex dictionary} are the same up to at most a sign. To see this, it is convenient to consider the tetrahedron dual to the vertex, as shown in \fref{tetrahedron labelled}. 
%=================================================================
%=================================================================
Explicitly, the four edges converging at the vertex correspond to the following superpotential terms
\beq
\begin{array}{ccl}
W & \supset & s_{j_{13}} \Lambda_{13} X_{34} X_{41} + s_{e_{13}} \bar{\Lambda}_{13} X_{12} X_{23} \\[.1cm]
& + & s_{j_{24}} \Lambda_{24} X_{41} X_{12} + s_{e_{24}} \bar{\Lambda}_{24} X_{23} X_{34}
\end{array}
\eeq
where $s_{j_{13}}$, $s_{e_{13}}$, $s_{j_{24}}$ and $s_{e_{24}}$ are the signs of the terms, which are encoded in the edge coloring. As before, each of the chiral fields $X_{ij}$ can be replaced by a product of consecutive chiral arrows without affecting our argument. Then, \eref{vertex dictionary} becomes
\beq
s_{j_{13}} s_{e_{13}} X_{12} X_{23} X_{34} X_{41} + s_{j_{24}} s_{e_{24}} X_{12} X_{23} X_{34} X_{41} \, .
\label{trace_condition_vertex_0}
\eeq
The two terms cancel each other if
\beq
s_{j_{13}} s_{e_{13}} = - s_{j_{24}} s_{e_{24}}  \ \ \ \Leftrightarrow \ \ \ s_{j_{13}} s_{e_{13}} s_{j_{24}} s_{e_{24}} = - 1
\label{trace_condition_vertex}
\eeq
Therefore, the trace condition is automatically satisfied if every vertex has either three positive and one negative edges, or three negative and one positive edges. \fref{two types of vertices} illustrates the resulting two possible types of vertices. We will refer to this constraint on the coloring of edges as the {\it vertex trace condition}.
%=================================================================
%=================================================================
Interestingly, this property is satisfied by all the BBMs constructed from $D1$-branes at toric CY$_4$'s \cite{Franco:2015tna,Franco:2015tya,Franco:2016nwv,Franco:2016qxh,Franco:2016fxm,Franco:2017cjj,Franco:2018qsc}. However, it has never been noted or emphasized before. Here we have elevated it to a fundamental condition that needs to be satisfied BFT$_2$'s, which guarantees that the trace condition is automatically fulfilled. Quite likely, this property is a key ingredient for other combinatorial objects in BFT$_2$'s.

\begin{figure}[H]
	\centering
	\includegraphics[width=10cm]{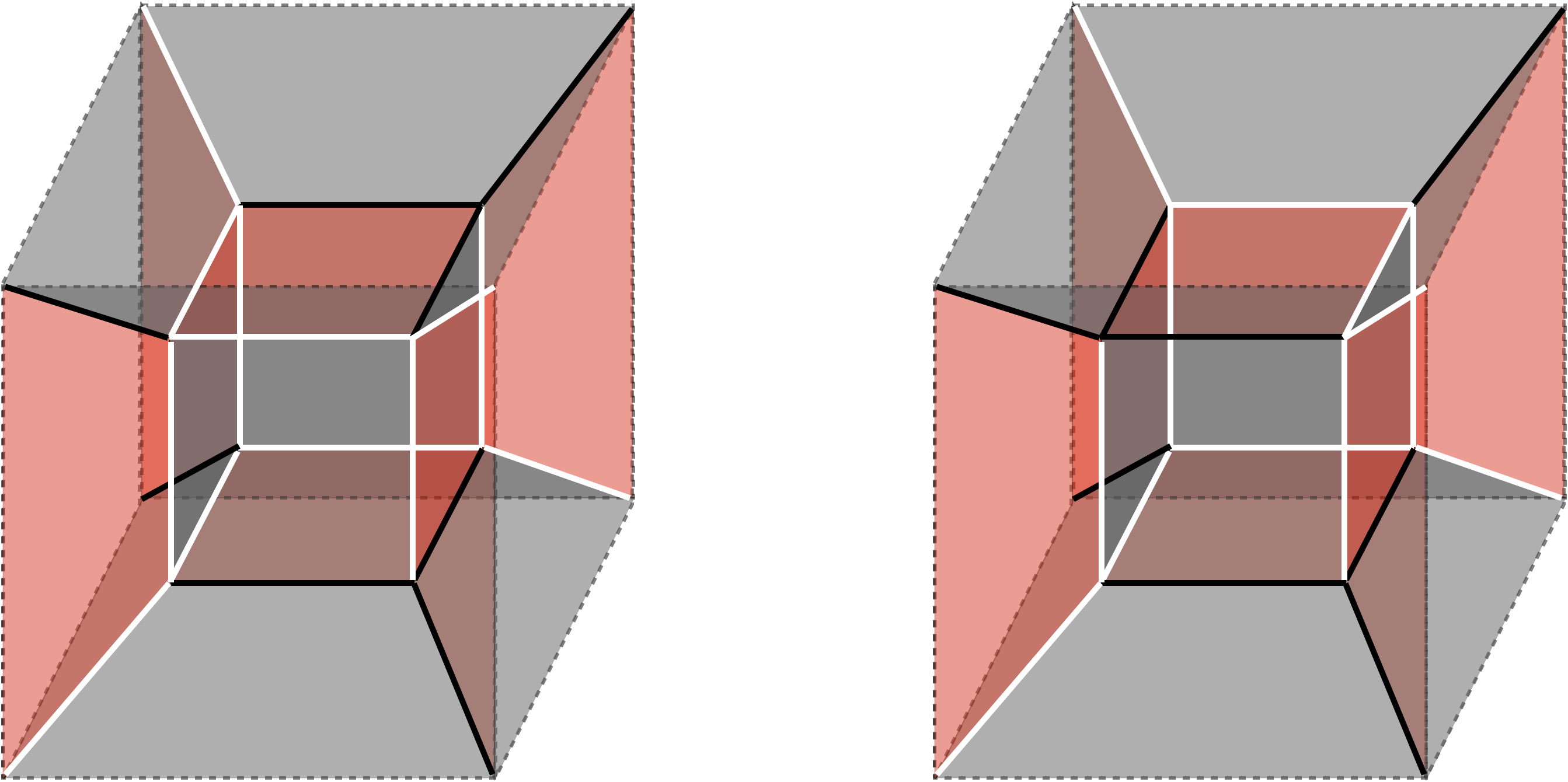}
	\caption{Two possible edge colorings for the cube model, which are equivalent upon field redefinitions.}
	\label{two edges colorings of the cube model}
\end{figure}

It is interesting to notice that, in general, combining the rule that opposite edges of Fermis must have opposite colors and the combinatorial trace condition, does not uniquely fix the coloring of edges. In fact, theories as simple as the cube model admit multiple edge colorings consistent with both conditions, as shown in \fref{two edges colorings of the cube model}. In this example, it is straightforward to verify that both color choices are equivalent upon field redefinitions. It is natural to ask whether such differences can always be absorbed by field redefinitions and, if so, whether there is a ``canonical” coloring. While these questions certainly deserve further study, they are beyond the scope of this paper. In particular, all the notions we will introduce are insensitive to this freedom. 

%=================================================================
%=================================================================

%=================================================================
\subsection{Summary}
%=================================================================

The table below summarizes the correspondence between a 2-dimensional CW complex $G$ in a 3-manifold $M_3$ and a BFT$_2$ and the basic rules for constructing this class of theories.

%=================================================================
\begin{center}
	\begin{tabular}{|p{0.35 \textwidth}|p{0.6 \textwidth}|}
	\hline
	\rowcolor{myBlue2} \textbf{2d CW Complex} & \textbf{BFT$_2$}\\
	\hline \hline
	Internal brick & $U(N)$ gauge group \\
	\hline
	External brick & $U(N)$ global group\\
	\hline \hline
	Gray face (even sided) separating two bricks $i$ and $j$ & Chiral field in the bifundamental representation of the groups $i$ and $j$ (adjoint if $i=j$) \\ \hline
	Red face (4-sided) separating two bricks $i$ and $j$ & Fermi field in the bifundamental representation of the groups $i$ and $j$ (adjoint if $i=j$) \\
	\hline \hline
	k-valent edge (involving one Fermi and $(k-1)$ chirals) & Monomial in the superpotential associated to a $J$- or $E$-term coupling a Fermi and $(k-1)$ chiral fields. The sign of the term is $(+/-)$ for a (white/black) edge \\
	\hline\hline 
	Vertex shared by two red faces $a$ and $b$ & Vanishing sum of the products of two $J$- and two $E$-terms intersecting at the vertex coming from Fermi fields $a$ and $b$\\
	\hline
	\end{tabular}
\end{center}
%=================================================================

In addition, the following conditions must be satisfied:

\begin{enumerate}

\item Every external edge is connected to a single face.

\item {\bf Anomaly cancellation.} Every internal brick is bounded by $n$ Fermi faces and $(n+2)$ chiral faces, with $n\ge 0$.

\item {\bf Toric condition.} Every Fermi face has four edges, where opposite internal edges have opposite colors. 

\item {\bf Combinatorial trace condition.} Every internal vertex is shared by four edges and is of one of two types: three white and one black edges or three black and one white edges.

\end{enumerate}

%=================================================================
\section{Combinatorial Tools}
%=================================================================

\label{section_combinatorial_tools}

In this section we introduce various combinatorial objects which are useful in connecting CW complexes, gauge theory and toric geometry in the context of BFT$_2$'s

%=================================================================
\subsection{Perfect Matchings}
%=================================================================

\label{section_perfect_matchings}

The first class of objects are the {\it almost perfect matchings} of a BFT$_2$. They were first introduced in \cite{Franco:2015tya} for BBMs, where they were called {\it brick matchings}. An alternative, equivalent definition was presented in \cite{Franco:2019bmx}, where they were also extended to $m$-dimers. Here we generalize them to general 3-manifolds with boundaries. For brevity, we will refer to them simply as {\it perfect matchings}. 

\smallskip

%=================================================================
\begin{center}
\fbox{\begin{minipage}{0.95 \textwidth}
\begin{center}
\begin{minipage}{0.92 \textwidth}
\noindent{\bf Definition 1.} A perfect matching $p$ is a collection of chiral fields such that for every Fermi field $\Lambda_a$, the chiral fields in $p$ cover \emph{either} all the monomials in $J^a$ \emph{or} all the monomials in $E_a$ exactly once.
\end{minipage}
\end{center}
\end{minipage}}
\end{center}
%=================================================================

\smallskip

Clearly, the toric condition of the superpotential of BFT$_2$'s is crucial for defining perfect matchings.

As explained in \cite{Franco:2015tya}, it is possible, and often illuminating, to extend the definition of perfect matchings such that they also include Fermi fields. It is however sufficient to specify their chiral field content, since it uniquely fixes the Fermi fields. A systematic method for determining the perfect matchings of BBMs, generalizing the Kasteleyn matrix approach of bipartite graphs, was introduced in \cite{Franco:2019bmx}. This method can be extended to general BFT$_2$'s, although we will not discuss it in this paper. 

Remarkably, there is an alternative definition due to Richard Kenyon, which is identical the one for perfect matchings of bipartite graphs underlying ordinary BFTs (see e.g. \cite{Franco:2012mm}).\footnote{We thank Richard Kenyon for private discussions leading to this insight. These conversations took place during a meeting of the NSF FRG in the Mathematical Sciences shared with Sebasti\'an Franco, and benefitted from ideas from the other members of the group: Gregg Musiker, David  Speyer and Lauren Williams.}\footnote{Interestingly, given the map between vertices and the {\it chiral cycles} introduced in \cite{Franco:2019bmx}, it is possible to show that this definition of perfect matchings coincides with the one given in \cite{Franco:2019bmx}.}

\smallskip

%=================================================================
\begin{center}
\fbox{\begin{minipage}{0.95 \textwidth}
\begin{center}
\begin{minipage}{0.92 \textwidth}
\noindent{\bf Definition 2.} A perfect $p$ is a subset of the chiral faces in $G$ such that:
\vspace{-.2cm}\begin{enumerate}
	\item Every internal vertex is covered exactly once by a chiral face in $p$
	\item Every external vertex belongs to either one or zero chiral faces in $p$.
\end{enumerate} 
\end{minipage}
\end{center}
\end{minipage}}
\end{center}
%=================================================================

\smallskip

Let us now show the equivalence between the two definitions. Consider a Fermi field $\Lambda_a$ and its corresponding $J$- and $E$-terms. Let us focus on a chiral field $Y$ contained in a perfect matching. Without loss of generality, following the first definition we can assume that $Y$ participates in the two monomials in $J^a$ (the case in which it participates in the two monomials in $E_a$ is identical). We can then write 
\beq
\label{brick matching in superpotential}
\begin{array}{ccc}
		J^a & \ \ \ \ \ \ & E_a\\
		X_{i_1}X_{i_2}\cdots \textcolor{blue}{Y}\cdots X_{i_{n-1}}X_{i_n}-X_{j_1}X_{j_2}\cdots \textcolor{blue}{Y}\cdots X_{j_{n-1}}X_{j_n} & &  E_{a+}(X)-E_{a-}(X)
\end{array}
\eeq

According to the dictionary, $\Lambda_a$ is translated into a 4-sided red face in $G$, in which the two monomials in $J^a$ and $E_a$ are represented by opposite edges. Moreover, as discussed in \sref{section_combinatorics_trace_condition}, each of the four vertices of $\Lambda_a$, which we call $\alpha$, $\beta$, $\gamma$ and $\delta$, gives rise to a contribution to the trace condition of the form \eref{trace_condition_vertex_0}, which in this case become
\beq
\label{brick matching in vertex}
\begin{array}{rrccl}
\alpha : & X_{i_1}X_{i_2}\cdots \textcolor{blue}{Y}\cdots X_{i_{n-1}}X_{i_n}\cdot E_{a+}(X) & - & X_{i_1}X_{i_2}\cdots \textcolor{blue}{Y}\cdots X_{i_{n-1}}X_{i_n}\cdot E_{a+}(X) & =0 \\[.05cm]
\beta : & -X_{i_1}X_{i_2}\cdots \textcolor{blue}{Y}\cdots X_{i_{n-1}}X_{i_n}\cdot E_{a-}(X) & + & X_{i_1}X_{i_2}\cdots \textcolor{blue}{Y}\cdots X_{i_{n-1}}X_{i_n}\cdot E_{a-}(X)& =0 \\[.05cm]
\gamma : & -X_{j_1}X_{j_2}\cdots \textcolor{blue}{Y}\cdots X_{j_{n-1}}X_{j_n}\cdot E_{a+}(X) & + & X_{j_1}X_{j_2}\cdots \textcolor{blue}{Y}\cdots X_{j_{n-1}}X_{j_n}\cdot E_{a+}(X)& =0 \\[.05cm]
\delta : & X_{j_1}X_{j_2}\cdots \textcolor{blue}{Y}\cdots X_{j_{n-1}}X_{j_n}\cdot E_{a-}(X) & - & X_{j_1}X_{j_2}\cdots \textcolor{blue}{Y}\cdots X_{j_{n-1}}X_{j_n}\cdot E_{a-}(X) & =0
\end{array}
\eeq
For each vertex, the first term comes from the Tr[$J\cdot E$] for $\Lambda_a$ in \eref{brick matching in superpotential}, while the second term comes from Tr[$J\cdot E$] products for other Fermi faces that intersect with $\Lambda_a$ on that vertex.

From \eref{brick matching in vertex}, we easily conclude that for the perfect matching $p$, each vertex is covered exactly once by the chiral field $Y$ in $p$, proving the equivalence between the two definitions. We present explicit examples of perfect matchings in \sref{section_BFT2s_and_toric_geometry} and \sref{section_additional_examples}.

We will consider the following map between chiral fields $X_i$ and perfect matchings $p_\mu$
\beq\label{map between chirals and brick matchings}
	X_i=\prod ^c_{\mu=1}p_{\mu}^{P_{i\mu}} \, ,
\eeq
where $c$ is the total number of perfect matchings, and $P_{i\mu}$ is equal to 1 if the face in $G$ associated to the chiral field $X_i$ is contained in $p_\mu$ and zero otherwise \cite{franco20152d}, i.e.
\beq
	P_{i\mu}=\left\{ 
    \begin{split}
    &1, ~X_i\in p_{\mu}\\
    &0, ~X_i\notin p_{\mu}
    \end{split}
    \right.
\label{P-matrix_definition}
\eeq
We will see in the coming sections that the {\it $P$-matrix} defined above plays an important role when computing the moduli space of the BFT$_2$'s, i.e. in connecting them to toric geometry.

%=================================================================
\subsection{Flows}
%=================================================================

\label{section_flows}

Another interesting concept for BFT$_2$'s is the one of {\it flows}, which are in one-to-one correspondence with perfect matchings. While perfect matchings are collections of faces in the CW complex, flows are oriented surfaces or, equivalently, collections of oriented edges giving rise to oriented paths in it. 

The flow $\mathfrak{p}_\mu$ associated to a perfect matching $p_\mu$ is defined as follows:
\begin{enumerate}
	\item Pick a reference perfect matching $p_{\text{ref}}$.
	\item For every perfect matching, orient chiral faces according to the corresponding bifundamental field. 
	\item For any perfect matching $p_\mu$, define the associated flow as $\mathfrak{p}_\mu=p_\mu-p_{\text{ref}}$, i.e. where the orientations of the chiral faces in $p_{\text{ref}}$ are reversed, and chiral faces present in both $p_\mu$ and $p_{\text{ref}}$ cancel each other.
\end{enumerate}
The surfaces resulting from this construction can be naturally completed by including the missing Fermis.

Alternatively, flows can be represented as oriented paths on $G$. This is simply achieved by orienting the edges of the chiral faces in a perfect matching according to the right-hand rule. The orientations of the edges in $p_{\text{ref}}$ are reversed when constructing $\mathfrak{p}_\mu=p_\mu-p_{\text{ref}}$.

Finally, we can also define the flows without referring to their graphical representation as follows:
\begin{enumerate}
	\item Pick a reference perfect matching $p_{\text{ref}}$.
	\item Associate to every perfect matching $p_\mu$ a variable $\tilde{p}_\mu$ which is the product of the chiral fields it contains. 
	\item The corresponding flow $\mathfrak{p}_\mu$ can then be written as 
	\beq 
	\label{flows from brick matching variables}
		\mathfrak{p}_\mu=\frac{\tilde{p}_\mu}{\tilde{p}_\text{ref}}.
	\eeq
 \end{enumerate}
In \sref{section_geometry_from_flows}, we will present explicit examples of this construction.

%=================================================================
\subsection{Brick Variables}
%=================================================================

Another useful concept is the notion of {\it brick variables}, which are the natural generalizations of the face variables of ordinary BFTs \cite{Franco:2012mm}. The definition is rather intuitive, they are given by the surface around a brick (gauge or global), with orientation. While it is natural to include Fermi fields in brick variables, for later applications it is sufficient to focus on their chiral field content. 

Focusing on the chiral fields, the brick variable $W_i$ associated to the brick $i$ is given by
\beq
W_i = {\prod_j X_{ij} \over \prod_j X_{ji} }
\label{bricks_in_terms_of_chirals}
\eeq
i.e. by the product of the outgoing chirals divided by the product of the incoming chirals.\footnote{Adjoint chiral fields $X_{ii}$ would appear in both the numerator and denominator, hence do not participate in brick variables.}

Not all the brick variables are independent, since they satisfy\footnote{Assuming a single component $M_3$.} 
\beq\label{product of brick variables}
	\prod_i^n W_i=1 \, .
\eeq

\begin{figure}[ht]
	\centering
	\includegraphics[width=13.5cm]{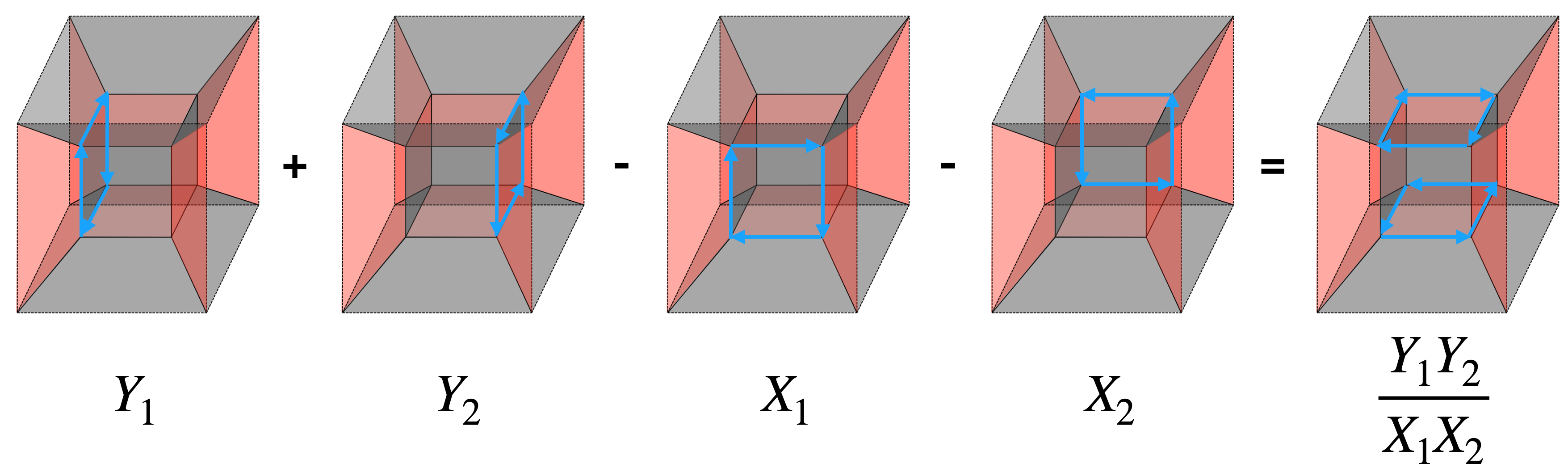}
	\caption{Example of the brick variable for the internal brick of the cube model in terms of oriented edges. When combining oriented faces, we will interchangeably talk about their addition/subtraction and multiplication/division depending on the context. We will use the former when referring to the combination of the faces in $G$ and the latter when discussing the corresponding variables.}
	\label{equivalence between two definitions of brick variables}
\end{figure}

Proceeding as in \sref{section_flows} to translate the orientations of chiral faces into the orientations of their edges, it is possible to represent brick variables in terms of oriented edges. \fref{equivalence between two definitions of brick variables} illustrates this construction. Further examples will be presented in \sref{section_geometry_from_flows}

%=================================================================
%=================================================================

%=================================================================
\section{Modifications of the CW Complex from a BFT$_2$ Perspective}
%=================================================================

\label{section_graph_modifications}

In this section we discuss the field theory interpretation of various possible modifications of the CW complex underlying a BFT$_2$. All this operations have been previously discussed in the context of BBMs \cite{Franco:2015tya}. We include them here for completeness and to further elaborate on some of them.

%=================================================================
\subsection{Higgsing}
%=================================================================

\begin{figure}[H]
	\centering
	\includegraphics[width=10cm]{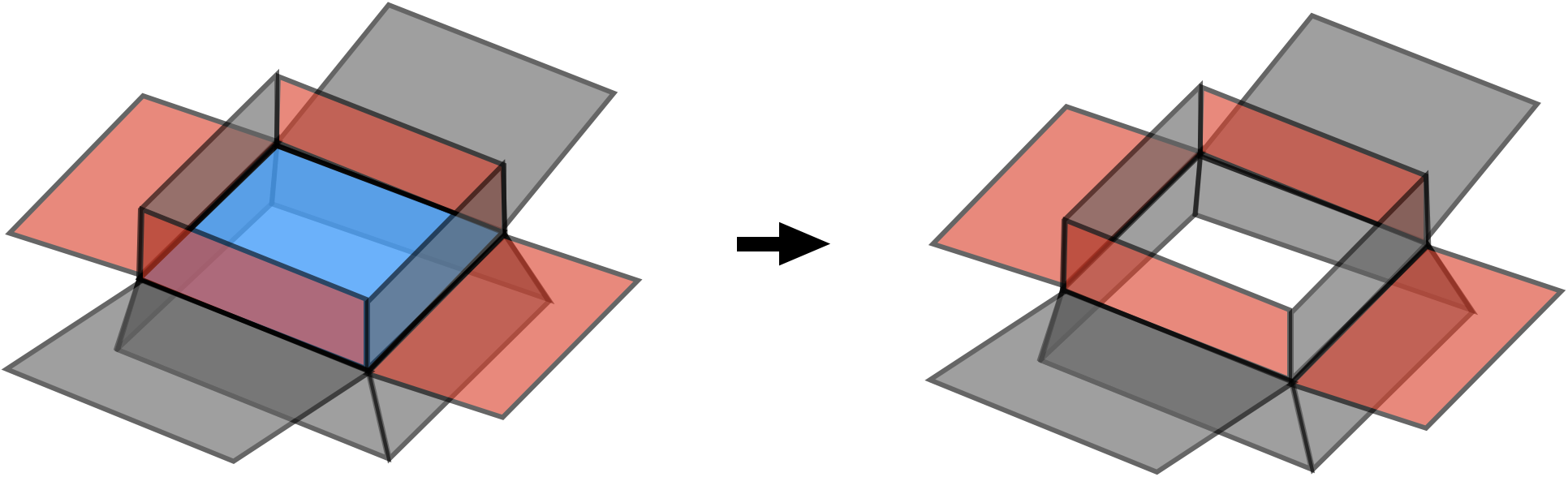}
	\caption{Giving a non-zero VEV to a chiral field translates into deleting the corresponding face, which is colored blue here. This results in the two adjacent bricks recombining into a single one.}
	\label{higgsing}
\end{figure}

Turning on a non-zero VEV for the scalar component of a bifundamental chiral field $X_{ij}$ translates into the removal of the corresponding face in $G$, which leads to the combination of the bricks $i$ and $j$ into a single brick, as illustrated in \fref{higgsing}. 

%=================================================================
%=================================================================

Depending on the types of bricks that are combined, there are three possibilities, which in turn have different gauge theory interpretations:\footnote{Keeping in mind the peculiarities of $2d$ gauge theories, here we have in mind the classical higgsing and symmetry breaking processes. They will become important when connecting these theories to geometry.}
\begin{itemize}
\item Internal-internal: higgsing of the corresponding $U(N) \times U(N)$ factor of the gauge group to the diagonal $U(N)$ subgroup.
\item Internal-external: color-flavor locking of the $U(N)_{\rm gauge} \times U(N)_{\rm global}$ symmetry associated to the bricks to the diagonal subgroup, which corresponds to a global symmetry.
\item External-external: breaking of a $U(N) \times U(N)$ factor of the global symmetry to the diagonal subgroup.
\end{itemize}
For simplicity, we will collectively refer to all of them as higgsing, while keeping in mind the fact that they may not always involve two gauge groups. In addition, Deleting the face for $X_{ij}$ nicely implements the effect of replacing it by its VEV, which without loss of generality is set to be equal to 1, in the superpotential.

%=================================================================
\subsection{Massive Fields}
%=================================================================

\label{section_massive_fields}
 
Massive fields correspond to Fermi-chiral pairs connecting the same pair of nodes in the quiver for which the corresponding quadratic term is present in the superpotential (equivalently, such that either the $J$- or $E$-term for the Fermi field contains a term that is linear in the chiral field). Such {\it mass terms} translate to edges in $G$ that connect a single Fermi face to a single chiral face.

%=================================================================
\paragraph{Integrating Out Massive Fields.}
%=================================================================

Without loss of generality, let us consider a massive pair consisting of fields $\Lambda_{ij}$ and $X_{ji}$, with a $J$-type mass term $\Lambda_{ij} X_{ji}$ in the superpotential. The case of an $E$-term mass is completely analogous. At low energies, $\Lambda_{ij}$ and $X_{ji}$ can be integrated out. The $J$-term associated to $\Lambda_{ij}$ takes the general form
\beq
J_{ji}: \ \ X_{ji} - f_{ji}(X) \ \ \ \longrightarrow \ \ \ X_{ji}=f_{ji}(X) \, .
\label{massive_J_term}
\eeq
where the linear term in $X_{ji}$ corresponds to the mass term and $f_{ji}(X)$ represents a product of chiral fields associated to a concatenation of arrows in the quiver going from node $i$ to node $j$. When integrating out the massive fields, we remove them from the theory, set $J_{ji}=0$ (therefore replacing $X_{ji} \to f_{ji}(X)$ as in \eref{massive_J_term}) and eliminate $E_{ij}$, the $E$-term for $\Lambda_{ij}$.

\begin{figure}[ht]
	\centering
	\includegraphics[width=8cm]{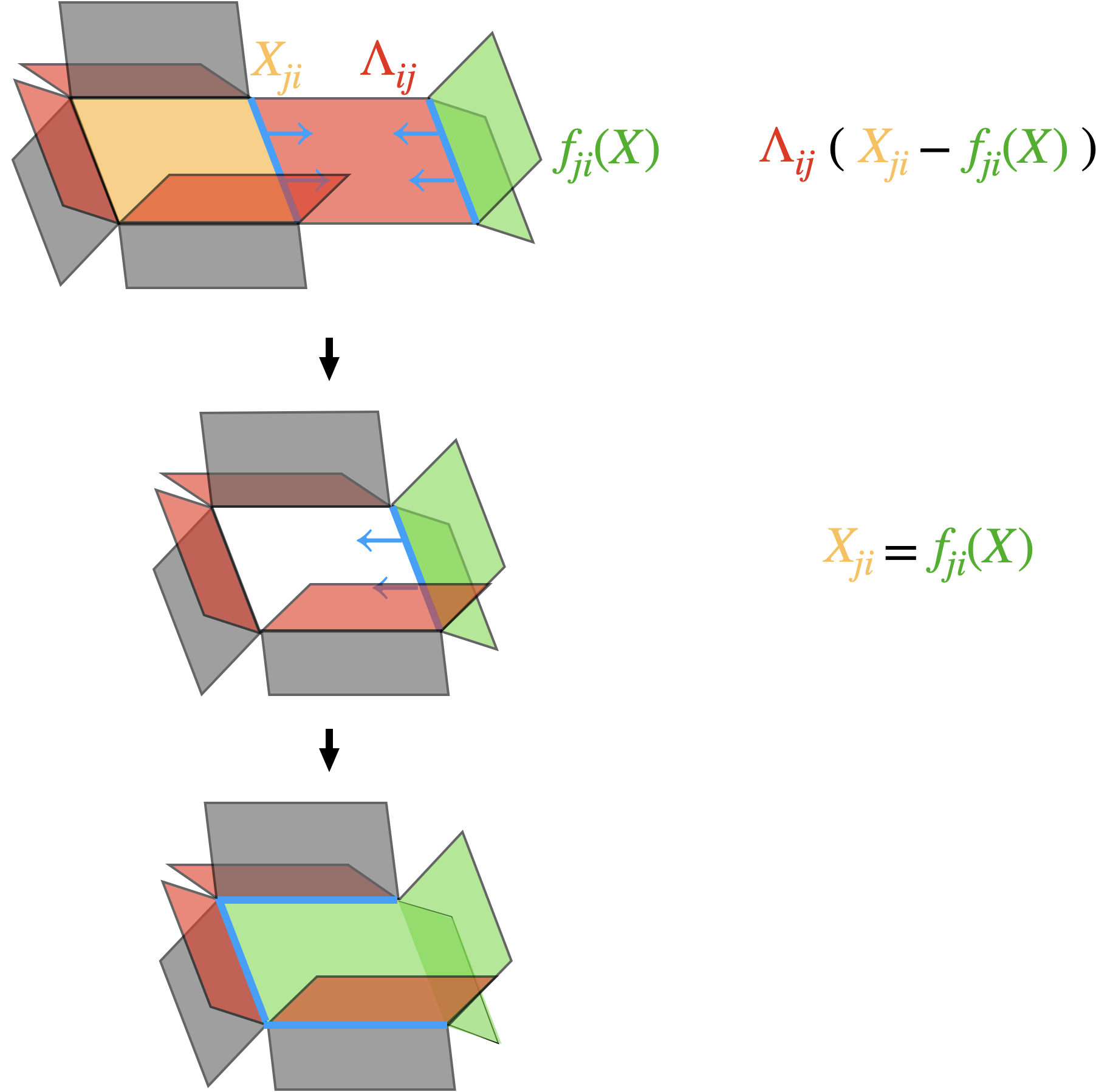}
	\caption{Integrating out massive fields in a BFT$_2$.}
	\label{integrating_out_massive}
\end{figure}

This process is beautifully captured by a simple transformation of $G$, as shown in \fref{integrating_out_massive}. As any other chirall field, $X_{ji}$ has an even number of edges, which we denote $2n$. \fref{integrating_out_massive} corresponds to $n=2$, but the picture and the discussion below extend to general $n$. For visualization, it is useful to divide the process into two stages. In the first step, we shrink the face associated to $\Lambda_{ij}$ until the massive edge merges with the opposite one into a single edge associated with $f_{ji}(X)$. This makes the Fermi face disappear. In the second step, the equation $X_{ji}=f_{ji}(X)$ removes the chiral face $X_{ji}$ and glues its edges to the one for $f_{ji}(X)$, as shown in the last step. 

%=================================================================
%=================================================================

Let us consider this process in further detail. Focusing for concreteness on the $n=2$ case as in \fref{integrating_out_massive}, the terms of the superpotential containing the massive fields $ \Lambda_{ij}$ and $X_{ji}$ have the general form
\beq
\begin{array}{ccl}
W & = & \Lambda_{ij} X_{ji} - \Lambda_{ij} f_{ji}(X)+\overline{\Lambda}_{ij} E^+_{ij}(X) -\overline{\Lambda}_{ij} E^-_{ij}(X) \\[.1cm]
 & + & \Lambda_1 X_{ji} g_1 (X) + \Lambda_2 X_{ji} g_2 (X) + \Lambda_3 X_{ji} g_3 (X) + \ldots \, ,
\end{array}
\label{W_massive_1}
\eeq
where $\Lambda_a$, $a=1,\ldots,2n-1$, are the Fermis connected to the extra edges of $X_{ji}$ and $g_a(X)$ are products of chiral fields. Without loss of generality, we have picked the orientation of the $\Lambda_a$'s such that their couplings to $X_{ij}$ are $J$-terms. 

In \fref{integrating_out_massive}, the chiral faces shown in green, which account for the fields in $f_{ji}(X)$, end up connected to $(2n-1)$ consecutive edges of the $2n$-sided massive $X_{ji}$ face, capturing the fact that the $X_{ji} \to f_{ji}(X)$ replacement occurs in $(2n-1)$ terms. The superpotential \eref{W_massive_1} becomes 
\beq
\begin{array}{ccl}
W & = & \Lambda_1 f_{ji}(X) g_1 (X) + \Lambda_2 f_{ji}(X) g_2 (X) + \Lambda_3 f_{ji}(X) g_3 (X) + \ldots
\end{array}
\eeq
It is interesting to note that for this to be possible in general, it might be necessary for some of the faces to be curved. This is not an issue for this paper, since we are exclusively concerned with combinatorial properties of BFT$_2$'s. Moreover, this feature is evaded if the $(2n-1)$ edges are collinear or if additional fields become simultaneously massive. The latter is indeed the case in several explicit examples previously studied (see e.g. \cite{Franco:2015tya,Franco:2016nwv}), where integrating out all massive fields leads to configurations without multiple chiral faces glued to $(2n-1)$ consecutive edges.

%=================================================================
\paragraph{Integrating In Massive Fields.}
%=================================================================

The inverse process, namely integrating in massive fields, is described by going through \fref{integrating_out_massive} in the opposite direction. This procedure has not been systematically discussed in the literature. Of course, there are in general multiple ways of introducing massive fields into a BFT$_2$. Doing so, we can reduce the order of superpotential terms.

In order for integrating in fields to be possible, a necessary condition is for some chiral operator $f_{ij}(X)$, which accounts for a single chiral field or a collection of chiral fields, to participate in $2n-1$ consecutive edges. Due to this condition, integrating in massive fields in BFT$_2$'s is more restrictive than the ordinary BFT counterpart. Therefore, it would be interesting to investigate whether it is possible to make a general statement regarding the maximal attainable reduction in the order of the superpotential. For example, in ordinary BFTs it is always possible to reduce the superpotential to a form with only order 2 and order 3 terms \cite{Franco:2012mm}. This question is beyond the scope of this paper.

\begin{figure}
	\centering
	\includegraphics[width=12cm]{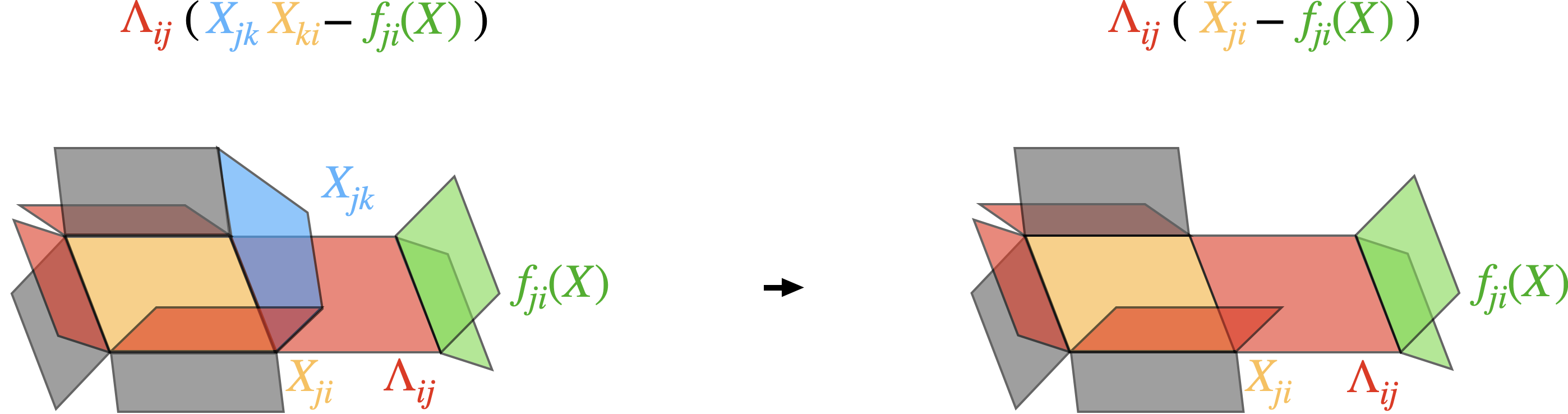}
	\caption{Generating a massive chiral-Fermi pair by higgsing.}
	\label{massive_from_higgsing_new}
\end{figure}

%=================================================================
\paragraph{Mass Terms from Higgsing.}
%=================================================================
Higgsing is one of various ways in which mass terms can be generated. In this case, a massive pair arises when an originally cubic term in the superpotential becomes quadratic after turning on a VEV for a chiral field. In terms of $G$, such a mass term corresponds to an edge which is initially attached to a Fermi and two chiral faces. When the face associated to the field acquiring the non-zero VEV is removed, a 2-valent edge representing a mass term is generated, as shown in \fref{massive_from_higgsing_new}. We refer the interested reader to \cite{Franco:2015tya} for further discussion of the generation of mass terms via higgsing in BFT$_2$'s.

%=================================================================
%=================================================================

%=================================================================
\subsection{Triality}
%=================================================================

Triality is an IR equivalence among $2d$ $\mathcal{N}=(0,2)$ gauge theories \cite{Gadde:2013lxa}. In this section we explain how, for the class of BFT$_2$ theories, triality is beautifully captured by certain transformations of $G$.

%=================================================================
\subsubsection{Triality for Quiver Theories}
%=================================================================

Let us first discuss triality for general quiver theories, i.e. not necessarily BFT$_2$'s. When a gauge node $k$ is dualized, the theory is transformed as described below.\footnote{We will often refer to the application of a triality transformation as {\it dualization}.}

%===================================================================
\paragraph{Ranks.} 
%===================================================================

The rank of the dualized node transforms according to
\beq
N'_k = \sum_{j\neq k} n_{jk}^\chi N_j - N_k \,, 
\label{rank-rule}
\eeq
where $n^\chi_{jk}$ is the number of chiral fields from node $j$ to node $k$. The ranks of all other nodes remain the same.

%===================================================================
\paragraph{Quiver.} 
%===================================================================
The field content around node $k$ is modified according to the following rules:
\begin{itemize}
\item[(Q1)]  {\bf Dual flavors.} Replace each of $(\rightarrow k)$, $(\leftarrow k)$, $(\textcolor{red}{\text{ --- }}  k)$ by  $(\leftarrow k)$, $( \textcolor{red}{\text{ --- }}  k)$, $(\rightarrow k)$, respectively. For later use, we generically refer to such fields in the original theory as $X_{ik}$, $Y_{kj}$, $\Lambda_{kl}$, and to the corresponding duals as $\tilde{Y}_{ki}$, $\tilde{\Lambda}_{kj}$, $\tilde{X}_{lk}$.
\item[(Q2)] {\bf Chiral mesons.} For every subquiver $i\rightarrow k \rightarrow j$, add a new chiral field $i\rightarrow j$. I.e. for every pair $X_{ik}$, $Y_{kj}$ in the original theory, include the meson $M_{ij}=X_{ik} Y_{kj}$ in the dual.
\item[(Q3)] {\bf Fermi mesons.} For every subquiver $i\rightarrow k \textcolor{red}{\text{ --- }} j$, add a new Fermi field $i \textcolor{red}{\text{ --- }}  j$. I.e. for every pair $X_{ik}$, $\Lambda_{kj}$ in the original theory, include the meson $\Psi_{ij}=X_{ik} \Lambda_{kj}$ in the dual.
\item[(Q4)] Remove chiral-Fermi massive pairs generated in the previous steps.\footnote{In order to determine which fields are massive, knowledge of the superpotential is required. The transformation of the superpotential is explained below. Since we are interested in the low energy limit of the theory, integrating out massive fields is natural, albeit optional.}
\end{itemize}

%===================================================================
\paragraph{Superpotential.} 
%===================================================================

The superpotential transforms as follows:
\begin{itemize}
\item[(W1)] Replace every $X_{ik} Y_{kj}$ product in the original superpotential by the corresponding meson $M_{ij}$. Similarly, replace all $X_{ik} \Lambda_{kj}$ products by $\Psi_{ij}$.

\item[(W2)] Introduce new cubic couplings between the dual flavors and mesons
\beq
		\Delta{W}_1= M_{ij} \overline{\tilde{\Lambda}}_{jk} \tilde{Y}_{ki} + \Psi_{ij} \tilde{X}_{jk} \tilde{Y}_{ki} \, .
\eeq

\item[(W3)] Replace every $\overline{\Lambda}_{lk}Y_{kj}$ product in the original superpotential by the corresponding product of dual flavors $\tilde{X}_{lk} \tilde{\Lambda}_{kj}$. Namely, transform terms containing such product as follows 
\beq
\overline{\Lambda}_{lk}Y_{kj} \mathcal{O}_{jl} \to \tilde{X}_{lk} \tilde{\Lambda}_{kj} \mathcal{O}_{jl} \, ,
\eeq
where $\mathcal{O}_{jl}$ is a holomorphic product of chiral fields connecting nodes $j$ and $l$. 

\item[(W4)] If the superpotential terms in (W3) are present, they also lead to new terms of the form
\beq
\overline{\Lambda}_{lk}Y_{kj} \mathcal{O}_{jl} \to \Delta{W}_2 = \overline{\Psi}_{l i}M_{ij} \mathcal{O}_{jl} \, .
\eeq
for every incoming chiral $X_{ik}$ at the dualized node. The existence of at least one chiral field $X_{ik}$, which is necessary for generating the mesons $\bar{\Psi}_{ik}$ and $M_{kj}$, does not need to be independently assumed in physically consistent theories. If this was not the case, other things would fail, e.g. the rank of the dual gauge group would be negative, according to \eref{rank-rule}.\footnote{Relaxing this condition might be interesting from a more formal/mathematical point of view.}

\end{itemize}
For a more detailed discussion of the transformation of the superpotential, see \cite{Franco:2017lpa}.

\bigskip

Triality is a duality of order 3, which means that acting with it three consecutive times on the same gauge node takes us back to the original theory. Interestingly, requiring that the rank of the dualized gauge group returns to itself in this process is equivalent to the $SU(N_k)^2$ anomaly cancellation condition for that node \cite{Franco:2017lpa}. Generically, trialities on different nodes do not commute, which leads to a much richer space of dual theories in the case of multiple gauge nodes.

%=================================================================
\subsubsection{Triality for BFT$_2$ Theories: Dual Quivers and CW Complexes}
%=================================================================

Any gauge node/internal brick in a BFT$_2$ can be dualized by applying the prescription outlined above. However, we will focus on dualizations of nodes which starting from a BFT$_2$ result in another BFT$_2$ on the same 3-manifold, i.e. in another theory described by a CW complex (or equivalently by its dual quiver) on the original 3-manifold.\footnote{Whether triality can sometimes lead to a theory of BFT$_2$ type but on a different 3-manifold is an interesting question that we will not explore in this paper.} We will refer to such nodes as {\it toric nodes}. The closely related problem of triality in BBMs has been studied at length in \cite{Franco:2016nwv,Franco:2016qxh}. Since triality is a local transformation, the discussion extends from BBMs to general BFT$_2$’s without changes.

%=================================================================
\paragraph{Toric nodes.}
%=================================================================
For a node $k$ to be toric, a necessary condition is that starting from a toric phase, in which all ranks are equal to $N$, all ranks remain the same after dualizing it. Specializing \eref{rank-rule} to a node $k$ in a toric phase, we get
\beq
N_k'=n_{k,in}^\chi N-N\, .
\eeq
We conclude that to obtain $N_k'=N$, we need $n_{k,in}^\chi=2$, i.e. toric nodes must have exactly two incoming chiral field arrows.

In the case of all equal ranks, the cancellation of the $SU(N)^2$ anomaly \eref{anomaly cancellation}, combined with $n_{k,in}^\chi=2$, implies that a toric node must have
\beq
n_{k,out}^\chi=n_k^F \, .
\eeq 
In summary, toric nodes have $n_{k,in}^\chi=2$ and $n_{k,out}^\chi=n_k^F \geq 2$. The lower bound is necessary to avoid SUSY breaking. Similarly, by reversing the orientation of chiral fields, we obtain toric nodes under the action of inverse triality, i.e. $n_{k,out}^\chi=2$ and $n_{k,in}^\chi=n_k^F \geq 2$.

\begin{figure}[H]
	\centering
	\includegraphics[width=9cm]{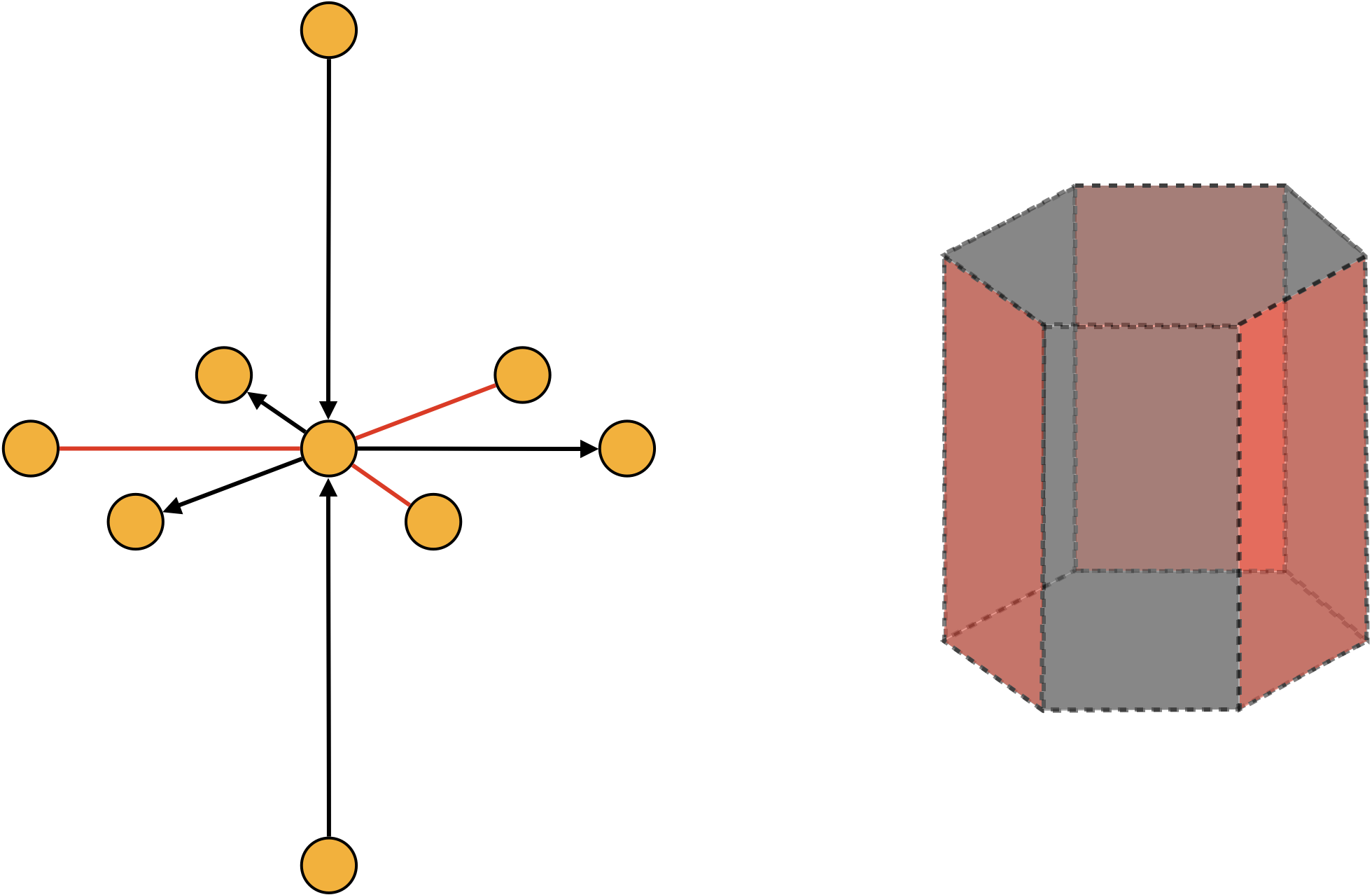}
	\caption{Local configuration for a toric node with $n_{k,out}^\chi=n_{k}^F=3$ and its dual cylindrical brick. The dotted lines may be attached to other faces and represent superpotential terms.}
	\label{cylinder_quiver}
\end{figure}

\begin{figure}[H]
	\centering
	\includegraphics[width=10cm]{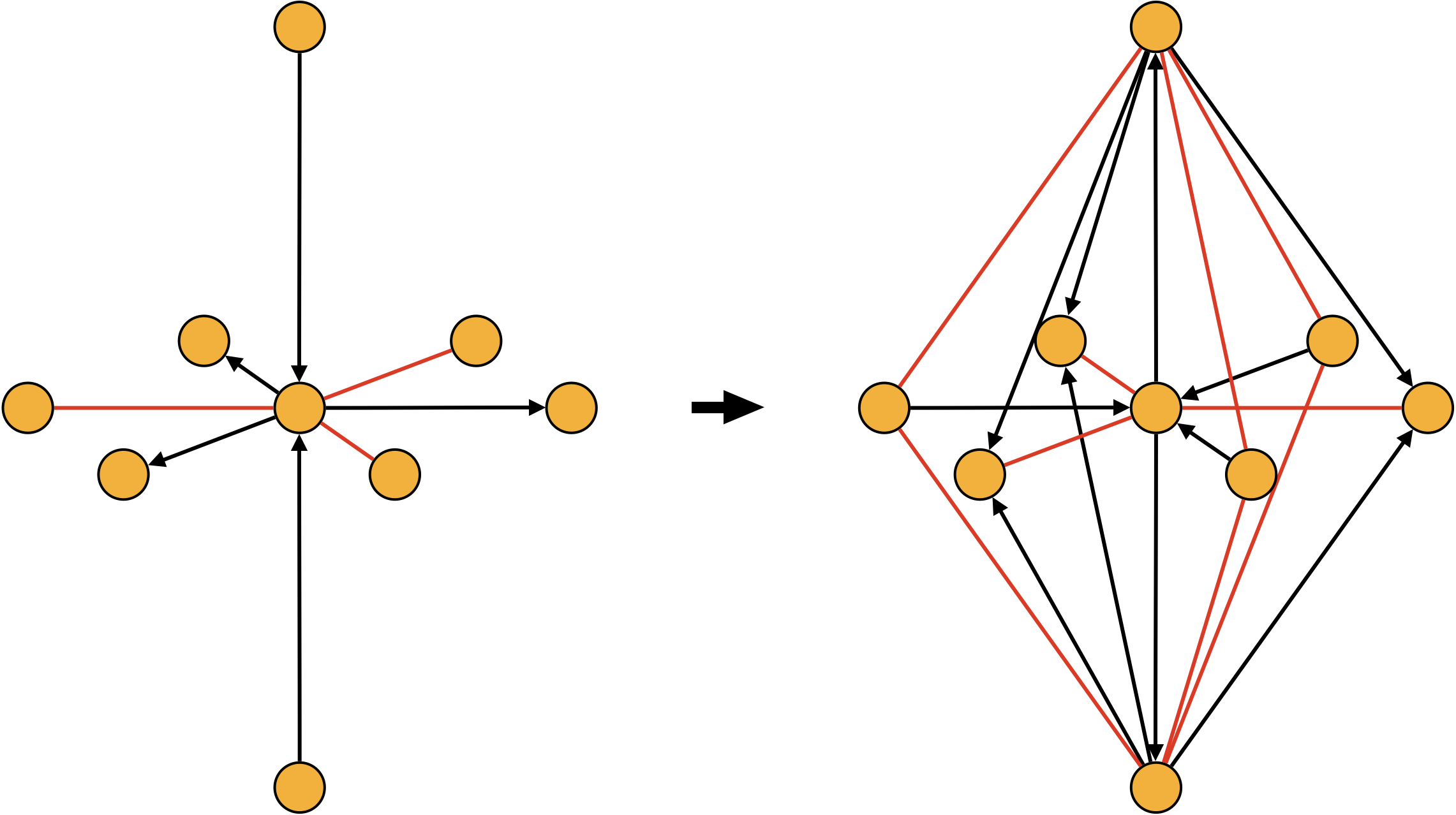}
	\caption{Local transformation of the quiver under triality for a node with $n^\chi_{k,out} = n^F_k = 3$. The initial configuration is the dual quiver of the cylindrical in \fref{cylinder_quiver}.}
	\label{basic_node_triality}
\end{figure}

Finally, for the dual theory to be a BFT$_2$ on the same 3-manifold, triality should not generate mesons that cross over the dualized node or interlaced loops of fields in the dual quiver. A natural solution to this problem is given by a local configuration in the dual quiver of the general form shown in \fref{basic_node_triality} for the case of $n_{k,out}^\chi = n_k^F =3$. Such a node is dual to a cylindrical brick with a $2 \, n_{k,out}^\chi$-sided base as illustrated in \fref{cylinder_quiver}.

%=================================================================
%=================================================================

The class of local transformations of the dual quiver illustrated in \fref{basic_node_triality} and the corresponding transformations $G$ implement the appropriate modifications of the superpotential.

%=================================================================
%=================================================================

\begin{figure}[H]
	\centering
	\includegraphics[width=9cm]{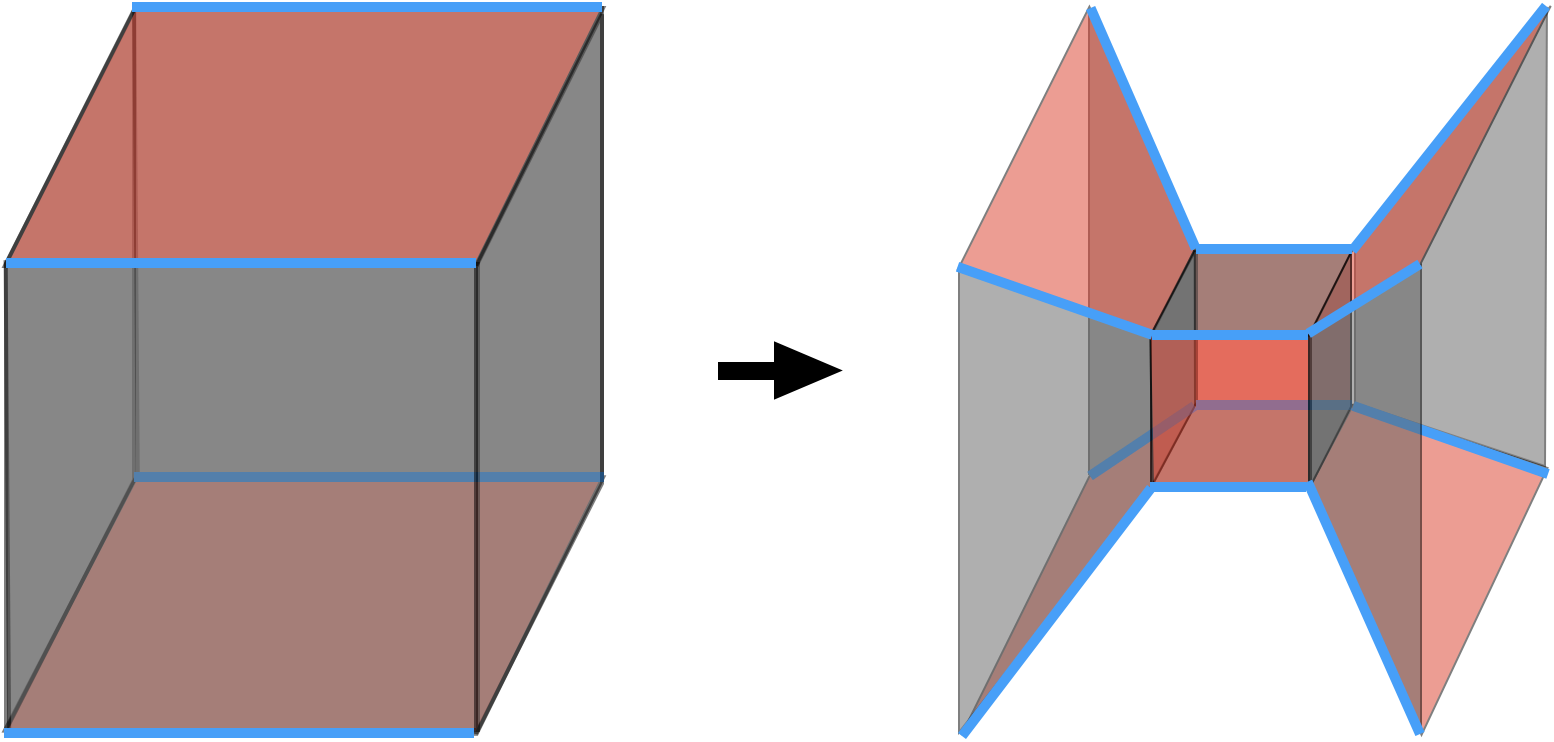}
	\caption{The cube move.}
	\label{basic_cube_move}
\end{figure}

It is interesting to consider the simplest version of a toric node, which corresponds to $n_{k,out}^\chi = n_k^F =2$, in further detail. \fref{basic_cube_move} shows the local transformation of $G$ associated to triality on such a node, to which we refer as a {\it cube move}. The edges of the cube have been colored to indicate how they deform under triality, together with the faces in $G$ outside of the cube that are glued to them.

%=================================================================
%=================================================================

The cube move is the analogue of the square move or urban renewal for ordinary BFTs (see \cite{Franco:2012mm} and references therein). As we explained in \sref{section_dictionary}, while Fermi faces are always 4-sided, chiral faces can in principle have an arbitrary number of even sides. For this reason, it is important to keep in mind that the chiral faces of the cube, which we have drawn as squares in \fref{basic_cube_move}, can in principle have more than four sides. The additional vertices we added to some of the edges are intended as a reminder of this fact. 

\begin{figure}[H]
	\centering
	\includegraphics[width=14.5cm]{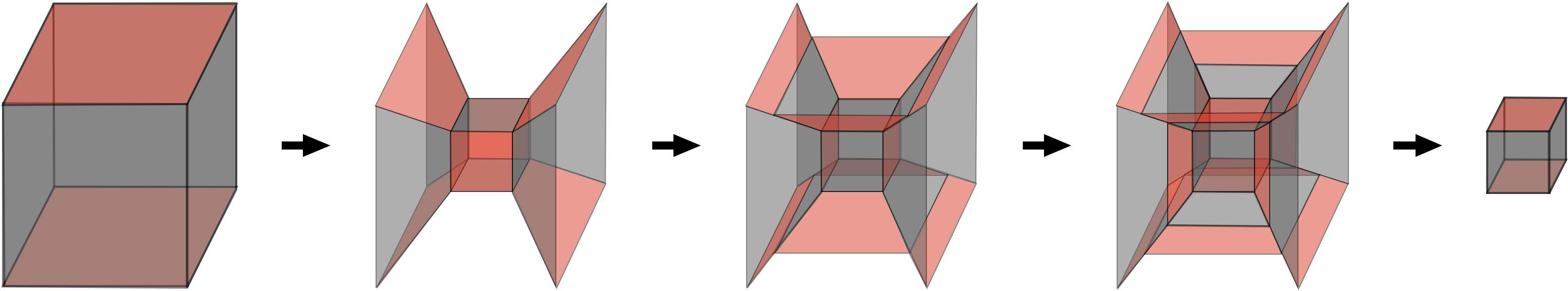}
	\caption{Three consecutive cube moves on a single cubic brick amount to the identity (the relative sizes of the cubes are irrelevant). Massive fields are preserved during triality transformation and are integrated out at the last step.}
	\label{three_consecutive_trialities}
\end{figure}

Cubic bricks are special in that they remain toric after triality. On the contrary, toric nodes with $n_{k,out}^\chi>2$ become non-toric after one dualization. \fref{three_consecutive_trialities} illustrates how three consecutive dualizations on a cubic brick amount to the identity. In this figure, we have decided to integrate out massive fields only at the final step.

%=================================================================
%=================================================================

%=================================================================
\subsection{Reduction}
%=================================================================

\label{section_reduction}

The final transformation is a BFT$_2$ analogue of {\it bubble reduction} \cite{Postnikov:2006kva,Arkani-Hamed:2016byb}. Such process was introduced in \cite{Franco:2018qsc}, by considering the field theory interpretation of BFT bubble reduction and finding its natural generalization.

In terms of the dual quiver, a BFT bubble corresponds to a node with one incoming and one outgoing bifundamental arrows. If all ranks are equal, this corresponds to an $N_f=N_c$ gauge group. Bubble reduction is therefore equivalent to formally applying Seiberg duality to such a node. When doing so, the dualized node disappears, there are no dual flavors and only the corresponding mesons are left. A BFT$_2$ bubble is therefore naturally defined as a gauge node/internal brick that would disappear when formally acting on it with triality or inverse triality.

\begin{figure}[ht]
	\centering
	\includegraphics[width=11cm]{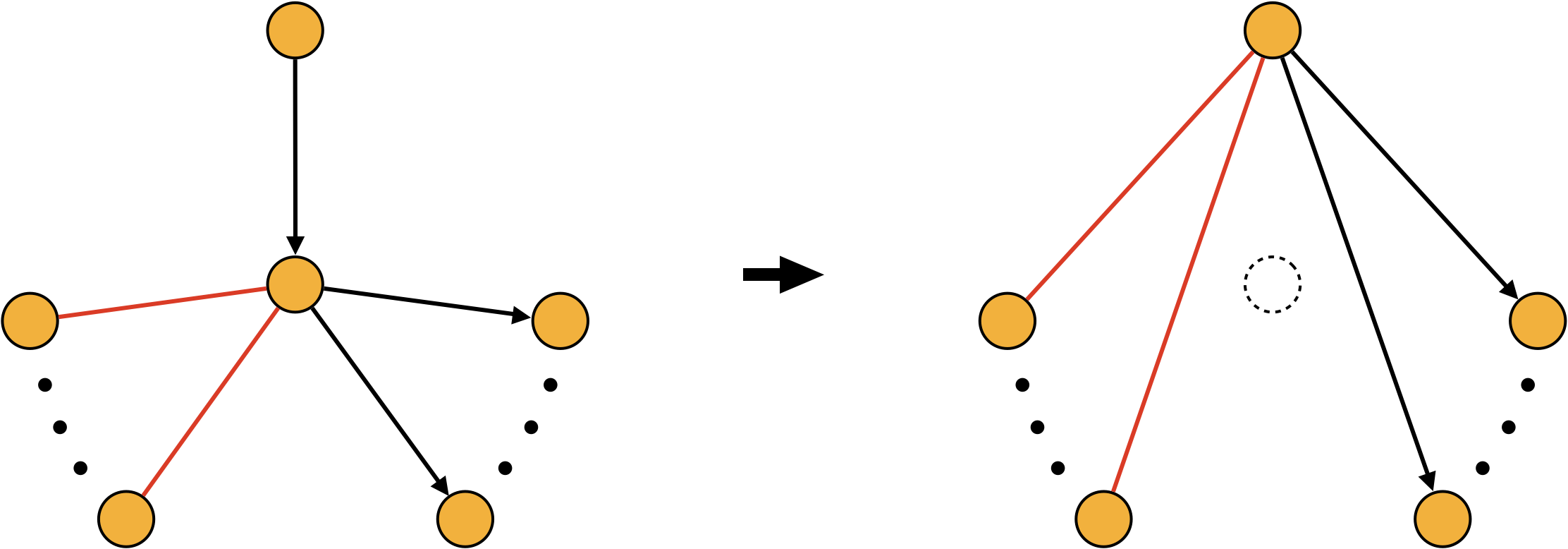}
\caption{A node with a single incoming chiral field is a BFT$_2$ bubble. Due to anomaly cancellation, $n^\chi_{k,out}= n^F_{k}+1$. The bubble is removed by triality. This process an also be interpreted as higgsing with a non-zero VEV for the incoming chiral.}
	\label{2d_bubble_reduction}
\end{figure}

More precisely, a bubble that disappears under triality corresponds to a node with a single incoming chiral arrow, i.e. $n^\chi_{k,in}=1$, and $n^\chi_{k,out}= n^F_{k}+1$. Similarly, a bubble that disappears under inverse triality corresponds to a node with $n^\chi_{k,out}=1$ and $n^\chi_{k,in}= n^F_{k}+1$. \fref{2d_bubble_reduction} shows a bubble and its elimination using triality. The node associated to the bubble disappears, while mesons connecting the other nodes are created. Interestingly, the same theory is obtained by higgsing with a VEV for the incoming chiral, which results in the merging of the bubble and the $in$ node. This is analogous to bubble reduction in BFTs, which can also be thought of as the higgsing associated to deleting one of the two edges of the bubble. The elimination of a bubble with inverse triality is completely analogous. Various explicit examples of bubble reduction will be presented in \sref{section_triality_and_moduli_space} and \sref{section_reduction_CY}.

%=================================================================
%=================================================================

%=================================================================
\paragraph{Reduction and Higgsing.}
%=================================================================

In \cite{Franco:2018qsc}, reducibility was related to the notion of inconsistency, which in the context of BBMs can be defined as the number of gauge groups (bricks) being larger than the normalized volume of the toric diagram of the underlying CY$_4$. 

As for ordinary BFTs, it is natural to expect that alternative definitions of reducibility exist. In particular, generalizing what occurs for BFTs \cite{Franco:2012mm}, it is natural to also define reducibility as the existence of a moduli space (to be discussed in \sref{section_BFT2s_and_toric_geometry}) preserving higgsing. In general, there might be multiple sets of chiral VEVs that reduce a given BFT$_2$. The identification of such VEVs can be systematized using the correspondence between perfect matchings and fields in the quiver of \eref{map between chirals and brick matchings} and \eref{P-matrix_definition}, in conjunction with geometric ideas that will be introduced in \sref{section_BFT2s_and_toric_geometry}. The previously mentioned equivalence between bubble reduction and higgsing is a case of this mechanism. However, explicit examples suggest that there can be BFT$_2$’s which are reducible by higgsing, but for which it is impossible to make bubbles explicit by sequences of trialities. It would be interesting to investigate whether the two definitions of reducibility are equivalent or if, instead, reduction by higgsing is more general.

Bubbles and higgsing might become manifest only after a series of trialities. General criteria for establishing the reducibility of BFT$_2$’s are desirable, but exceed the scope of this paper.

%=================================================================
\section{BFT$_2$'s and Toric Calabi-Yaus: Moduli Spaces}
%=================================================================

\label{section_BFT2s_and_toric_geometry}

A remarkable property of BFT$_2$’s is that they are intimately related to toric CYs, which arise as their {\it master} and {\it moduli spaces}. Moreover, there is a one-to-one correspondence between perfect matchings and gauged linear sigma model (GLSM) fields in the toric description of these geometries, which significantly simplifies this map.

\begin{figure}[ht]
	\centering
	\includegraphics[width=13cm]{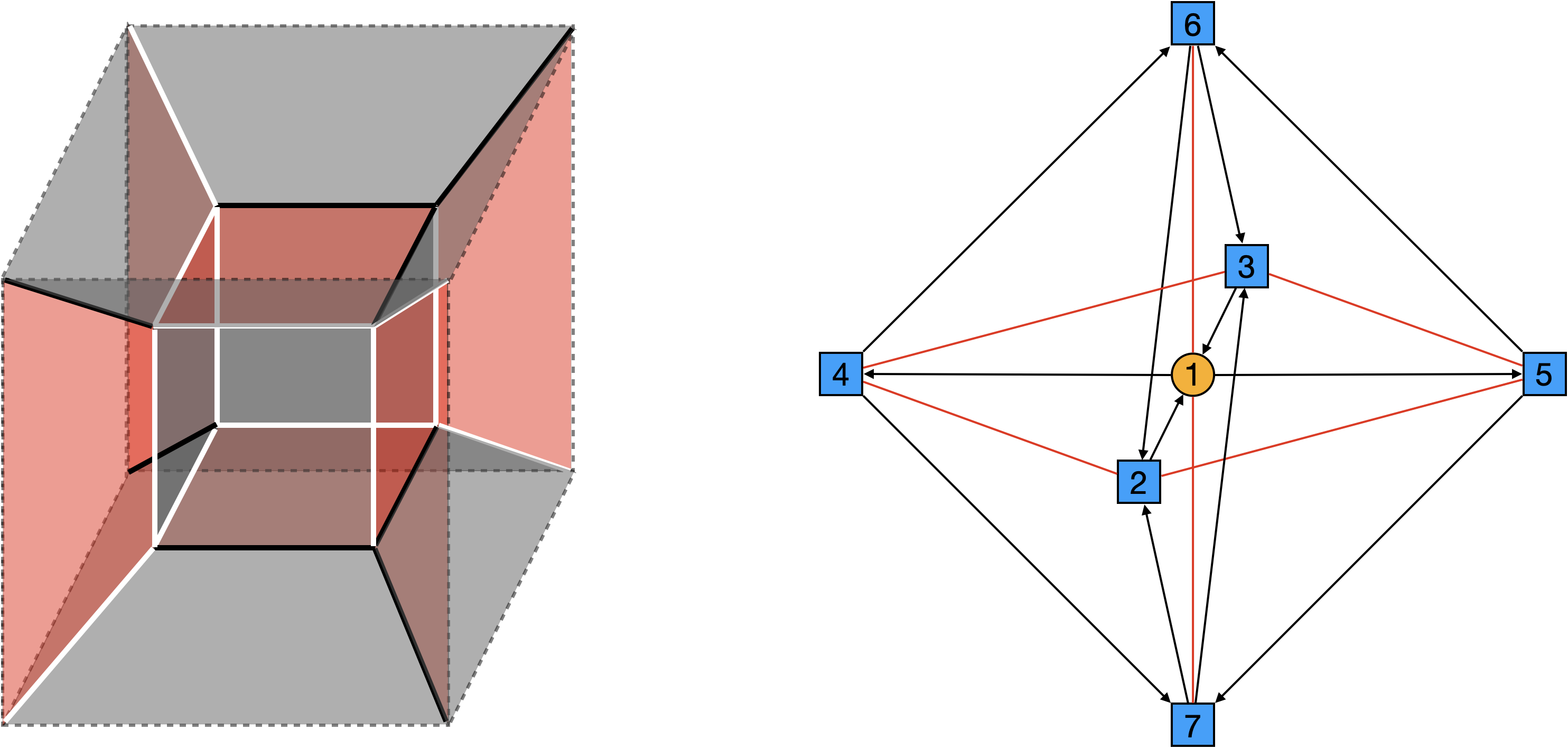}
	\caption{The cube model and its dual quiver.}
	\label{cube_model_quiver}
\end{figure}

We will illustrate the ideas introduced in this section with the simple cube model, which for convenience we reproduce in \fref{cube_model_quiver}, together with its dual quiver. 

%=================================================================
%=================================================================

The corresponding $J$- and $E$-terms are
\beq
\label{J and E-terms cube model}
\begin{array}{rcccc}
		& & J &\ \ \ \ & E\\
		\Lambda_{16}: & & X_{62}X_{21}-X_{63}X_{31} &&  X_{14}X_{46}-X_{15}X_{56}\\
		\Lambda_{17}: & & X_{73}X_{31}-X_{72}X_{21} &&  X_{14}X_{47}-X_{15}X_{57}\\
		\Lambda_{25}: & & X_{56}X_{62}-X_{57}X_{72} &&  \textcolor{myorange}{X_{21}X_{15}}\\
		\Lambda_{34}: & & X_{46}X_{63}-X_{47}X_{73} &&  \textcolor{myorange}{X_{31}X_{14}}\\
		\Lambda_{35}: & & X_{57}X_{73}-X_{56}X_{63} &&  \textcolor{myorange}{X_{31}X_{15}}\\
		\Lambda_{24}: & & X_{47}X_{72}-X_{46}X_{62} &&  \textcolor{myorange}{X_{21}X_{14}}
\end{array}
\eeq
where we have highlighted the single terms (in this case $E$-terms) associated to the unpaired edges of external Fermis in orange.

%=================================================================
\subsection{$J$- and $E$-flatness and Perfect Matchings}
%=================================================================

\label{section_J_and_E_flatness_PMs}

The map between chiral fields and perfect matchings introduced in \eref{map between chirals and brick matchings} is such that $J$- and $E$-term equations are automatically satisfied, as we now review. This fact was originally established for BBMs in \cite{Franco:2015tya}.

Given an internal Fermi $\Lambda_a$, the toric condition on the superpotential \eref{toric_superpotential} implies that the $J$- and $E$-term equations take the form
\beq
\begin{array}{lcccc}
J^a=0 & \ \ \ \Leftrightarrow \ \ \ & J^a_+ = J^a_-  \\
E_a=0 & \ \ \ \Leftrightarrow \ \ \ & E_{a+} = E_{a-}
\end{array}
\eeq

\begin{figure}[H]
	\centering
	\includegraphics[width=11cm]{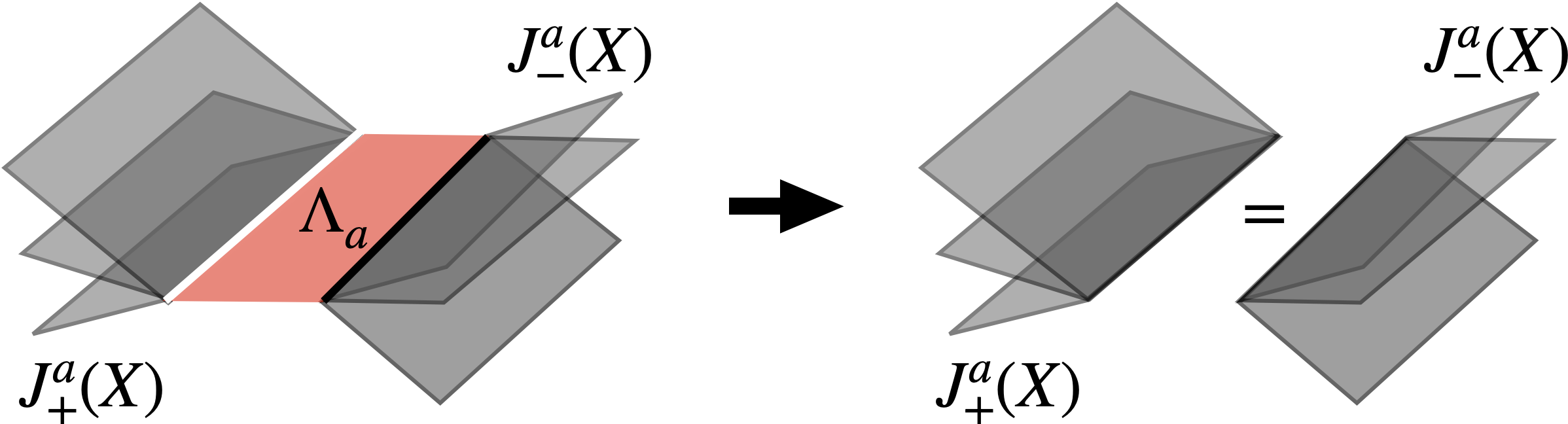}
	\caption{Graphical representation of the $J$-term equations for a Fermi. The $E$-term equations are completely analogous.}
	\label{graph representation of J-flatness}
\end{figure}

Each of these two equations has a simple graphical interpretation as shown in \fref{graph representation of J-flatness}. The product of the chiral fields attached to an edge of a Fermi must be equal to the product of the chiral fields attached to the opposite edge.

%=================================================================
%=================================================================

Using the map between chiral fields and perfect matchings in \eref{map between chirals and brick matchings}, these equations become
\beq
\begin{array}{lcccc}
J^a=0 & \ \ \ \Leftrightarrow \ \ \ & {\displaystyle \prod_{X_i\in J^a_+}\prod_\mu  p_\mu^{P_{i \mu} }}  & = & {\displaystyle \prod_{X_i\in J^a_-}\prod_\mu  p_\mu^{P_{i \mu}}} \\[.6cm]
E_a=0 & \ \ \ \Leftrightarrow \ \ \ & {\displaystyle \prod_{X_i\in E_{a+}}\prod_\mu  p_\mu^{P_{i \mu} }} & = & {\displaystyle \prod_{X_i\in E_{a-}}\prod_\mu  p_\mu^{P_{i \mu}}} 
\end{array}
\label{J_E_pms}
\eeq
From the definition of perfect matchings, it is clear that these equations are automatically satisfied, since every perfect matching appearing on the L.H.S. of one of these equations, also appears on its R.H.S. This is particularly manifest in Definition 1 from \sref{section_perfect_matchings}.

%=================================================================
\paragraph{External Fermis.}
%=================================================================

For external Fermis, we will only impose the vanishing of the equation associated to the two non-external opposite edges. Requiring that the single terms associated to the unpaired edges (e.g. the orange terms in \eref{J and E-terms cube model} for the cube model) vanish would eliminate the entire moduli space. In the coming sections, we will see that this prescription leads to interesting geometries as the BFT$_2$ moduli spaces. Physically, we can regard this choice as attaching a single auxiliary chiral field (or several of them) to every external edge, which takes the necessary VEV to compensate the product of the chiral fields at the opposite edge. Finally, we note that imposing, albeit partially, the vanishing of $J$- or $E$-terms for external Fermis is in line with interpreting them as dynamical fields, as discussed in \sref{section_dictionary}. 

Repeating the reasoning above, but with only one of the two equations in \eref{J_E_pms}, we conclude that the map \eref{map between chirals and brick matchings} solves all $J$- and $E$-term equations in the theory.

%=================================================================
\subsection{Master Space}
%=================================================================

\label{section_master_space}

The first interesting geometry associated to a BFT$_2$ is its {\it master space}, which we define as the space of solutions to $J$- and $E$-term equations. The concept of master space was first introduced in the context of $4d$ $\mathcal{N}=1$ gauge theories, where it corresponds to the space of solutions of $F$-term equations \cite{Forcella:2008bb}. It was later extended to $2d$ $\mathcal{N}=(0,2)$ gauge theories in \cite{franco20152d}. Since the vanishing of $D$-terms is not imposed, the master space can be regarded as the full moduli space including baryonic directions.

\begin{figure}[ht]
	\centering
	\includegraphics[width=13cm]{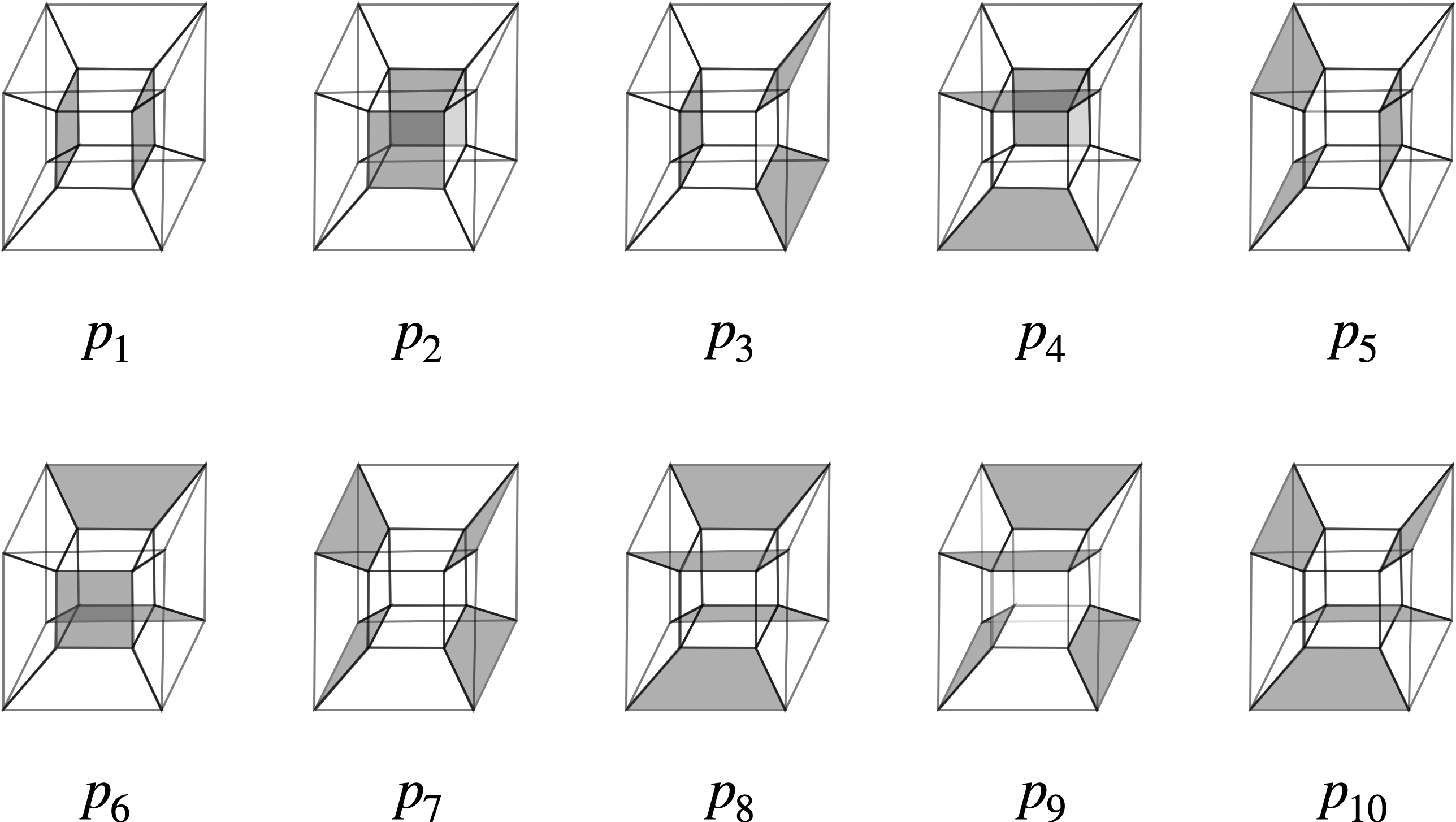}
	\caption{The ten perfect matchings for the cube model. For clarity, we have only colored the chiral faces in them.}
	\label{pms_cube_model}
\end{figure}

Following the previous discussion, the master space of a BFT$_2$ can be efficiently parameterized in terms of perfect matchings. The cube model contains ten perfect matchings, which are shown in \fref{pms_cube_model}.

%=================================================================
%=================================================================

The $P$-matrix \eref{P-matrix_definition} encoding the map between chiral fields and the perfect matchings in \fref{pms_cube_model} is 
\beq
P=\left(
\begin{array}{c|cccccccccc}
& p_1 & p_2 & p_3 & p_4 & p_5 & p_6 & p_7 & p_8 & p_9 & p_{10} \\ \hline
X_{14} \ & 1 & 0 & 1 & 0 & 0 & 0 & 0 & 0 & 0 & 0 \\
X_{15} \ & 1 & 0 & 0 & 0 & 1 & 0 & 0 & 0 & 0 & 0 \\
X_{21} \ & 0 & 1 & 0 & 0 & 0 & 1 & 0 & 0 & 0 & 0 \\
X_{31} \ & 0 & 1 & 0 & 1 & 0 & 0 & 0 & 0 & 0 & 0 \\
X_{46} \ & 0 & 0 & 0 & 0 & 1 & 0 & 1 & 0 & 0 & 1 \\
X_{47} \ & 0 & 0 & 0 & 0 & 1 & 0 & 1 & 0 & 1 & 0 \\
X_{56} \ & 0 & 0 & 1 & 0 & 0 & 0 & 1 & 0 & 0 & 1 \\
X_{57} \ & 0 & 0 & 1 & 0 & 0 & 0 & 1 & 0 & 1 & 0 \\
X_{62} \ & 0 & 0 & 0 & 1 & 0 & 0 & 0 & 1 & 1 & 0 \\
X_{63} \ & 0 & 0 & 0 & 0 & 0 & 1 & 0 & 1 & 1 & 0 \\
X_{72} \ & 0 & 0 & 0 & 1 & 0 & 0 & 0 & 1 & 0 & 1 \\
X_{73} \ & 0 & 0 & 0 & 0 & 0 & 1 & 0 & 1 & 0 & 1 \\
\end{array}
\right)
\label{P-matrix_cube_model}
\eeq

The master space of a BFT$_2$ is toric, i.e. it can be described by a GLSM. In addition, there is a one-to-one correspondence between perfect matchings and GLSM fields. In this language, the $J$- and $E$-term equations are translated into $U(1)$ charges for the perfect matchings, which can be summarized in a charge matrix $Q_{JE}$ defined as follows
\beq
	Q_{JE}=Ker \, P \, .
\eeq

From \eref{P-matrix_cube_model}, we get for the cube model that
\beq
Q_{JE} = \left(
\begin{array}{cccccccccc}
p_1 & p_2 & p_3 & p_4 & p_5 & p_6 & p_7 & p_8 & p_9 & p_{10} \\ \hline
 1 & 1 & -1 & -1 & -1 & -1 & 0 & 0 & 1 & 1 \\
 0 & 1 & 0 & -1 & 0 & -1 & 0 & 1 & 0 & 0 \\
 1 & 0 & -1 & 0 & -1 & 0 & 1 & 0 & 0 & 0 \\
\end{array}
\right)
\eeq

The toric diagram of the master space is given by $Ker \, Q_{JE}$, which is actually $P$. Every column of the $P$-matrix is the position of the corresponding perfect matching in the toric diagram. Moreover, every perfect matching is mapped to a different point. Given the analogous construction for plabic graphs \cite{MR2525057} and its extension to general BFTs \cite{Franco:2012mm,Franco:2012wv,Franco:2014nca,Franco:2013nwa}, it is natural to also refer to the toric diagram of the master space as the {\it matching polytope}. 

For a simpler visualization of the geometry of the toric diagram, it is convenient to consider the row-reduced version of $P$. For \eref{P-matrix_cube_model}, it becomes
\beq
G_{mast}=\left(
\begin{array}{cccccccccc}
p_1 & p_2 & p_3 & p_4 & p_5 & p_6 & p_7 & p_8 & p_9 & p_{10} \\ \hline
 1 & 0 & 0 & 0 & 0 & 0 & -1 & 0 & 0 & -1 \\
 0 & 1 & 0 & 0 & 0 & 0 & 0 & -1 & 0 & -1 \\
 0 & 0 & 1 & 0 & 0 & 0 & 1 & 0 & 0 & 1 \\
 0 & 0 & 0 & 1 & 0 & 0 & 0 & 1 & 0 & 1 \\
 0 & 0 & 0 & 0 & 1 & 0 & 1 & 0 & 0 & 1 \\
 0 & 0 & 0 & 0 & 0 & 1 & 0 & 1 & 0 & 1 \\
 0 & 0 & 0 & 0 & 0 & 0 & 0 & 0 & 1 & -1
\end{array}
\right)
\label{G_mast_cube}
\eeq
The master space is thus 7-complex dimensional, with a toric diagram consisting of ten different points. The entries in every column of $G_{mast}$ add up to 1, which furthermore implies that the master space is Calabi-Yau. Remarkably, the master spaces of all theories considered in this paper satisfy the CY condition. It is natural to expect that this is a general property of all BFT$_2$’s. Given that the toric diagram lives on a 6-dimensional hyperplane at distance 1 from the origin, it is sufficient to consider a projection onto it. The facts that the master space is $7d$ and that its toric diagram can be projected to $6d$ are not coincidences. As we will see in \sref{section_geometry_from_flows}, these features were expected, since the cube model lives on a 3-ball and contains seven bricks, six of which are independent. This is analogous to what occurs for ordinary BFTs \cite{Franco:2012mm}.

%=================================================================
\subsection{Mesonic Moduli Space}
%=================================================================

\label{section_moduli_space}

The second geometry naturally associated to a BFT$_2$ is its {\it mesonic moduli space}. For brevity, we will often refer to it just as the {\it moduli space}. For any $2d$ $\mathcal{N}=(0,2)$ gauge theory, the moduli space is the space of solutions to vanishing $J$-, $E$- and $D$-terms. Therefore, it corresponds to projecting the master space onto the subspace of vanishing $D$-terms.

There is a D-term for every gauge group in the BFT$_2$, i.e. for every internal brick. The transformation properties of the chiral fields under the gauge symmetry can be summarized in the charge matrix $\Delta$.\footnote{In this paper, we focus on the simple case of $U(1)$ gauge groups when computing master and moduli spaces.} This matrix has a row for every gauge group and a column for every chiral field.  The non-zero entries on the column associated to a field $X_{ij}$ are a $+1$ on the $i^{th}$ row and a $-1$ on the $j^{th}$ row. All entries in columns for adjoint fields $X_{ii}$ vanish. For the cube model, there is a single internal brick, which corresponds to node 1 in the dual quiver shown in \fref{cube_model_quiver}, and therefore $\Delta$ has a single row. It is given by
\beq
\Delta= \left(\begin{array} {cccccccccccc}
X_{14} & X_{15} & X_{21} & X_{31} & X_{46} & X_{47} & X_{56} & X_{57} & X_{62} & X_{63} & X_{72} & X_{73} \\ \hline
1 & 1 & -1 & -1 & 0 & 0 & 0 & 0 & 0 & 0 & 0 & 0
\end{array}\right) \, .
\eeq

Given the map between chiral fields and perfect matchings, we can regard the charges of chiral fields as resulting from charges of perfect matchings. Such charges are encoded in a matrix $Q_D$, which satisfies
\beq
\Delta=Q_{D} \cdot P^T \, .
\label{equation_QD}
\eeq
It is interesting to note that this equation does not uniquely fix $Q_{D}$. However, different solutions are equivalent for computing the moduli space. For the cube model, a possible choice is
\beq
Q_{D} = \left( \begin{array}{cccccccccc}
p_1 & p_2 & p_3 & p_4 & p_5 & p_6 & p_7 & p_8 & p_9 & p_{10} \\ \hline
1 & -1 & 0 & 0 & 0 & 0 & 0 & 0 & 0 & 0
\end{array} \right) \, .
\eeq

The next step is to combine $Q_{JE}$ and $Q_D$ into a single charge matrix $Q$
\beq
	Q=\begin{pmatrix}
		Q_{JE}\\
		Q_{D}
	\end{pmatrix} \,.
\eeq
The toric diagram of the moduli space is thus encoded in a matrix $G$ defined as
\beq
G=Ker \, Q \, .
\eeq

For the cube model, we get
\beq
G=\left(
\begin{array}{cccccccccc}
p_1 & p_2 & p_3 & p_4 & p_5 & p_6 & p_7 & p_8 & p_9 & p_{10} \\ \hline
 0 & 0 & 1 & 0 & 0 & 0 & 1 & 0 & 0 & 1 \\
 0 & 0 & 1 & 0 & 0 & 0 & 1 & 0 & 1 & 0 \\
 -1 & -1 & -2 & 0 & 0 & 0 & -1 & 1 & 0 & 0 \\
 1 & 1 & 1 & 0 & 0 & 1 & 0 & 0 & 0 & 0 \\
 0 & 0 & -1 & 0 & 1 & 0 & 0 & 0 & 0 & 0 \\
 1 & 1 & 1 & 1 & 0 & 0 & 0 & 0 & 0 & 0 \\
\end{array}
\right)
\label{G_cube_model}
\eeq
We conclude that the moduli space is a $6d$ toric variety. Similar to the master space, the toric diagram lies on a $5d$ hyperplane at distance 1 from the origin. This implies that the moduli space is a toric CY 6-fold. As we will explain in \sref{section_geometry_from_flows}, the fact that the moduli space is $6d$ with a $5d$ toric diagram follows from the cube model having six external bricks, five of which are actually independent. From \eref{G_cube_model}, we see that the toric diagram has nine distinct points and, interestingly, the perfect matchings $p_1$ and $p_2$ are mapped to the same point. This feature will also be addressed in this coming section from the perspective of flows. As we will illustrate in several examples, having non-trivial multiplicities of perfect matchings for points in the toric diagram, including extremal ones, is a generic property of BFT$_2$'s. Finally, it is interesting to mention that the toric diagram of the moduli space is the natural generalization of the {\it matroid polytope}, which is normally studied in the context of plabic graphs and BFTs \cite{MR2525057,Franco:2012mm,Franco:2012wv,Franco:2014nca,Franco:2013nwa}, to BFT$_2$'s. Based on the evidence provided by all the examples in this paper, it seems reasonable to expect that the moduli spaces of all BFT$_2$'s satisfy the CY condition. We therefore conjecture that

\medskip

%=================================================================
\begin{center}
\fbox{\begin{minipage}{0.95 \textwidth}
\begin{center}
\begin{minipage}{0.92 \textwidth}
\begin{center}
{\bf Conjecture} \\
The master and moduli spaces of every BFT$_2$ are toric Calabi-Yaus.
\end{center}
\end{minipage}
\end{center}
\end{minipage}}
\end{center}
%=================================================================

\medskip

\noindent We expect the existence of a simple proof of this statement based on the combinatorics of $Q_{JE}$ and $Q_D$.

\begin{figure}[ht]
	\centering
	\includegraphics[width=9cm]{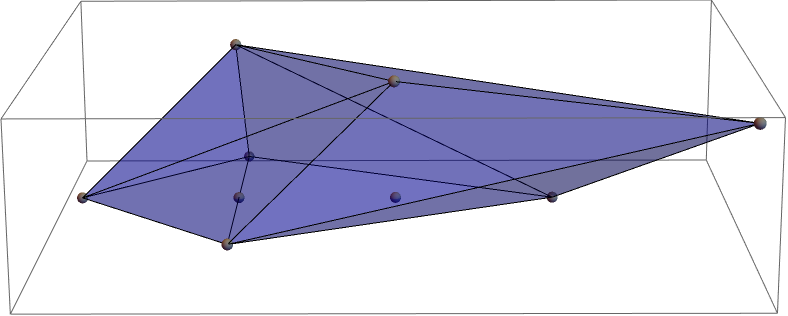}
	\caption{A $3d$ projection of the toric diagram for the moduli space, corresponding to \eref{G_cube_model}. The projection corresponds to the following combination of rows: $(G_1, G_3 - G_4, G_2 - G_5)$.}
	\label{toric_diagram_cube_model}
\end{figure}

Given its high dimensionality, visualizing the toric diagram defined by \eref{G_cube_model} is rather challenging. \fref{toric_diagram_cube_model} shows a $3d$ projection of it. While not particularly illuminating, this projection is special in that it does not make different points coincide.

%=================================================================
%=================================================================

The procedure introduced in \sref{section_master_space} and \sref{section_moduli_space} for the computation of the moduli space applies to arbitrary BFT$_2$ theories, regardless of the underlying 3-manifold. It is the generalization of the {\it fast forward algorithm} to BFT$_2$’s . We refer the interested reader to e.g. \cite{Franco:2005rj,Franco:2012mm,Franco:2015tya} for previous discussions of this algorithm in various contexts, including brane tilings, BFTs and BBMs.

%=================================================================
\section{Geometry from Flows}
%=================================================================

\label{section_geometry_from_flows}

In this section we introduce an alternative method for deriving the master and moduli spaces of BFT$_2$’s. More specifically, we will use flows and brick variables to map perfect matchings to points in the corresponding toric diagrams. Rather than presenting a lengthy discussion, we will simply illustrate the construction for the cube model.

%=================================================================
\subsection{Flows and Brick Variables}
%=================================================================

\begin{figure}[ht]
	\centering
	\includegraphics[width=9.5cm]{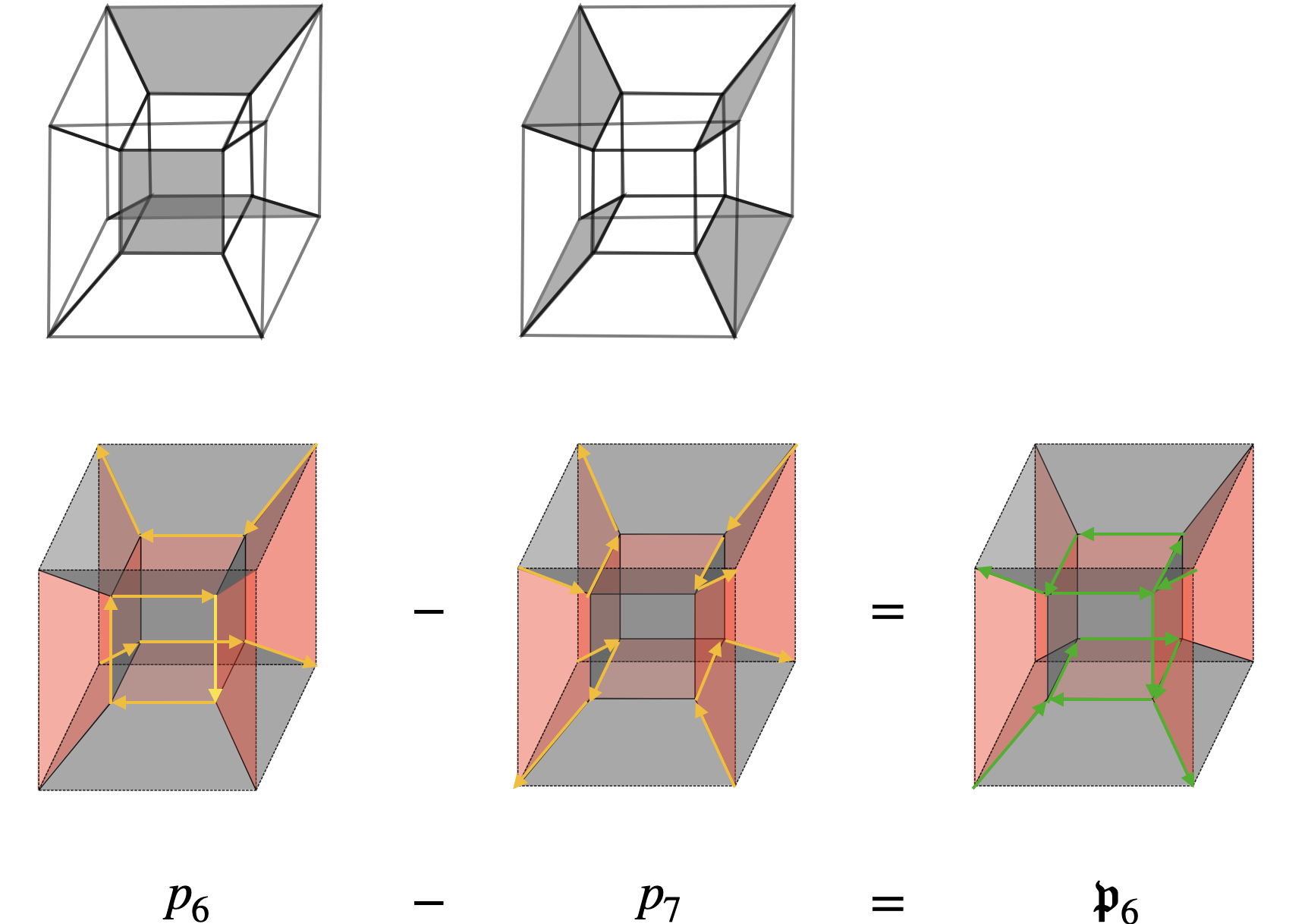}
	\caption{Flow $\mathfrak{p}_6$ associated to the perfect matching $p_6$, using  $p_7$ as the reference perfect matching.}	\label{flow_from_brick_matching_example}
\end{figure}

As explained in \sref{section_flows}, we can associate a flow $\mathfrak{p}_\mu$ to every perfect matching $p_\mu$, after picking a reference perfect matching. \fref{flow_from_brick_matching_example} shows the flow for $p_6$ in the cube model, using $p_7$ as reference. As previously mentioned, flows can be equivalently regarded as oriented surfaces or as collections of oriented edges. 

%=================================================================
%=================================================================

\begin{figure}[ht]
	\centering
	\includegraphics[width=15cm]{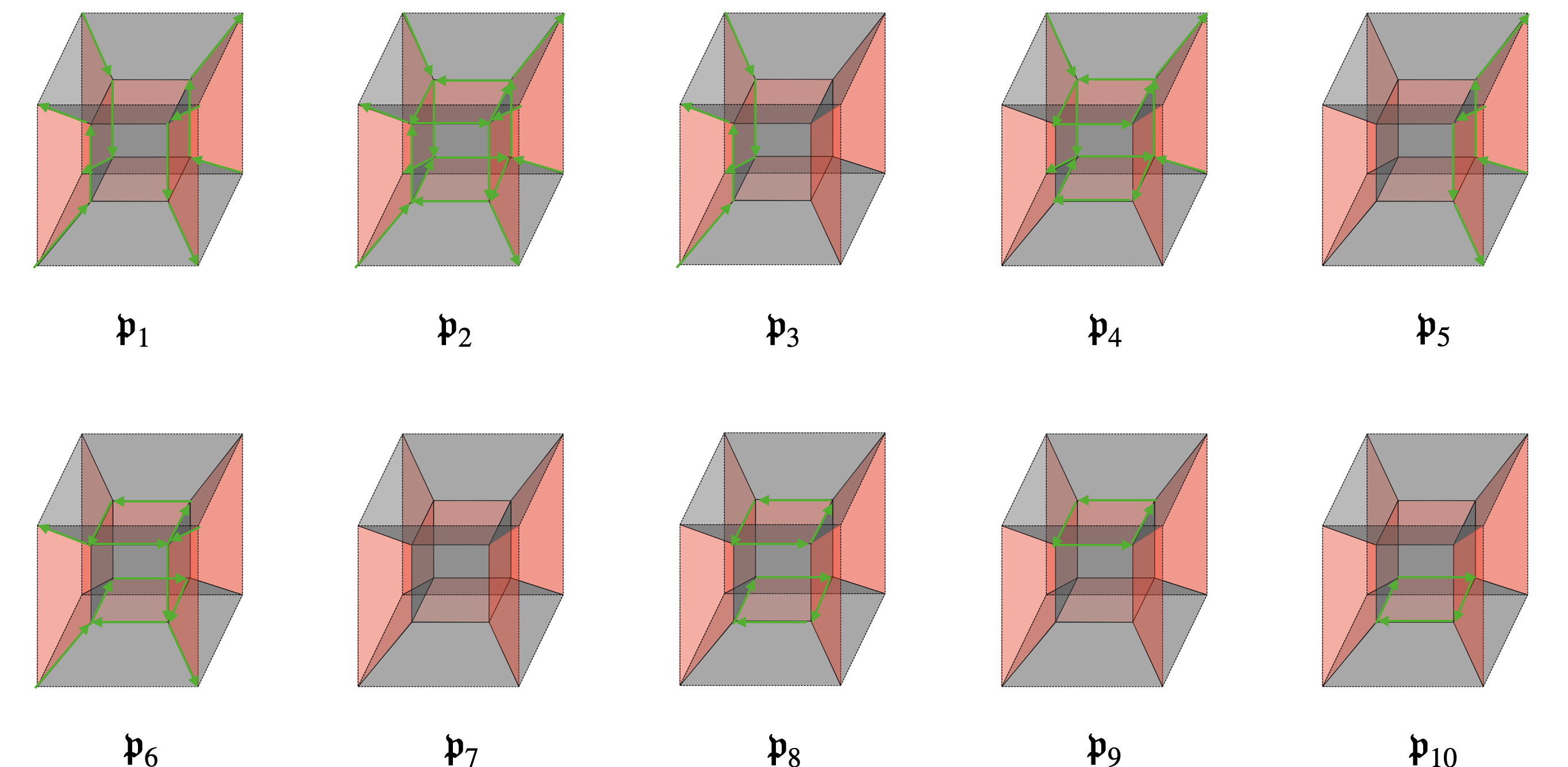}
	\caption{Flows for the ten perfect matchings of the cube model shown in \fref{pms_cube_model} using $p_7$ as the reference, whose associated flow is therefore trivial.}
	\label{flows_of_cube_models}
\end{figure}

Repeating this procedure, we can construct the flows associated to the ten perfect matchings of the cube model, which are shown in \fref{flows_of_cube_models}.

%=================================================================
%=================================================================

Equation \eref{bricks_in_terms_of_chirals} naturally generalizes to any oriented surface, including flows. Expressing the flows in terms of chiral fields, we obtain
\beq
\begin{array}{cclcccl}
	\mathfrak{p}_1&=& \frac{X_{14}X_{15}}{X_{46}X_{56}X_{47}X_{57}} & \ \ \ \ \ \ & \mathfrak{p}_6&=& \frac{X_{21}X_{63}X_{73}}{X_{46}X_{56}X_{47}X_{57}} \\[.2cm]
	\mathfrak{p}_2&=& \frac{X_{21}X_{31}}{X_{46}X_{56}X_{47}X_{57}} & & \mathfrak{p}_7&=& 1 \\[.2cm]
	\mathfrak{p}_3&=& \frac{X_{14}}{X_{46}X_{47}} & & \mathfrak{p}_8&=& \frac{X_{62}X_{63}X_{72}X_{73}}{X_{46}X_{56}X_{47}X_{57}} \\[.2cm]
	\mathfrak{p}_4&=& \frac{X_{31}X_{62}X_{72}}{X_{46}X_{56}X_{47}X_{57}} & & \mathfrak{p}_9&=& \frac{X_{62}X_{63}}{X_{46}X_{56}} \\[.2cm]
    \mathfrak{p}_5&=& \frac{X_{15}}{X_{56}X_{57}} & & \mathfrak{p}_{10}&=& \frac{X_{72}X_{73}}{X_{47}X_{57}}  	
\end{array}
 \label{flows of the cube model in terms of chiral fields}
 \eeq
 
\begin{figure}[ht]
	\centering
	\includegraphics[width=13cm]{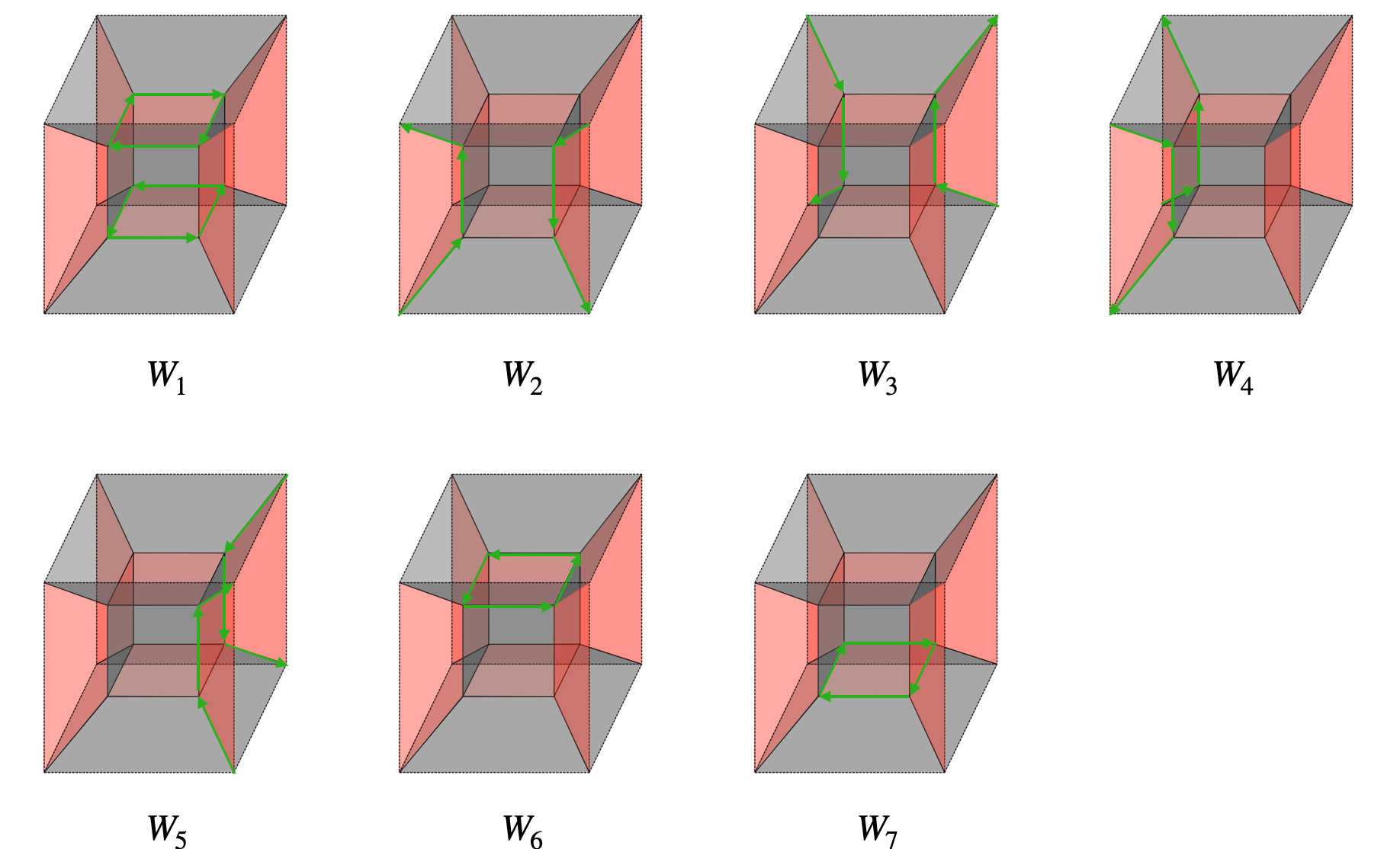}
	\caption{Graphic representation of the seven brick variables (one internal and six external) for the cube model.}
	\label{brick_variables_cube_model}
\end{figure}

In terms of chiral fields, the seven brick variables of the cube model become
\beq
\begin{array}{cclcccl}
W_1&= & \frac{X_{14}X_{15}}{X_{21}X_{31}} &  \ \ \ \ \ \ & W_5&=& \frac{X_{56}X_{57}}{X_{15}}\\[.2cm]
		W_2&=& \frac{X_{21}}{X_{62}X_{72}} & & W_6&=& \frac{X_{62}X_{63}}{X_{46}X_{56}}\\[.2cm]
		W_3&=& \frac{X_{31}}{X_{63}X_{73}} & & W_7&=&\frac{X_{72}X_{73}}{X_{47}X_{57}}\\[.2cm]
		W_4&=& \frac{X_{46}X_{47}}{X_{14}} & & 		
\end{array}
\label{brick variables in chiral fields in cube model}
\eeq
For completeness, \fref{brick_variables_cube_model} shows the graphical representation of these brick variables in terms of oriented edges.

%=================================================================
%=================================================================

Flows can be fully specified by expanding them in terms of a basis. For BFT$_2$'s on a 3-ball such as the cube model, brick variables provide a convenient basis in which any oriented surface can be expanded. More general 3-manifolds require additional ``surface variables”, analogous to the ones that arise for BFTs on general Riemann surfaces of higher genus and/or with multiple boundary components \cite{Franco:2013nwa}. We will not discuss such variables in further detail in this paper, since the fast forward algorithm of \sref{section_BFT2s_and_toric_geometry} automatically produces all required additional variables. BBMs provide concrete examples in which there are three additional variables associated to the fundamental directions of $\mathbb{T}^3$, which can be regarded as the origin of the $3d$ toric diagrams of their CY$_4$ moduli spaces \cite{Franco:2015tna,Franco:2015tya,Franco:2016nwv,Franco:2016qxh,Franco:2016fxm,Franco:2017cjj,Franco:2018qsc}.

Let us momentarily distinguish explicitly between internal bricks $W_i^{(int)}$, $i=1,\ldots,B_i$, and external ones $W_j^{(ext)}$, $j=1,\ldots,B_e$, with $B_i$ and $B_e$ the numbers of internal and external bricks, respectively. We refer to the total number of bricks as $B=B_i+B_e$. These variables are subject to the constraint $\prod_{i=1}^{B_i} W_{(int)i} \prod_{j=1}^{B_e} W_{(ext)j}=1$. This implies that one of them is  redundant which, without loss of generality, we can take to be one of the external bricks. This will turn out to be the manifestation, in the language of bricks, of the extra coordinate in the toric diagrams discussed in \sref{section_BFT2s_and_toric_geometry}.

In terms of brick variables, the flows for the cube become
\beq
\begin{array}{cclcccl}
	\mathfrak{p}_1&=& W_1W_2W_3W_6W_7 &  \ \ \ \ \ \ & \mathfrak{p}_6&=& W_2W_6W_7\\[.2cm]
	\mathfrak{p}_2&=& W_2W_3W_6W_7 & & \mathfrak{p}_7&=& 1\\[.2cm]	
	\mathfrak{p}_3&=& W_4^{-1} & & \mathfrak{p}_8&=& W_6W_7\\[.2cm]
	\mathfrak{p}_4&=& W_3W_6W_7 & & \mathfrak{p}_9&=& W_6\\[.2cm]
	\mathfrak{p}_5&=& W_1W_2W_3W_4W_6W_7 & & \mathfrak{p}_{10}&=& W_7
\end{array}
\label{flows in terms of brick variables in cube model}
\eeq
We can either find these expansions directly or by comparing \eref{flows of the cube model in terms of chiral fields} and \eref{brick variables in chiral fields in cube model}, which express flows and brick variables in terms of chiral fields. In \eref{flows in terms of brick variables in cube model}, we have chosen to remove $W_5$ by expressing it in terms of the other brick variables.

%=================================================================
\subsection{Master and Moduli Spaces from Flows and Brick Variables}
%=================================================================

Following our previous discussion, in the case of 3-balls, flows $\mathfrak{p}_\mu$ are mapped to points in a $(B-1)$-dimensional space with integer coordinates, according to:\footnote{As already mentioned, more general 3-manifolds require additional variables and, hence, extra coordinates.}
\beq
\mathfrak{p}_\mu=\prod_{i=1}^{B_i} W_{(int)i}^{a_{i,\mu}} \prod_{j=1}^{B_e-1} W_{(ext)j}^{b_{j,\mu}}  \ \ \ \ \mapsto \ \ \ \ \begin{array}{c}{\rm \underline{Coordinates}:} \\ (a_{1,\mu},\ldots,a_{B_i,\mu},b_{1,\mu},\ldots,b_{B_e-1,\mu}) \end{array}
\label{flows_from_bricks}
\eeq

These coordinates can be organized as columns in a matrix. From \eref{flows in terms of brick variables in cube model}, we obtain for the cube model
\beq
G_{mast}=\left(
\begin{array}{c|cccccccccc}
& \mathfrak{p}_1 & \mathfrak{p}_2 & \mathfrak{p}_3 & \mathfrak{p}_4 & \mathfrak{p}_5 & \mathfrak{p}_6 & \mathfrak{p}_7 & \mathfrak{p}_8 & \mathfrak{p}_9 & \mathfrak{p}_{10} \\ \hline
W_1 \ & 1 & 0 & 0 & 0  & 1 & 0 & 0 & 0 & 0 & 0 \\ \hline
W_2 \ & 1 & 1 & 0 & 0  & 1 & 1 & 0 & 0 & 0 & 0 \\
W_3 \ & 1 & 1 & 0 & 1  & 1 & 0 & 0 & 0 & 0 & 0 \\
W_4 \ & 0 & 0 & -1 & 0 & 1 & 0 & 0 & 0 & 0 & 0 \\
W_6 \ & 1 & 1 & 0 & 1  & 1 & 1 & 0 & 1 & 1 & 0 \\
W_7 \ & 1 & 1 & 0 & 1  & 1 & 1 & 0 & 1 & 0 & 1 
\end{array}
\right) \, ,
\label{master space of cube model from flows}
\eeq
which defines a $6d$ polytope with ten distinct points. Indeed, it is possible to show that this matrix is $SL(6,\mathbb{Z})$ equivalent to \eref{G_mast_cube}. Flows therefore provide an alternative way of computing the toric diagram for the master space of a BFT$_2$. This is a general result, since this procedure generates the linear relations between points in the toric diagram that encode the corresponding relations between GLSM fields.

In the language of flows, the further projection onto vanishing of $D$-terms which is necessary for obtaining the moduli space simply translates into discarding the coordinates associated to the internal bricks. Thus, for the cube model we delete the $W_1$ row in \eref{master space of cube model from flows}, obtaining
\beq
G=\left(
\begin{array}{c|cccccccccc}
& \mathfrak{p}_1 & \mathfrak{p}_2 & \mathfrak{p}_3 & \mathfrak{p}_4 & \mathfrak{p}_5 & \mathfrak{p}_6 & \mathfrak{p}_7 & \mathfrak{p}_8 & \mathfrak{p}_9 & \mathfrak{p}_{10} \\ \hline
W_2 \ & 1 & 1 & 0 & 0  & 1 & 1 & 0 & 0 & 0 & 0 \\
W_3 \ & 1 & 1 & 0 & 1  & 1 & 0 & 0 & 0 & 0 & 0 \\
W_4 \ & 0 & 0 & -1 & 0 & 1 & 0 & 0 & 0 & 0 & 0 \\
W_6 \ & 1 & 1 & 0 & 1  & 1 & 1 & 0 & 1 & 1 & 0 \\
W_7 \ & 1 & 1 & 0 & 1  & 1 & 1 & 0 & 1 & 0 & 1 
\end{array}
\right) \, .
\eeq
Up to an $SL(5,\mathbb{Z})$ transformation, this is equivalent to (the $5d$ projection of) \eref{G_cube_model} and hence defines the toric diagram of the moduli space. In this language, $\mathfrak{p}_1$ and $\mathfrak{p}_2$ descend to the same point in the toric diagram because they only differ by a contribution coming from the internal brick variable $W_1$.

%=================================================================
\section{Additional Examples}
%=================================================================

\label{section_additional_examples}

The literature contains a vast list of explicit examples of BFT$_2$’s, in the form of BBMs (see e.g. \cite{Franco:2015tna,Franco:2015tya,Franco:2016nwv,Franco:2016qxh,Franco:2016fxm,Franco:2017cjj,Franco:2018qsc}). Such theories correspond to BFT$_2$'s on $\mathbb{T}^3$ without boundaries. In this section, we illustrate our ideas with additional examples, emphasizing the novel features of different 3-manifolds and the presence of boundaries. We pay particular attention to the geometry of the corresponding master and moduli spaces, their determination in terms of perfect matchings and the invariance of the moduli space under triality.

%=================================================================
\subsection{A Hexagonal Prism-Cube Model on a 3-Ball}
%=================================================================

\label{section_hexagona_model}

\begin{figure}[ht]
	\centering
	\includegraphics[width=13cm]{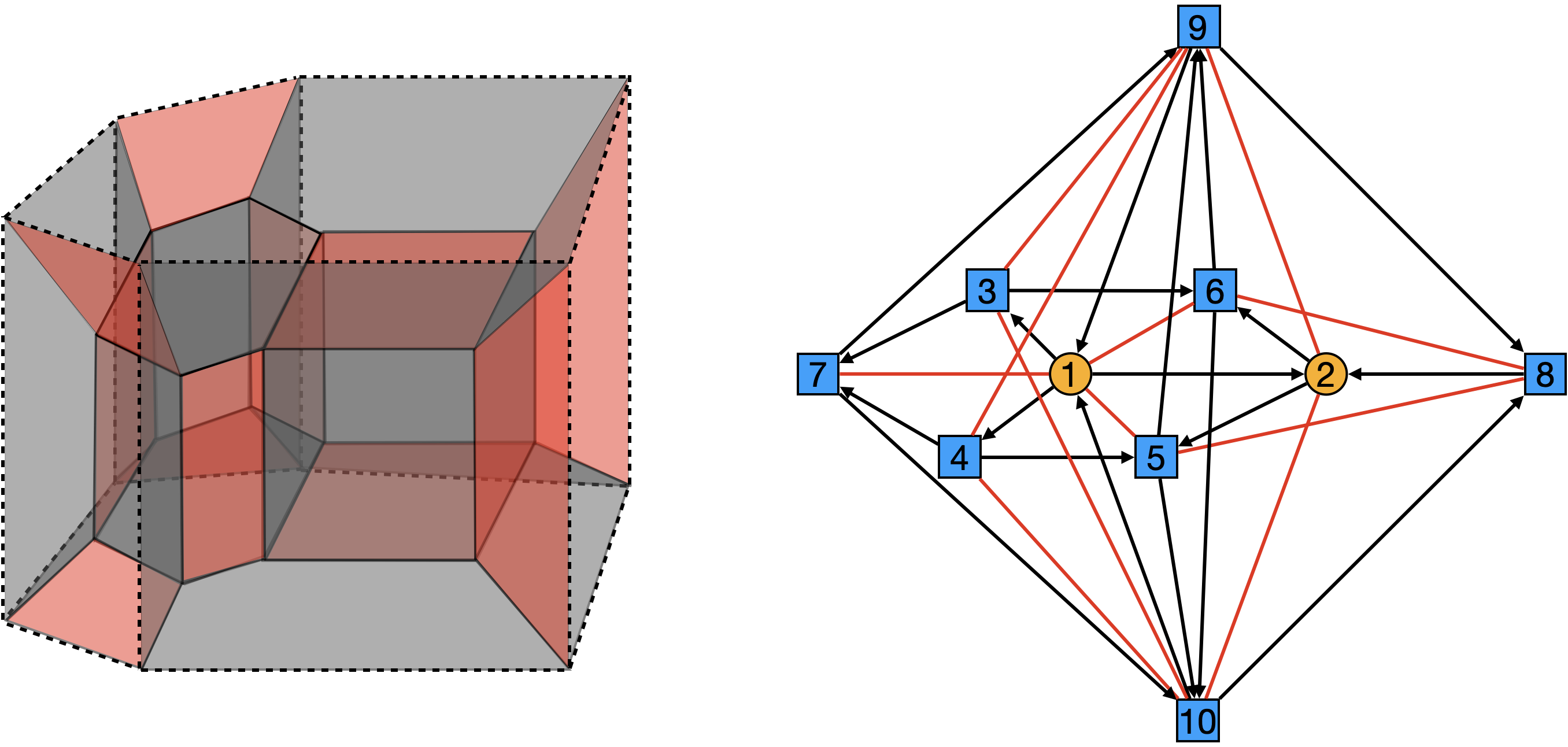}
	\caption{A hexagonal prism-cube model and its dual quiver.} 
	\label{hexagonal_prism_cube}
\end{figure}

Let us consider the example shown in \fref{hexagonal_prism_cube}. $G$ lives on a $3$-ball, the internal bricks are a hexagonal prism and a cube, and it has eight external bricks.

%=================================================================
%=================================================================

This theory has 27 perfect matchings, which are encoded in the following $P$-matrix

%=================================================================
{\tiny
\beq
P=
\left(
\begin{array}{c|ccccccccccccccccccccccccccc}
	& p_{1} & p_{2} & p_{3} & p_{4} & p_{5} & p_{6} & p_{7} & p_{8} & p_{9} & p_{10} & p_{11} & p_{12} & p_{13} & p_{14} & p_{15} & p_{16} & p_{17} & p_{18} & p_{19} & p_{20} & p_{21} & p_{22} & p_{23} & p_{24} & p_{25} & p_{26} & p_{27}\\
	\hline 
 X_{12} & 0 & 1 & 0 & 0 & 1 & 0 & 0 & 0 & 0 & 0 & 0 & 1 & 0 & 1 & 1 & 1 & 1 & 0 & 0 & 0 & 0 & 0 &
   1 & 0 & 1 & 0 & 1 \\
 X_{13} & 0 & 1 & 1 & 0 & 1 & 1 & 0 & 0 & 0 & 0 & 0 & 0 & 0 & 0 & 0 & 1 & 0 & 0 & 0 & 0 & 0 & 0 &
   0 & 1 & 1 & 0 & 0 \\
 X_{14} & 0 & 1 & 1 & 0 & 1 & 0 & 0 & 0 & 0 & 1 & 0 & 0 & 0 & 0 & 1 & 0 & 0 & 0 & 0 & 0 & 0 & 1 &
   1 & 0 & 0 & 0 & 0 \\
 X_{91} & 1 & 0 & 0 & 1 & 0 & 0 & 0 & 0 & 0 & 0 & 0 & 0 & 0 & 0 & 0 & 0 & 0 & 0 & 1 & 0 & 0 & 0 &
   0 & 0 & 0 & 0 & 0 \\
 X_{10,1} & 1 & 0 & 0 & 1 & 0 & 0 & 0 & 0 & 0 & 0 & 0 & 0 & 0 & 0 & 0 & 0 & 0 & 1 & 0 & 0 & 0 & 0 &
   0 & 0 & 0 & 0 & 0 \\
X_{25} &  0 & 0 & 1 & 0 & 0 & 0 & 0 & 0 & 0 & 1 & 1 & 0 & 1 & 0 & 0 & 0 & 0 & 0 & 0 & 1 & 0 & 1 &
   0 & 1 & 0 & 1 & 0 \\
X_{26} &  0 & 0 & 1 & 0 & 0 & 1 & 0 & 0 & 1 & 0 & 0 & 0 & 1 & 0 & 0 & 0 & 0 & 0 & 0 & 0 & 1 & 1 &
   0 & 1 & 0 & 1 & 0 \\
X_{82} &  1 & 1 & 0 & 0 & 0 & 0 & 0 & 0 & 0 & 0 & 0 & 1 & 0 & 0 & 0 & 0 & 0 & 0 & 0 & 0 & 0 & 0 &
   1 & 0 & 1 & 0 & 1 \\
 X_{36} & 0 & 0 & 0 & 0 & 0 & 0 & 0 & 0 & 1 & 0 & 0 & 1 & 1 & 1 & 1 & 0 & 1 & 0 & 0 & 0 & 1 & 1 &
   1 & 0 & 0 & 1 & 1 \\
 X_{37} & 0 & 0 & 0 & 0 & 0 & 0 & 1 & 0 & 0 & 1 & 0 & 0 & 0 & 0 & 1 & 0 & 1 & 0 & 0 & 1 & 1 & 1 &
   1 & 0 & 0 & 1 & 1 \\
 X_{45} & 0 & 0 & 0 & 0 & 0 & 0 & 0 & 0 & 0 & 0 & 1 & 1 & 1 & 1 & 0 & 1 & 1 & 0 & 0 & 1 & 0 & 0 &
   0 & 1 & 1 & 1 & 1 \\
 X_{47} & 0 & 0 & 0 & 0 & 0 & 1 & 1 & 0 & 0 & 0 & 0 & 0 & 0 & 0 & 0 & 1 & 1 & 0 & 0 & 1 & 1 & 0 &
   0 & 1 & 1 & 1 & 1 \\
 X_{59} & 0 & 0 & 0 & 0 & 0 & 1 & 1 & 1 & 1 & 0 & 0 & 0 & 0 & 0 & 0 & 0 & 0 & 1 & 0 & 0 & 1 & 0 &
   0 & 0 & 0 & 0 & 0 \\
 X_{69} & 0 & 0 & 0 & 0 & 0 & 0 & 1 & 1 & 0 & 1 & 1 & 0 & 0 & 0 & 0 & 0 & 0 & 1 & 0 & 1 & 0 & 0 &
   0 & 0 & 0 & 0 & 0 \\
 X_{79} & 0 & 0 & 0 & 0 & 0 & 0 & 0 & 1 & 1 & 0 & 1 & 1 & 1 & 1 & 0 & 0 & 0 & 1 & 0 & 0 & 0 & 0 &
   0 & 0 & 0 & 0 & 0 \\
 X_{98} & 0 & 0 & 0 & 1 & 1 & 0 & 0 & 0 & 0 & 0 & 0 & 0 & 0 & 1 & 1 & 1 & 1 & 0 & 1 & 0 & 0 & 0 &
   0 & 0 & 0 & 0 & 0 \\
 X_{5,10} & 0 & 0 & 0 & 0 & 0 & 1 & 1 & 1 & 1 & 0 & 0 & 0 & 0 & 0 & 0 & 0 & 0 & 0 & 1 & 0 & 1 & 0 &
   0 & 0 & 0 & 0 & 0 \\
 X_{6,10} & 0 & 0 & 0 & 0 & 0 & 0 & 1 & 1 & 0 & 1 & 1 & 0 & 0 & 0 & 0 & 0 & 0 & 0 & 1 & 1 & 0 & 0 &
   0 & 0 & 0 & 0 & 0 \\
 X_{7,10} & 0 & 0 & 0 & 0 & 0 & 0 & 0 & 1 & 1 & 0 & 1 & 1 & 1 & 1 & 0 & 0 & 0 & 0 & 1 & 0 & 0 & 0 &
   0 & 0 & 0 & 0 & 0 \\
 X_{10,8} & 0 & 0 & 0 & 1 & 1 & 0 & 0 & 0 & 0 & 0 & 0 & 0 & 0 & 1 & 1 & 1 & 1 & 1 & 0 & 0 & 0 & 0 &
   0 & 0 & 0 & 0 & 0 \\
\end{array}
\right).
\eeq
}
%=================================================================
While this matrix is not particularly illuminating, we include it to emphasize that it is straightforward to be explicit about the perfect matchings.

Row-reducing the $P$-matrix, we obtain the toric diagram for the master space
%=================================================================
{\tiny \beq
G_{mast}=\left(
\begin{array}{ccccccccccccccccccccccccccc}
	p_{1} & p_{2} & p_{3} & p_{4} & p_{5} & p_{6} & p_{7} & p_{8} & p_{9} & p_{10} & p_{11} & p_{12} & p_{13} & p_{14} & p_{15} & p_{16} & p_{17} & p_{18} & p_{19} & p_{20} & p_{21} & p_{22} & p_{23} & p_{24} & p_{25} & p_{26} & p_{27}\\
	\hline 
 1 & -1 & -2 & 0 & -2 & -1 & 0 & 0 & 0 & -1 & 0 & 1 & 0 & 0 & -1 & -1 & 0 & 0 & 0 & 0 & 0
   & -1 & 0 & -1 & 0 & 0 & 1 \\
 1 & 0 & 1 & 1 & 0 & 0 & -1 & -1 & 0 & 0 & 0 & 0 & 1 & 0 & 0 & 0 & 0 & 0 & 0 & 0 & 0 & 1
   & 0 & 1 & 0 & 1 & 0 \\
 0 & 1 & 1 & 0 & 1 & 1 & 0 & 0 & 0 & 0 & 0 & 0 & 0 & 0 & 0 & 1 & 0 & 0 & 0 & 0 & 0 & 0 &
   0 & 1 & 1 & 0 & 0 \\
 0 & 1 & 1 & 0 & 1 & 0 & 0 & 0 & 0 & 1 & 0 & 0 & 0 & 0 & 1 & 0 & 0 & 0 & 0 & 0 & 0 & 1 &
   1 & 0 & 0 & 0 & 0 \\
 -1 & 0 & 0 & -1 & 0 & 1 & 1 & 1 & 1 & 0 & 0 & 0 & 0 & 0 & 0 & 0 & 0 & 0 & 0 & 0 & 1 & 0
   & 0 & 0 & 0 & 0 & 0 \\
 -1 & 0 & 0 & -1 & 0 & 0 & 1 & 1 & 0 & 1 & 1 & 0 & 0 & 0 & 0 & 0 & 0 & 0 & 0 & 1 & 0 & 0
   & 0 & 0 & 0 & 0 & 0 \\
 1 & 0 & 0 & 1 & 0 & 0 & 0 & 0 & 0 & 0 & 0 & 0 & 0 & 0 & 0 & 0 & 0 & 0 & 1 & 0 & 0 & 0 &
   0 & 0 & 0 & 0 & 0 \\
 1 & 0 & 0 & 1 & 0 & 0 & 0 & 0 & 0 & 0 & 0 & 0 & 0 & 0 & 0 & 0 & 0 & 1 & 0 & 0 & 0 & 0 &
   0 & 0 & 0 & 0 & 0 \\
 0 & 0 & 0 & 1 & 1 & 0 & 0 & -1 & -1 & 0 & -1 & -1 & -1 & 0 & 1 & 1 & 1 & 0 & 0 & 0 & 0 &
   0 & 0 & 0 & 0 & 0 & 0 \\
 -1 & 0 & 0 & -1 & 0 & 0 & 0 & 1 & 1 & 0 & 1 & 1 & 1 & 1 & 0 & 0 & 0 & 0 & 0 & 0 & 0 & 0
   & 0 & 0 & 0 & 0 & 0 
\end{array}
\right)
\eeq
}
%=================================================================
As expected, the master space is a CY 10-fold, since this theory lives on a 3-ball and has 10 bricks.

Finally, the moduli space is a CY 8-fold with toric diagram given by

%=================================================================
{\small
\beq
G=\left(
\begin{array}{cccccccccccccccccccc}
		-1&-2&-1&-1&2&3&-1&3&4&3&-1&-1&0&0&0&0&0&0&0&1\\
		1&1&1& 0&-1& -1& 0& -1& -1&-1& 0& 1& 0& 0& 0& 0& 0& 0& 1& 0\\
		1& 1& 0& 0& -1& -1& 1& -1& -1& -1& 1& 0& 0& 0& 0& 0& 0& 1& 0& 0\\
		0& 0& 1& 1& 0&0& 0& -1& -1& -1& 0& 0& 0& 0& 0& 0& 1& 0& 0& 0\\
		0&0& 0& 1&0& -1& 1& 0&-1& -1& 0& 0& 0& 0& 0& 1& 0& 0& 0& 0\\
		0& 0&
   0& 0& 1& 1& 0& 1& 1& 1& 0& 0& 0& 0& 1& 0& 0& 0& 0& 0\\ 
   0& 0& 0& 0&
   1& 1& 0& 1& 1& 1& 0& 0
   & 0& 1& 0& 0& 0& 0& 0& 0\\
   0& 1& 0& 0& 
  -1& -1& 0&-1& -1& 0& 1& 1& 1&0& 0& 0& 0& 0& 0& 0\\
 	\hline
		\ \,  \textbf{3}\ \,  &\ \,  \textbf{2}\ \,  &\ \,  \textbf{1}\ \,  &\ \,  \textbf{1}\ \,  &\ \,  \textbf{1}\ \,  &\ \,  \textbf{1}\ \,  &\ \,  \textbf{1}\ \,  &\ \,  \textbf{1}\ \,  &\ \,  \textbf{2}\ \,  &\ \,  \textbf{1}\ \,  &\ \,  \textbf{1}\ \,  &\ \,  \textbf{1}\ \,  &\ \,  \textbf{1}\ \,  &\ \,  \textbf{1}\ \,  &\ \,  \textbf{1}\ \,  &\ \,  \textbf{1}\ \,  &\ \,  \textbf{1}\ \,  &\ \,  \textbf{2}\ \,  &\ \,  \textbf{2}\ \,  &\ \,  \textbf{2}\ \,  
\end{array}
\right)
\label{G_hexagon_cube}
\eeq
}
%=================================================================
For simplicity, we have introduced a notation in which columns indicate the positions of points in the toric diagram and the last row shows the multiplicities of perfect matchings associated to each of them. Since this BFT$_2$ lives on a 3-ball, the dimensionality of the moduli space is equal to the number of external bricks.

%=================================================================
\subsection{A Model on $\mathbb{T}^2 \times I$}
%=================================================================

\label{section_model_T2_segment}

\begin{figure}[ht]
	\centering
	\includegraphics[width=13cm]{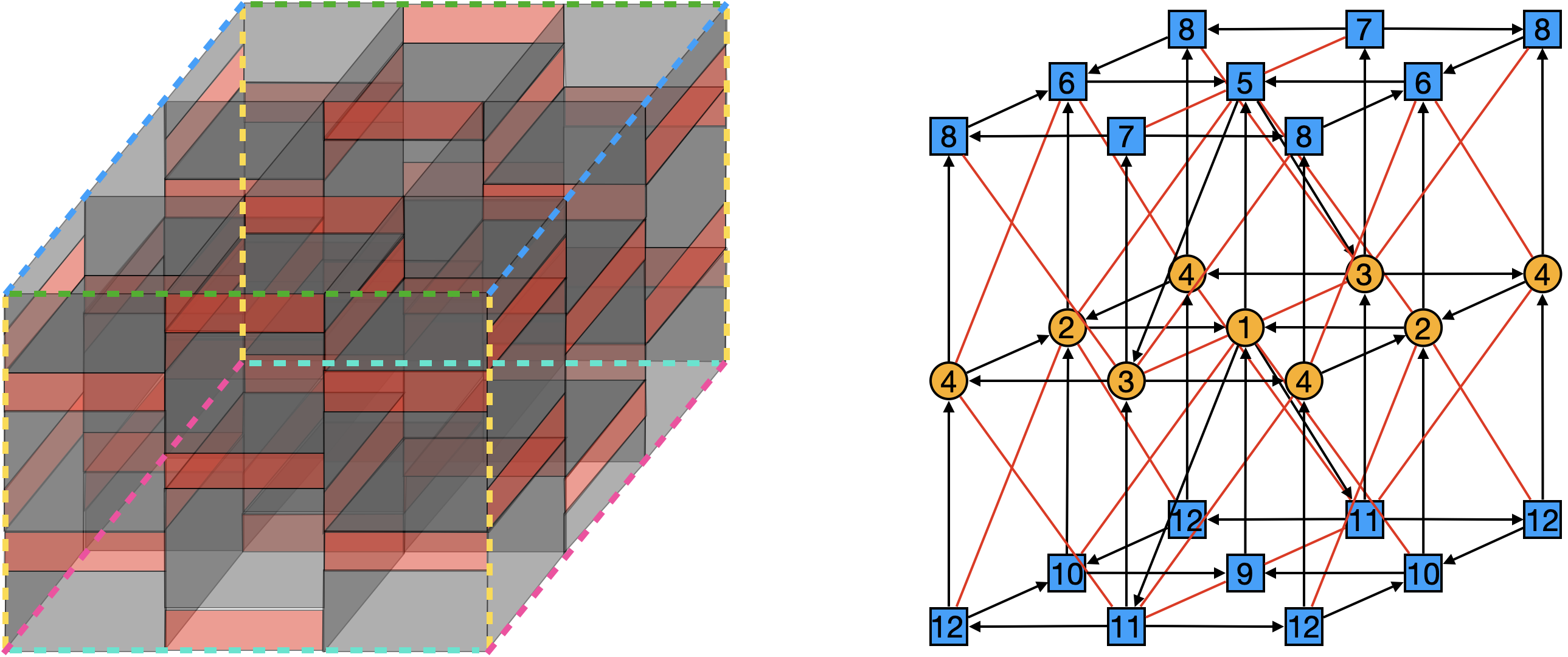}
	\caption{BFT$_2$ defined on $\mathbb{T}^2 \times I$ and its dual quiver. We have included colored dotted lines to emphasize the $\mathbb{T}^2$ identifications on the horizontal plane.}
	\label{T2_segment}
\end{figure}

Consider the BFT$_2$ on $\mathbb{T}^2$ times an interval shown in \fref{T2_segment}.

%=================================================================
%=================================================================

This theory has 54 perfect matchings. The corresponded $P$-matrix is presented in Appendix \ref{section_appendix_P_matrix}. The moduli space is a CY 11-fold with toric diagram given by

%=================================================================
{\tiny
\begin{equation}
	G=\left(
\begin{array}{cccccccccccccccccccccccccccccccccc}
 -1 & -1 & -1 & 0 & -1 & -1 & 0 & -2 & 0 & -1 & -1 & -1 & 0 & -1 & 0 & -1 & 0 & -1 & -1
   & 0 & -1 & -1 & 1 & 0 & 0 & 0 & 0 & 0 & 0 & 0 & 0 & 0 & 0 & 1 \\
 0 & 0 & 1 & 0 & 0 & 0 & 0 & 1 & 0 & 0 & 0 & 0 & 0 & 0 & 0 & 0 & 0 & 0 & 1 & 1 & 1 & 1 &
   0 & 0 & 0 & 0 & 0 & 0 & 0 & 0 & 0 & 0 & 1 & 0 \\
 -1 & 0 & -1 & -1 & -1 & 0 & -1 & -2 & -1 & -1 & 0 & 0 & 1 & 0 & -1 & 1 & 2 & 1 & 0 & 1
   & 0 & 0 & 1 & 0 & 0 & 0 & 0 & 0 & 0 & 0 & 0 & 1 & 0 & 0 \\
 1 & 0 & 0 & 0 & 1 & 0 & 0 & 1 & 0 & 1 & 0 & 1 & 1 & 1 & 1 & 0 & 0 & 0 & 0 & 0 & 0 & 0 &
   0 & 0 & 0 & 0 & 0 & 0 & 0 & 0 & 1 & 0 & 0 & 0 \\
 1 & 1 & 1 & 1 & 1 & 1 & 1 & 2 & 2 & 1 & 1 & 0 & -1 & 0 & 0 & 0 & -1 & 0 & 0 & -1 & 0 &
   0 & -1 & 0 & 0 & 0 & 0 & 0 & 0 & 1 & 0 & 0 & 0 & 0 \\
 0 & 0 & 0 & 0 & 0 & 0 & 0 & 0 & -1 & 0 & 0 & 0 & 0 & 0 & 0 & 0 & 0 & 0 & 0 & 0 & 0 & 0
   & 0 & 0 & 0 & 0 & 0 & 0 & 1 & 0 & 0 & 0 & 0 & 0 \\
 0 & 0 & 0 & 0 & 0 & 0 & 0 & 1 & 0 & 0 & 1 & 0 & 0 & 0 & 0 & 0 & 0 & 0 & 0 & 0 & 0 & 1 &
   0 & 0 & 0 & 0 & 0 & 1 & 0 & 0 & 0 & 0 & 0 & 0 \\
 1 & 1 & 1 & 1 & 0 & 0 & 0 & 1 & 0 & 1 & 0 & 0 & 0 & 1 & 0 & 0 & 0 & 1 & 0 & 0 & 1 & 0 &
   0 & 0 & 0 & 0 & 1 & 0 & 0 & 0 & 0 & 0 & 0 & 0 \\
 0 & 0 & 0 & 0 & 1 & 1 & 1 & 0 & 1 & 0 & 0 & 1 & 0 & 0 & 0 & 1 & 0 & 0 & 1 & 0 & 0 & 0 &
   0 & 0 & 0 & 1 & 0 & 0 & 0 & 0 & 0 & 0 & 0 & 0 \\
 0 & 0 & 0 & 0 & 0 & 0 & 0 & 0 & 0 & 0 & 0 & 0 & 0 & 0 & 1 & 0 & 0 & 0 & 0 & 0 & 0 & 0 &
   0 & 0 & 1 & 0 & 0 & 0 & 0 & 0 & 0 & 0 & 0 & 0 \\
 0 & 0 & 0 & 0 & 0 & 0 & 0 & -1 & 0 & 0 & 0 & 0 & 0 & 0 & 0 & 0 & 0 & 0 & 0 & 0 & 0 & 0
   & 0 & 1 & 0 & 0 & 0 & 0 & 0 & 0 & 0 & 0 & 0 & 0 \\
   \hline 
\ \textbf{3} \ &\ \textbf{4} \ &\ \textbf{1} \ &\ \textbf{2} \ &\ \textbf{2} \ &\ \textbf{3} \ &\ \textbf{1} \ &\ \textbf{1} \ &\ \textbf{1} \ &\ \textbf{2} \ &\ \textbf{2} \ &\ \textbf{1} \ &\ \textbf{1} \ &\ \textbf{1} \ &\ \textbf{1} \ &\ \textbf{1} \ &\ \textbf{1} \ &\ \textbf{1} \ &\ \textbf{1} \ &\ \textbf{1} \ &\ \textbf{1} \ &\ \textbf{1} \ &\ \textbf{1} \ &\ \textbf{1} \ &\ \textbf{1} \ &\ \textbf{1} \ &\ \textbf{1} \ &\ \textbf{1} \ &\ \textbf{1} \ &\ \textbf{1} \ &\ \textbf{4} \ &\ \textbf{4} \ &\ \textbf{2} \ &\ \textbf{3} \ \\
\end{array}
\right)
\end{equation}
}
%=================================================================

The dimensionality of the moduli space is a consequence of the eight external bricks, the two fundamental directions of $\mathbb{T}^2$ and the two boundary components. We plan to revisit the systematics of this correspondence in future work.

%=================================================================
\section{Triality and the Moduli Space}
%=================================================================

\label{section_triality_and_moduli_space}

It is natural to ask how to determine whether two BFT$_2$’s are related by sequences of cube moves and bubble reductions. This can be rather challenging since, if the theories are indeed connected, such transformations are generically not unique. Moreover, depending on the complexity of the theories, the necessary sequence of operations might be highly intricate. The moduli space, being invariant under both triality and reduction, is ideal for diagnosing whether two BFT$_2$ can be connected in this sense, without actually constructing the transformations that turn one theory into the other. To be precise, matching of the moduli space is a necessary condition. It would be interesting to establish whether it is also sufficient or if other moduli space-preserving transformations that go beyond triality and reduction exist. 

The multiple triality phases of BBMs represent explicit examples of how dual BFT$_2$'s have the same moduli space (see e.g. \cite{Franco:2016nwv,Franco:2016fxm,Franco:2018qsc}). Below we present a new example exhibiting a novel feature, a boundary. Further examples are presented in \sref{section_reduction_CY}.

%=================================================================
\subsection{The Triality Dual of the Hexagonal Prism-Cube Model}
%=================================================================

\begin{figure}[ht]
	\centering
	\includegraphics[width=13.5cm]{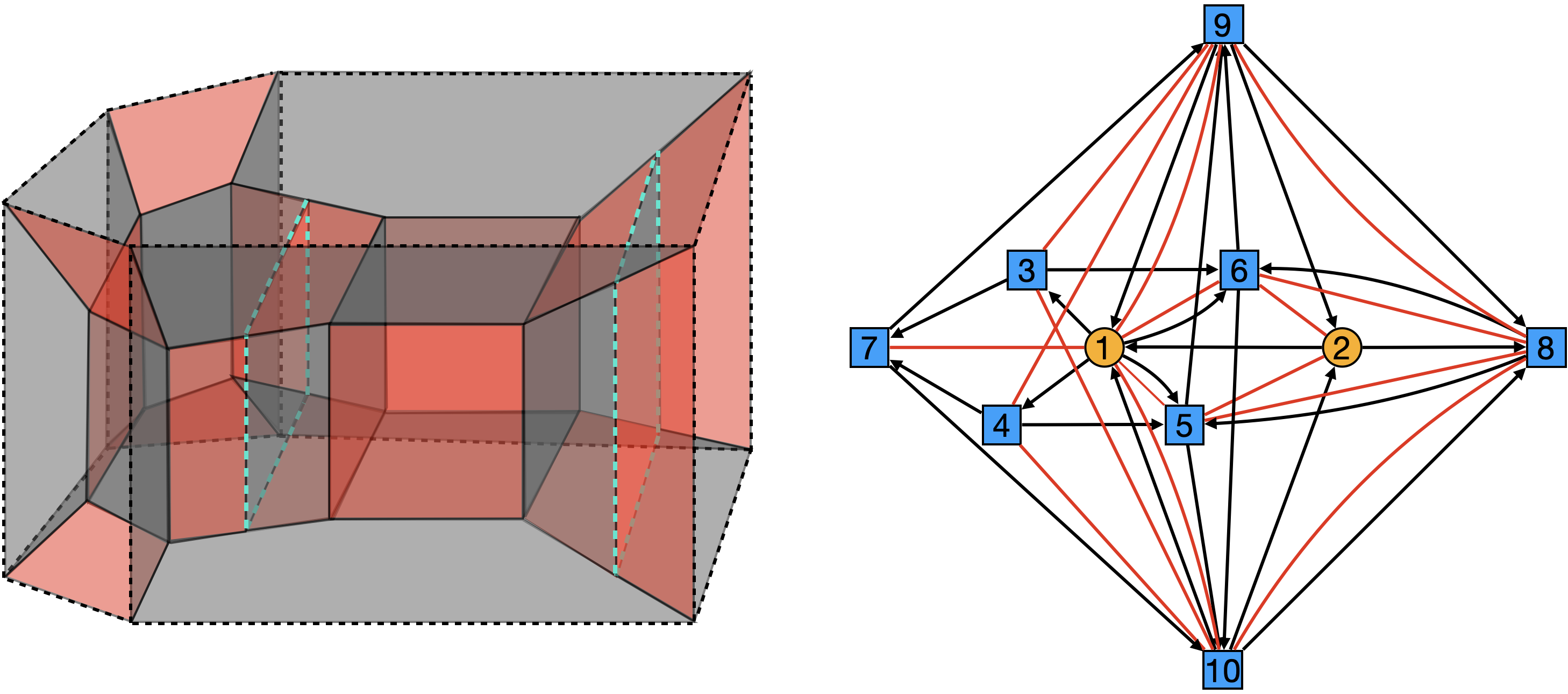}
	\caption{Model obtained by acting with triality on node 2 of the one in \fref{hexagonal_prism_cube}.}
	\label{dual_hexagonal_prism_cube}
\end{figure}

Acting with triality on gauge group 2 of the hexagonal prism-cube model of \sref{section_hexagona_model}, we obtain the theory shown in \fref{dual_hexagonal_prism_cube}. Dotted edges represent mass terms and we have not integrated out the corresponding massive fields.

%=================================================================
%=================================================================

This theory has 23 perfect matchings, which give rise to a moduli space with toric diagram
%=================================================================
{\small
\beq
G=\left(
\begin{array}{cccccccccccccccccccc}
		-1&-2&-1&-1&2&3&-1&3&4&3&-1&-1&0&0&0&0&0&0&0&1\\
		1&1&1& 0&-1& -1& 0& -1& -1&-1& 0& 1& 0& 0& 0& 0& 0& 0& 1& 0\\
		1& 1& 0& 0& -1& -1& 1& -1& -1& -1& 1& 0& 0& 0& 0& 0& 0& 1& 0& 0\\
		0& 0& 1& 1& 0&0& 0& -1& -1& -1& 0& 0& 0& 0& 0& 0& 1& 0& 0& 0\\
		0&0& 0& 1&0& -1& 1& 0&-1& -1& 0& 0& 0& 0& 0& 1& 0& 0& 0& 0\\
		0& 0&
   0& 0& 1& 1& 0& 1& 1& 1& 0& 0& 0& 0& 1& 0& 0& 0& 0& 0\\ 
   0& 0& 0& 0&
   1& 1& 0& 1& 1& 1& 0& 0
   & 0& 1& 0& 0& 0& 0& 0& 0\\
   0& 1& 0& 0& 
  -1& -1& 0&-1& -1& 0& 1& 1& 1&0& 0& 0& 0& 0& 0& 0\\
 	\hline
		\ \,  \textbf{3}\ \,  &\ \,  \textbf{2}\ \,  &\ \,  \textbf{1}\ \,  &\ \,  \textbf{1}\ \,  &\ \,  \textbf{1}\ \,  &\ \,  \textbf{1}\ \,  &\ \,  \textbf{1}\ \,  &\ \,  \textbf{1}\ \,  &\ \,  \textbf{1}\ \,  &\ \,  \textbf{1}\ \,  &\ \,  \textbf{1}\ \,  &\ \,  \textbf{1}\ \,  &\ \,  \textbf{1}\ \,  &\ \,  \textbf{1}\ \,  &\ \,  \textbf{1}\ \,  &\ \,  \textbf{1}\ \,  &\ \,  \textbf{1}\ \,  &\ \,  \textbf{1}\ \,  &\ \,  \textbf{1}\ \,  &\ \,  \textbf{1}\ \,  
\end{array}
\right) .
\label{G_hexagon_cube_dual}
\eeq
}
%=================================================================
This toric diagram agrees with the one for the original theory, given by \eref{G_hexagon_cube}, up to multiplicities of perfect matchings for some of its points. Several examples of theories with the same moduli space up to multiplicities of perfect matchings in the toric diagram can be found in the BBM literature \cite{Franco:2016nwv,Franco:2016fxm,Franco:2018qsc}. Further examples are presented in the coming section.

Interestingly, integrating out the massive fields connected to node 1, this node ends up having a single incoming arrow. Node 1 would thus disappear upon acting on it with triality. This implies that the hexagonal prism-cube model is actually reducible.

%=================================================================
\section{Reduction and Calabi-Yau Geometry}
%=================================================================

\label{section_reduction_CY}

\begin{figure}[ht]
	\centering
	\includegraphics[width=7cm]{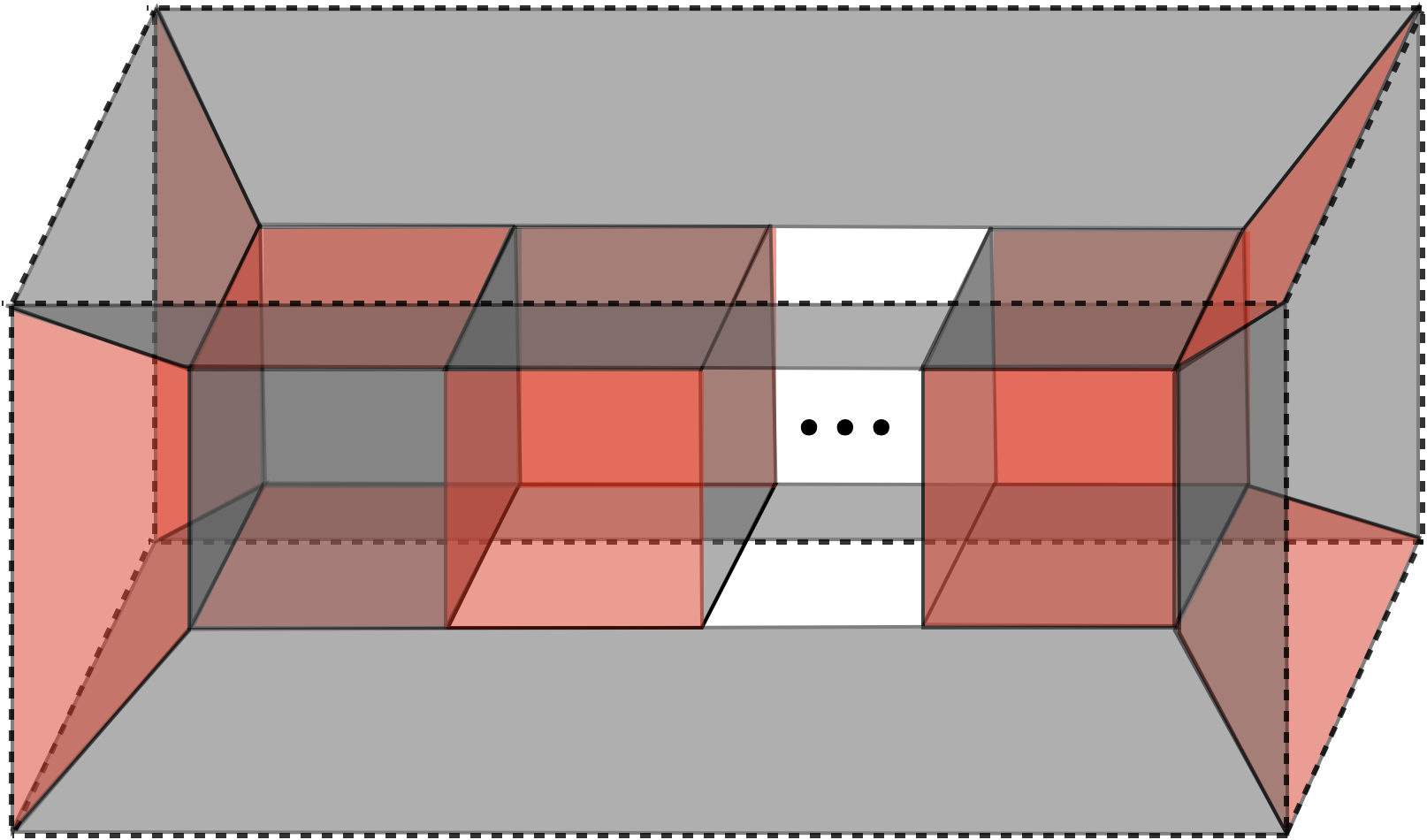}
	\caption{A multi-cube model. Internal bricks are separated by chiral faces. The 90$^\circ$ between the external faces at the two endpoints of this figure corresponds to having an even number of cubes.}
	\label{n-cube model even}
\end{figure}

We now investigate the moduli space of reducible BFT$_2$'s. For concreteness, we illustrate our discussion using the multi-cube family of theories schematically shown in \fref{n-cube model even}. The simplest theory in this class is the cube model we considered in previous sections. There are two families of these models, depending on whether the number of cubes is even or odd. For an even number of cubes, there is a relative 90$^\circ$ rotation of the external faces at the two endpoints, as shown in \fref{2-cube model}. This rotation is not present for an odd number of cubes, such as in the example in \fref{3-cube model}.

%=================================================================
%=================================================================

The multi-cube model with $n$ cubes lives on a 3-ball, and has $n$ internal and six external bricks. Following the discussion in previous sections, we thus expect their master spaces to be toric CY $(n+6)$-folds, with $(n+5)$-dimensional toric diagrams. Similarly, for any $n$, the moduli space should be a toric CY 6-fold with a $5d$ toric diagram. We will now explain that the agreement between the moduli spaces of all these theories extends beyond their dimension.

The single cube model was thoroughly studied in previous sections, with its master and moduli spaces captured by \eref{G_mast_cube} and \eref{G_cube_model}, respectively. Let us now consider the 2- and 3-cube models.

%=================================================================
\subsection{2-Cube Model}
%=================================================================

\begin{figure}[ht]
	\centering
	\includegraphics[width=13cm]{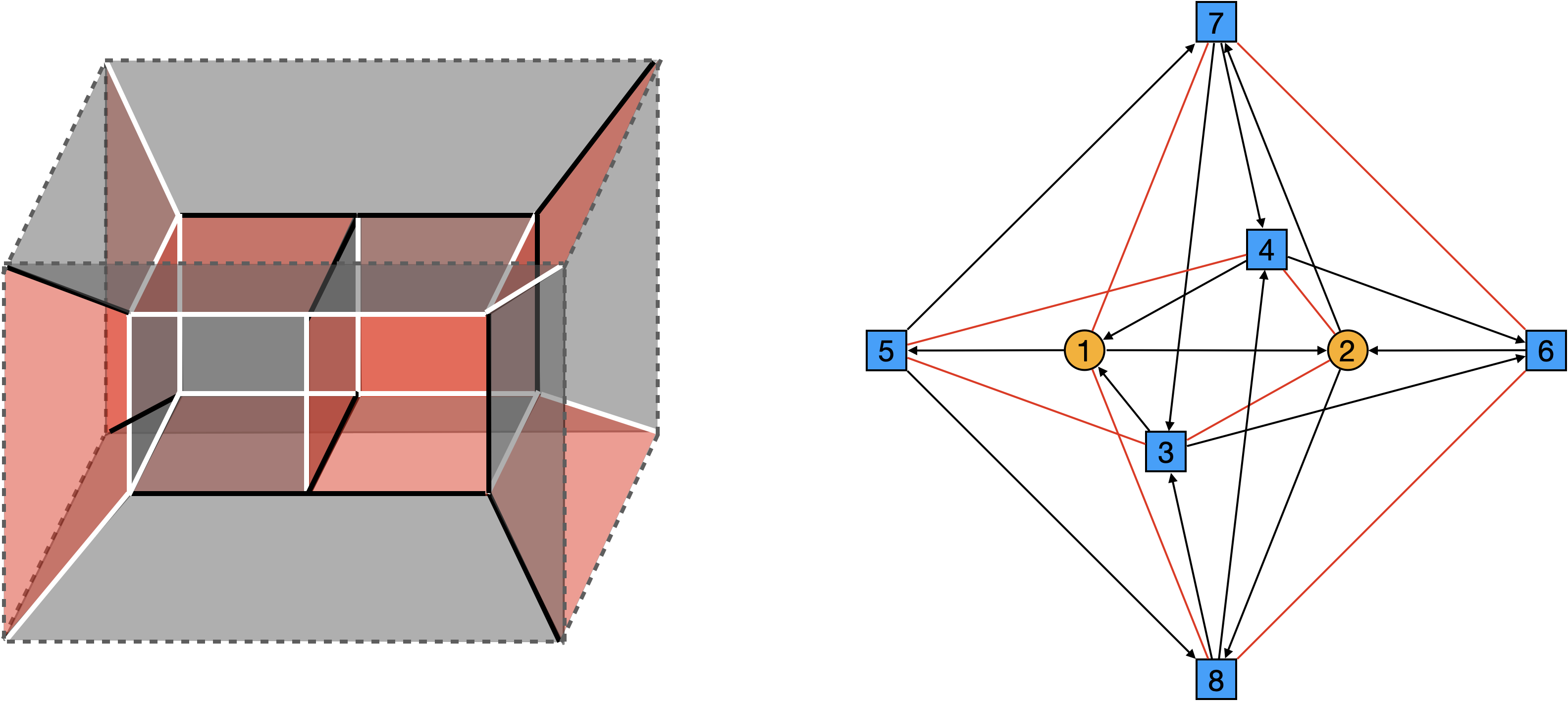}
	\caption{The 2-cube model and its dual quiver.}
	\label{2-cube model}
\end{figure}

The 2-cube model is shown in \fref{2-cube model}.

%=================================================================
%=================================================================

Following the procedure in \sref{section_BFT2s_and_toric_geometry}, we compute its moduli space. There are 13 perfect matchings and the moduli space parameterized by them is encoded in the toric diagram defined by
\beq
G=\left(
\begin{array}{ccccccccc}
 0 & 1 & 0 & 0 & 0 & 1 & 0 & 0 & 1 \\
 0 & 1 & 0 & 0 & 0 & 1 & 0 & 1 & 0 \\
 -1 & -2 & 0 & 0 & 0 & -1 & 1 & 0 & 0 \\
 1 & 1 & 0 & 0 & 1 & 0 & 0 & 0 & 0 \\
 0 & -1 & 0 & 1 & 0 & 0 & 0 & 0 & 0 \\
 1 & 1 & 1 & 0 & 0 & 0 & 0 & 0 & 0 \\ \hline
\ \,  {\bf 2} \ \, & \ \,  {\bf 3} \ \, & \ \,  {\bf 1} \ \, & \ \,  {\bf 1} \ \, & \ \,  {\bf 1} \ \, & \ \,  {\bf 2} \ \, & \ \,  {\bf 1} \ \, & \ \,  {\bf 1} \ \, & \ \,  {\bf 1} \ \, 
\end{array}
\right) \, .
\label{G_2-cube_model}
\eeq
Interestingly, despite the different multiplicities, the toric diagram coincides with the one for the single cube model in \eref{G_cube_model}. The 1- and 2-cube models have the same moduli space.

%=================================================================
\subsection{3-Cube Model}
%=================================================================

\begin{figure}[ht]
\centering
	\includegraphics[width=13cm]{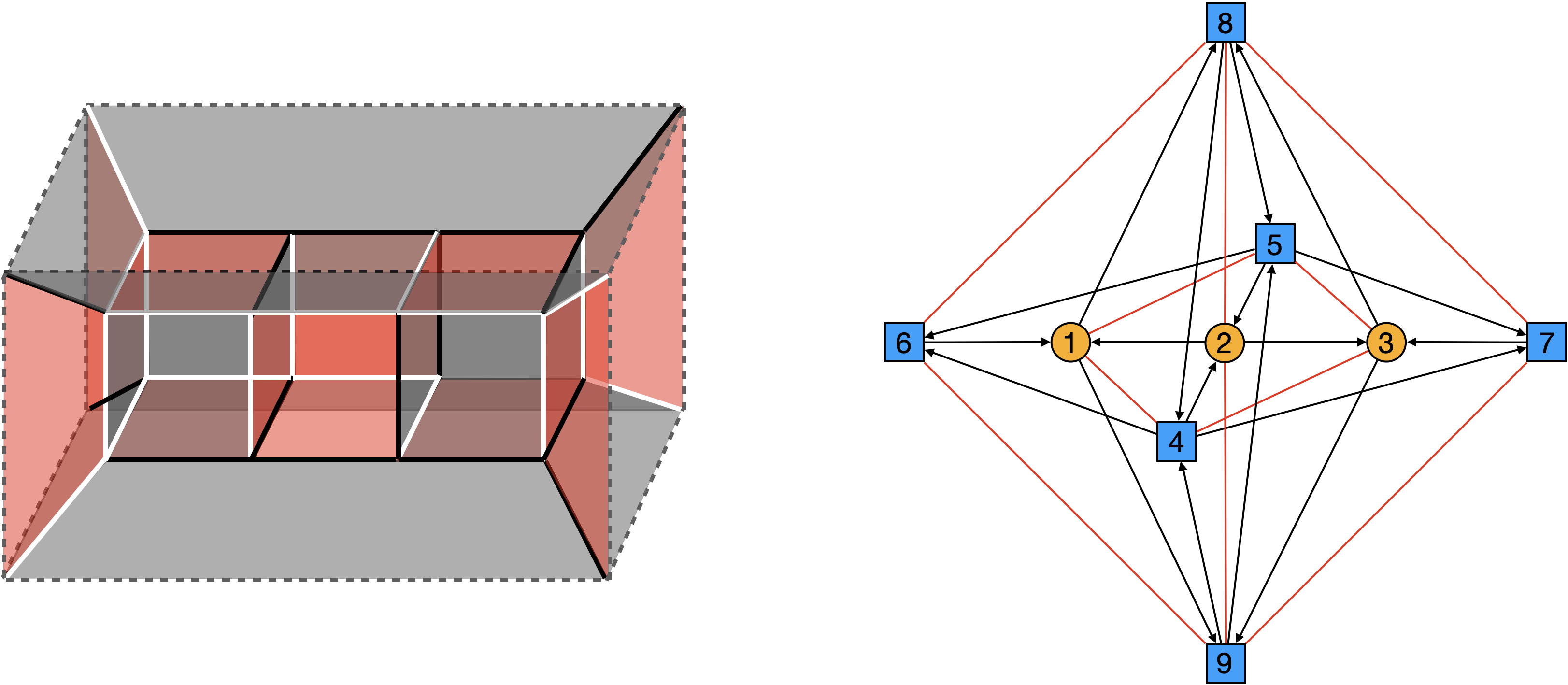}
	\caption{The 3-cube model and its dual quiver.}
	\label{3-cube model}
\end{figure}

The 3-cube model is shown in \fref{3-cube model}.

%=================================================================
%=================================================================

Computing its moduli space, we obtain 18 perfect matchings which form a toric diagram defined by the following matrix
\beq
G=\left(
\begin{array}{ccccccccc}
0 & 1 & 0 & 0 & 0 & 1 & 0 & 0 & 1 \\
0 & 1 & 0 & 0 & 0 & 1 & 0 & 1 & 0 \\
-1 & -2 & 0 & 0 & 0 & -1 & 1 & 0 & 0 \\
1 & 1 & 0 & 0 & 1 & 0 & 0 & 0 & 0 \\
0 & -1 & 0 & 1 & 0 & 0 & 0 & 0 & 0 \\
1 & 1 & 1 & 0 & 0 & 0 & 0 & 0 & 0 \\ \hline
\ \,  {\bf 5} \ \, & \ \,  {\bf 3} \ \, & \ \,  {\bf 1} \ \, & \ \,  {\bf 3} \ \, & \ \,  {\bf 1} \ \, & \ \,  {\bf 2} \ \, & \ \,  {\bf 1} \ \, & \ \,  {\bf 1} \ \, & \ \,  {\bf 1} \ \, 
\end{array}
\right) \, .
\label{G_3-cube_model}
\eeq
Once again, we obtain the same moduli space of the single cube model given by \eref{G_cube_model}, with different multiplicities of perfect matchings for some of the points in the toric diagram.

%=================================================================
\subsection{Reduction and Triality}
%=================================================================

The previous examples illustrate a more general fact: all $n$-cube models have the same moduli space, differing only by the multiplicities of perfect matchings. As we explain below, this fact can be understood as the result of a beautiful interplay between triality and reduction. As an aside, it would be interesting to investigate the combinatorics of the perfect matchings in this class of models and these multiplicities in further detail. 

\begin{figure}[ht]
	\centering
	\includegraphics[width=13cm]
	{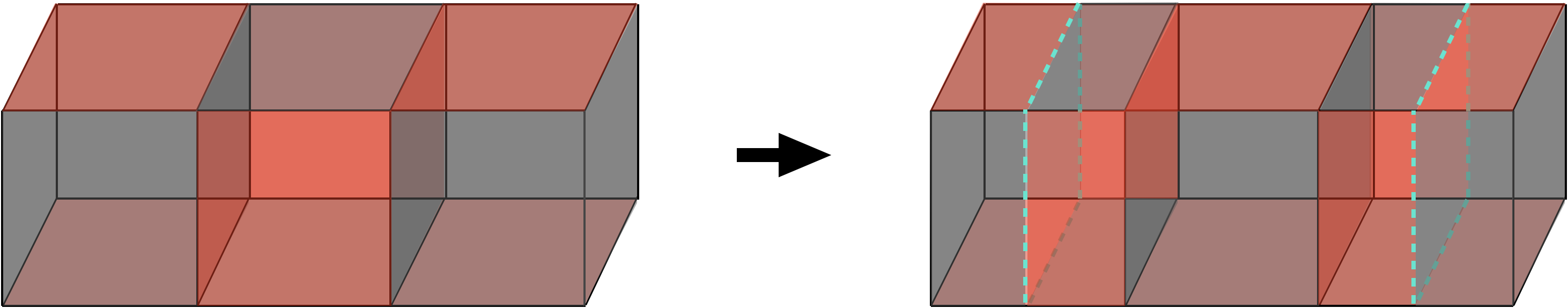}
	\caption{Triality on an intermediate brick of an $n$-cube model. Eight 2-valent edges (shown as dashed lines), i.e. mass terms, are generated during this process. To avoid confusion, it is important to emphasize that there are no vertical faces suspended from the dashed edges.}
	\label{cube reduction in bipartite graph}
\end{figure}

\begin{figure}[ht]
	\centering
	\includegraphics[width=14cm]{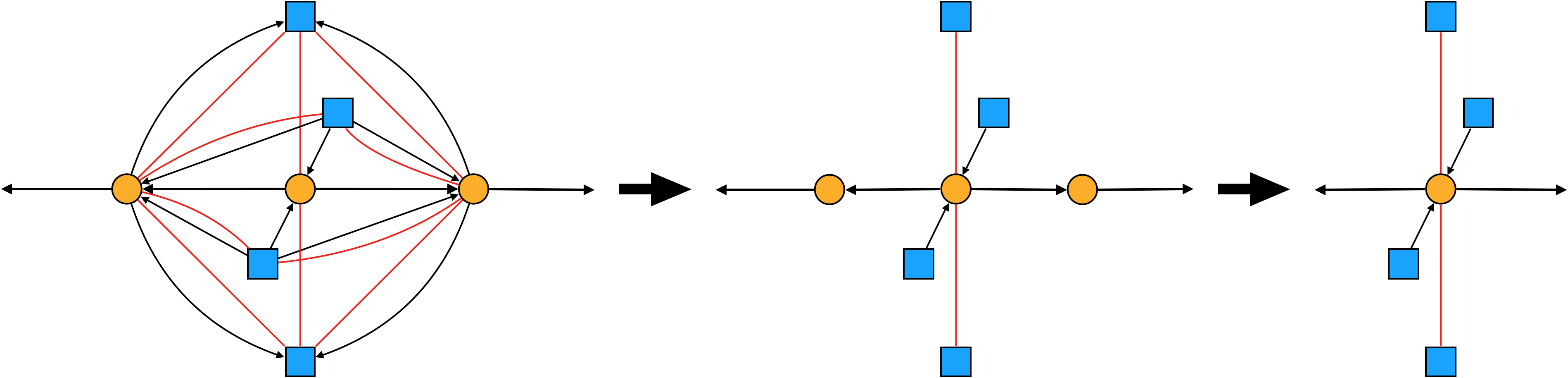}
	\caption{The dual quiver representing the reduction of the $n$-cube model. (a) Configuration with massive fields obtained after triality on an intermediate cube. (b) Two nodes associated to bubbles arise after integrating out massive fields. (c) Triality on the bubble nodes reduces the theory.}
	\label{cube reduction in quiver diagram}
\end{figure}

Let us start from an $n$-cube model with $n\geq 3$ and apply the following procedure:

\begin{enumerate}
	\item Perform a triality transformation in any of the cubes that is not at the endpoints. This dualization is implemented by a cube move, as shown in \fref{basic_cube_move}.
	
	\item The previous step generates eight 2-valent edges, i.e. mass terms in the superpotential. These mass terms are shown as dashed edges in \fref{cube reduction in bipartite graph}. Integrating out the corresponding massive fields translates into edge condensation as discussed in \sref{section_massive_fields}. Each of the two bricks adjacent to the dualized cube end up with only two faces which, furthermore, correspond to one incoming and one outgoing chiral fields.
	
	\item These two bricks are therefore bubbles, as defined in \sref{section_reduction}, and disappear upon acting on them with triality, as shown in the last step of \fref{cube reduction in quiver diagram}.
\end{enumerate}

%=================================================================
%=================================================================

%=================================================================
%=================================================================

This process turns an $n$-cube model into an $(n-2)$-cube one. Iterating this procedure, we can thus reduce odd and even $n$ models all the way down to the 1-cube model and 2-cube model, respectively. Finally, the connection between even and odd $n$ is also straightforward, albeit slightly more subtle. Dualizing any of the cubes in the 2-cube model we generate a model with a bubble, which after reduction, results in a single cube. In addition, there are some external massive faces which, due to the way in which we treat the corresponding fields, are equivalent to the single external faces of the basic 1-cube model. In summary, the $n$-cube models are reducible for $n>1$ and can be connected by triality and reduction to the 1-cube theory. All of them share the same moduli space, which is the toric CY 6-fold associated to the toric diagram \eref{G_cube_model}.

%=================================================================
\section{Conclusions and Outlook}
%=================================================================

\label{section_conclusions}

We introduced and initiated the study of BFT$_2$’s, a general class of $2d$ $\mathcal{N}=(0,2)$ gauge theories defined by $2$-dimensional CW complexes on oriented 3-manifolds. While we primarily focused on examples with boundaries, in order to illustrate their novel features, theories without boundaries are also extremely interesting.

The dynamics of BFT$_2$’s is neatly captured by transformations of the CW complexes. A salient feature of this family of theories is its deep connection to combinatorial and geometric objects. These objects, such as the toric moduli spaces, which in turn are computed in terms of perfect matchings, provide powerful tools for addressing questions like the characterization of triality equivalence classes or diagnosing reducibility.

We identified structures, such as the ones leading to the automatic implementation of the trace condition, which were unnoticed in earlier considerations of BFT$_2$’s in the context of BBMs. We expect some of these new insights may shed light on open problems like, for instance, a better understanding of the BFT$_2$ analogues of zig-zag paths. 

Our work suggests various natural directions for future investigation. Below we summarize some of them.

\begin{itemize}

\item Embedding additional sub-classes of BFT$_2$’s, beyond the well-studied BBMs, in string theory. Of particular interest are theories with boundaries. Natural scenarios for such constructions involve systems of (fractional) D1-branes and flavor D5-branes at toric CY 4-folds, and their $T$-dual configurations of NS5 and D4-branes.\footnote{It would also be interesting to engineer other classes of theories that are closely related to BFT$_2$’s, e.g. ones in which (some of) the external Fermis are non-dynamical. Such theories would be similar in spirit to the close relatives of ordinary BFTs of \cite{Xie:2012mr,Heckman:2012jh}.}

\item Understanding how the dimension of the moduli space depends on the topology of the underlying 3-manifold. While the methods introduced in \sref{section_BFT2s_and_toric_geometry} apply to any 3-manifold, it would be illuminating to systematically determine a convenient basis of flows depending on the topology.

\item Generalizations of dimer models based on $m$-graded quivers with superpotentials were introduced in \cite{Franco:2019bmx}, where they were referred to as $m$-dimers. They capture the open string sector of the topological $B$-model on toric CY $(m + 2)$-folds. BBMs, which BFT$_2$’s generalize, correspond to $m=2$. We expect similar generalizations of $m$-dimers, which we can generically call BFT $_m$, exist for $m>2$. It would be interesting to work out the details of such theories.

\item Ordinary BFTs have applications to a wide range of interesting systems, which goes from D-branes at CY 3-folds to scattering amplitudes \cite{Franco:2012mm}. BFT$_2$’s certainly include the gauge theories that arise on D-branes at CY 4-folds \cite{Franco:2015tya}. It would be interesting to determine whether their range of applicability is also broader.

\item Finally, another interesting, albeit rather speculative, question is whether BFT$_2$’s admit natural generalizations of perfect orientations and boundary measurements. If so, do they give rise to certain notion of positive objects in some special cases? Such theories would become the analogues of ordinary BFTs associated to plabic graphs.

\end{itemize}

We expect to address some of these questions in future work.

%=================================================================
\acknowledgments
%=================================================================

We would like to thank Amihay Hanany, Richard Kenyon, Gregg Musiker, Tadashi Okasaki, David Speyer and Lauren Willians for enjoyable discussions and correspondence on related topics. We are also grateful to Gregg Musiker for carefully reading the manuscript and providing useful feedback. The research of SF was supported by the U.S. National Science Foundation grants PHY-1820721 and DMS-1854179.

%=================================================================
\appendix

\newpage

%=================================================================
\section{Perfect Matchings for the BFT$_2$ on $\mathbb{T}^2 \times I$}
%=================================================================

\label{section_appendix_P_matrix}

The 54 perfect matchings for the BFT$_2$ on $\mathbb{T}^2 \times I$ presented in \sref{section_model_T2_segment} are summarized in the following $P$-matrix.

%=================================================================
{\tiny
\beq
P=\left(
\begin{array}{c|ccccccccccccccccccccccccccc}
&p_1 &p_2 &p_3 &p_4 &p_5 &p_6 &p_7 &p_8 &p_9 &p_{10} &p_{11} &p_{12} &p_{13} &p_{14} &p_{15} &p_{16} &p_{17} &p_{18} &p_{19} &p_{20} &p_{21} &p_{22} &p_{23} &p_{24} &p_{25} &p_{26} &p_{27} \\
\hline 
X_{21} & 0 & 0 & 0 & 0 & 0 & 0 & 0 & 0 & 0 & 0 & 0 & 0 & 0 & 0 & 0 & 0 & 0 & 1 & 0 & 0 & 1 & 1 & 1 & 1 & 1 & 1 & 1 \\ 
Y_{21} & 0 & 0 & 0 & 0 & 0 & 0 & 0 & 0 & 0 & 0 & 0 & 0 & 0 & 0 & 0 & 0 & 0 & 0 & 0 & 1 & 1 & 1 & 1 & 1 & 1 & 1 & 1 \\ 
X_{42} & 0 & 0 & 0 & 0 & 0 & 0 & 0 & 0 & 1 & 1 & 1 & 1 & 1 & 1 & 1 & 1 & 1 & 0 & 1 & 0 & 0 & 0 & 0 & 0 & 0 & 0 & 0\\ 
Y_{42} & 0 & 0 & 0 & 0 & 0 & 0 & 0 & 0 & 1 & 1 & 1 & 1 & 1 & 1 & 1 & 1 & 1 & 0 & 0 & 0 & 0 & 0 & 0 & 0 & 0 & 0 & 0\\ 
X_{34} & 1 & 1 & 1 & 1 & 1 & 1 & 1 & 1 & 0 & 0 & 0 & 0 & 0 & 0 & 0 & 0 & 0 & 1 & 0 & 0 & 0 & 0 & 0 & 0 & 0 & 0 & 0\\ 
Y_{34} & 1 & 1 & 1 & 1 & 1 & 1 & 1 & 1 & 0 & 0 & 0 & 0 & 0 & 0 & 0 & 0 & 0 & 0 & 0 & 1 & 0 & 0 & 0 & 0 & 0 & 0 & 0\\ 
Z_{15} & 0 & 0 & 0 & 0 & 0 & 0 & 0 & 0 & 0 & 0 & 0 & 0 & 0 & 0 & 0 & 0 & 0 & 0 & 0 & 0 & 0 & 0 & 0 & 0 & 0 & 0 & 0\\ 
Z_{26} & 0 & 0 & 0 & 0 & 0 & 0 & 0 & 0 & 0 & 0 & 0 & 0 & 0 & 0 & 0 & 0 & 0 & 0 & 0 & 0 & 1 & 1 & 1 & 0 & 1 & 1 & 1\\ 
Z_{37} & 1 & 0 & 1 & 0 & 1 & 0 & 1 & 0 & 0 & 1 & 0 & 0 & 1 & 0 & 0 & 1 & 0 & 0 & 0 & 0 & 0 & 1 & 0 & 0 & 0 & 1 & 0\\ 
Z_{48} & 0 & 0 & 0 & 0 & 0 & 0 & 0 & 0 & 0 & 1 & 1 & 0 & 1 & 1 & 0 & 1 & 1 & 0 & 0 & 0 & 0 & 1 & 1 & 0 & 0 & 1 & 1\\ 
Z_{91} & 0 & 0 & 1 & 1 & 0 & 0 & 0 & 0 & 0 & 0 & 0 & 1 & 1 & 1 & 0 & 0 & 0 & 0 & 0 & 0 & 0 & 0 & 0 & 0 & 1 & 1 & 1\\ 
Z_{10,2} & 1 & 1 & 1 & 1 & 0 & 0 & 0 & 0 & 1 & 1 & 1 & 1 & 1 & 1 & 0 & 0 & 0 & 0 & 0 & 0 & 0 & 0 & 0 & 0 & 0 & 0 & 0\\ 
Z_{10,3} & 0 & 0 & 0 & 0 & 0 & 0 & 0 & 0 & 0 & 0 & 0 & 0 & 0 & 0 & 0 & 0 & 0 & 0 & 0 & 0 & 0 & 0 & 0 & 0 & 0 & 0 & 0\\ 
Z_{12,4} & 1 & 1 & 1 & 1 & 0 & 0 & 1 & 1 & 0 & 0 & 0 & 0 & 0 & 0 & 0 & 0 & 0 & 0 & 0 & 0 & 0 & 0 & 0 & 0 & 0 & 0 & 0\\ 
X_{1,11} & 0 & 0 & 0 & 0 & 0 & 0 & 0 & 0 & 0 & 0 & 0 & 0 & 0 & 0 & 0 & 0 & 0 & 0 & 1 & 0 & 0 & 0 & 0 & 0 & 0 & 0 & 0\\ 
Y_{1,11} & 0 & 0 & 0 & 0 & 0 & 0 & 0 & 0 & 0 & 0 & 0 & 0 & 0 & 0 & 0 & 0 & 0 & 0 & 0 & 0 & 0 & 0 & 0 & 0 & 0 & 0 & 0\\ 
X_{53} & 0 & 0 & 0 & 0 & 0 & 0 & 0 & 0 & 0 & 0 & 0 & 0 & 0 & 0 & 0 & 0 & 0 & 0 & 1 & 0 & 0 & 0 & 0 & 0 & 0 & 0 & 0\\ 
Y_{53} & 0 & 0 & 0 & 0 & 0 & 0 & 0 & 0 & 0 & 0 & 0 & 0 & 0 & 0 & 0 & 0 & 0 & 0 & 0 & 0 & 0 & 0 & 0 & 0 & 0 & 0 & 0\\ 
X_{65} & 0 & 0 & 0 & 0 & 0 & 0 & 0 & 0 & 0 & 0 & 0 & 0 & 0 & 0 & 0 & 0 & 0 & 1 & 0 & 0 & 0 & 0 & 0 & 1 & 0 & 0 & 0\\ 
Y_{65} & 0 & 0 & 0 & 0 & 0 & 0 & 0 & 0 & 0 & 0 & 0 & 0 & 0 & 0 & 0 & 0 & 0 & 0 & 0 & 1 & 0 & 0 & 0 & 1 & 0 & 0 & 0\\ 
X_{86} & 0 & 0 & 0 & 0 & 0 & 0 & 0 & 0 & 1 & 0 & 0 & 1 & 0 & 0 & 1 & 0 & 0 & 0 & 1 & 0 & 1 & 0 & 0 & 0 & 1 & 0 & 0\\ 
Y_{86} & 0 & 0 & 0 & 0 & 0 & 0 & 0 & 0 & 1 & 0 & 0 & 1 & 0 & 0 & 1 & 0 & 0 & 0 & 0 & 0 & 1 & 0 & 0 & 0 & 1 & 0 & 0\\ 
X_{78} & 0 & 1 & 0 & 1 & 0 & 1 & 0 & 1 & 0 & 0 & 1 & 0 & 0 & 1 & 0 & 0 & 1 & 1 & 0 & 0 & 0 & 0 & 1 & 0 & 0 & 0 & 1\\ 
Y_{78} & 0 & 1 & 0 & 1 & 0 & 1 & 0 & 1 & 0 & 0 & 1 & 0 & 0 & 1 & 0 & 0 & 1 & 0 & 0 & 1 & 0 & 0 & 1 & 0 & 0 & 0 & 1\\ 
X_{10,9} & 1 & 1 & 0 & 0 & 0 & 0 & 0 & 0 & 1 & 1 & 1 & 0 & 0 & 0 & 0 & 0 & 0 & 1 & 0 & 0 & 1 & 1 & 1 & 1 & 0 & 0 & 0\\ 
Y_{10,9} & 1 & 1 & 0 & 0 & 0 & 0 & 0 & 0 & 1 & 1 & 1 & 0 & 0 & 0 & 0 & 0 & 0 & 0 & 0 & 1 & 1 & 1 & 1 & 1 & 0 & 0 & 0\\ 
X_{12,10} & 0 & 0 & 0 & 0 & 0 & 0 & 1 & 1 & 0 & 0 & 0 & 0 & 0 & 0 & 1 & 1 & 1 & 0 & 1 & 0 & 0 & 0 & 0 & 0 & 0 & 0 & 0\\ 
Y_{12,10} & 0 & 0 & 0 & 0 & 0 & 0 & 1 & 1 & 0 & 0 & 0 & 0 & 0 & 0 & 1 & 1 & 1 & 0 & 0 & 0 & 0 & 0 & 0 & 0 & 0 & 0 & 0\\ 
X_{11,12} & 0 & 0 & 0 & 0 & 1 & 1 & 0 & 0 & 0 & 0 & 0 & 0 & 0 & 0 & 0 & 0 & 0 & 1 & 0 & 0 & 0 & 0 & 0 & 0 & 0 & 0 & 0\\ 
Y_{11,12} & 0 & 0 & 0 & 0 & 1 & 1 & 0 & 0 & 0 & 0 & 0 & 0 & 0 & 0 & 0 & 0 & 0 & 0 & 0 & 1 & 0 & 0 & 0 & 0 & 0 & 0 & 0
\end{array}
\right. 
\nonumber
\eeq
}
%=================================================================

%=================================================================
{\tiny
\beq
\left.
\begin{array}{c|ccccccccccccccccccccccccccc}
& p_{28} &p_{29} &p_{30} &p_{31} &p_{32} &p_{33} &p_{34} &p_{35} &p_{36} &p_{37} &p_{38} &p_{39} & p_{40} &p_{41} &p_{42} &p_{43} &p_{44} &p_{45} &p_{46} &p_{47} &p_{48} &p_{49} & p_{50} &p_{51} &p_{52} &p_{53} &p_{54}\\
\hline 
X_{21} & 1 & 0 & 0 & 0 & 0 & 0 & 0 & 0 & 0 & 0 & 0 & 0 & 0 & 0 & 0 & 0 & 0 & 0 & 0 & 0 & 0 & 0 & 0 & 0 & 0 & 0 & 0 \\ 
Y_{21} & 1 & 0 & 0 & 0 & 0 & 0 & 0 & 0 & 0 & 0 & 0 & 0 & 0 & 0 & 0 & 0 & 0 & 0 & 0 & 0 & 0 & 0 & 0 & 0 & 0 & 0 & 0 \\ 
X_{42} & 0 & 0 & 0 & 0 & 0 & 0 & 0 & 0 & 0 & 0 & 0 & 0 & 0 & 0 & 0 & 0 & 0 & 0 & 0 & 0 & 0 & 0 & 0 & 0 & 0 & 0 & 0 \\ 
Y_{42} & 0 & 0 & 0 & 0 & 0 & 0 & 0 & 0 & 0 & 0 & 0 & 0 & 0 & 0 & 0 & 0 & 0 & 0 & 0 & 0 & 0 & 1 & 0 & 0 & 0 & 0 & 0 \\ 
X_{34} & 0 & 0 & 0 & 0 & 0 & 0 & 0 & 0 & 0 & 0 & 0 & 0 & 0 & 0 & 0 & 0 & 0 & 0 & 0 & 0 & 0 & 0 & 0 & 0 & 0 & 0 & 0 \\ 
Y_{34} & 0 & 0 & 0 & 0 & 0 & 0 & 0 & 0 & 0 & 0 & 0 & 0 & 0 & 0 & 0 & 0 & 0 & 0 & 0 & 0 & 0 & 0 & 0 & 0 & 0 & 0 & 0 \\ 
Z_{15} & 0 & 1 & 1 & 1 & 1 & 1 & 1 & 1 & 1 & 1 & 1 & 1 & 1 & 1 & 1 & 1 & 1 & 1 & 1 & 1 & 1 & 0 & 0 & 0 & 0 & 0 & 0 \\ 
Z_{26} & 0 & 1 & 1 & 1 & 0 & 1 & 1 & 1 & 0 & 1 & 1 & 1 & 0 & 1 & 1 & 1 & 0 & 1 & 1 & 1 & 0 & 0 & 0 & 0 & 0 & 0 & 0 \\ 
Z_{37} & 0 & 0 & 1 & 0 & 0 & 0 & 1 & 0 & 0 & 0 & 1 & 0 & 0 & 0 & 1 & 0 & 0 & 0 & 1 & 0 & 0 & 0 & 0 & 0 & 0 & 0 & 0 \\ 
Z_{48} & 0 & 0 & 1 & 1 & 0 & 0 & 1 & 1 & 0 & 0 & 1 & 1 & 0 & 0 & 1 & 1 & 0 & 0 & 1 & 1 & 0 & 0 & 0 & 0 & 0 & 0 & 0 \\ 
Z_{91} & 1 & 0 & 0 & 0 & 0 & 0 & 0 & 0 & 0 & 1 & 1 & 1 & 1 & 0 & 0 & 0 & 0 & 0 & 0 & 0 & 0 & 0 & 0 & 0 & 1 & 0 & 0 \\ 
Z_{10,2} & 0 & 0 & 0 & 0 & 0 & 1 & 1 & 1 & 1 & 1 & 1 & 1 & 1 & 0 & 0 & 0 & 0 & 0 & 0 & 0 & 0 & 0 & 0 & 1 & 1 & 0 & 0 \\ 
Z_{10,3} & 0 & 0 & 0 & 0 & 0 & 1 & 1 & 1 & 1 & 1 & 1 & 1 & 1 & 1 & 1 & 1 & 1 & 1 & 1 & 1 & 1 & 0 & 0 & 1 & 1 & 1 & 1 \\ 
Z_{12,4} & 0 & 0 & 0 & 0 & 0 & 1 & 1 & 1 & 0 & 1 & 1 & 1 & 0 & 0 & 0 & 0 & 0 & 1 & 1 & 1 & 1 & 0 & 0 & 1 & 1 & 0 & 1 \\ 
X_{1,11} & 0 & 1 & 1 & 1 & 1 & 0 & 0 & 0 & 0 & 0 & 0 & 0 & 0 & 0 & 0 & 0 & 0 & 0 & 0 & 0 & 0 & 0 & 1 & 0 & 0 & 0 & 0 \\ 
Y_{1,11} & 0 & 1 & 1 & 1 & 1 & 0 & 0 & 0 & 0 & 0 & 0 & 0 & 0 & 0 & 0 & 0 & 0 & 0 & 0 & 0 & 0 & 1 & 1 & 0 & 0 & 0 & 0 \\ 
X_{53} & 0 & 0 & 0 & 0 & 0 & 0 & 0 & 0 & 0 & 0 & 0 & 0 & 0 & 0 & 0 & 0 & 0 & 0 & 0 & 0 & 0 & 0 & 1 & 1 & 1 & 1 & 1 \\ 
Y_{53} & 0 & 0 & 0 & 0 & 0 & 0 & 0 & 0 & 0 & 0 & 0 & 0 & 0 & 0 & 0 & 0 & 0 & 0 & 0 & 0 & 0 & 1 & 1 & 1 & 1 & 1 & 1 \\ 
X_{65} & 1 & 0 & 0 & 0 & 1 & 0 & 0 & 0 & 1 & 0 & 0 & 0 & 1 & 0 & 0 & 0 & 1 & 0 & 0 & 0 & 1 & 0 & 0 & 0 & 0 & 0 & 0 \\ 
Y_{65} & 1 & 0 & 0 & 0 & 1 & 0 & 0 & 0 & 1 & 0 & 0 & 0 & 1 & 0 & 0 & 0 & 1 & 0 & 0 & 0 & 1 & 0 & 0 & 0 & 0 & 0 & 0 \\ 
X_{86} & 0 & 1 & 0 & 0 & 0 & 1 & 0 & 0 & 0 & 1 & 0 & 0 & 0 & 1 & 0 & 0 & 0 & 1 & 0 & 0 & 0 & 0 & 0 & 0 & 0 & 0 & 0 \\ 
Y_{86} & 0 & 1 & 0 & 0 & 0 & 1 & 0 & 0 & 0 & 1 & 0 & 0 & 0 & 1 & 0 & 0 & 0 & 1 & 0 & 0 & 0 & 1 & 0 & 0 & 0 & 0 & 0 \\ 
X_{78} & 0 & 0 & 0 & 1 & 0 & 0 & 0 & 1 & 0 & 0 & 0 & 1 & 0 & 0 & 0 & 1 & 0 & 0 & 0 & 1 & 0 & 0 & 0 & 0 & 0 & 0 & 0 \\ 
Y_{78} & 0 & 0 & 0 & 1 & 0 & 0 & 0 & 1 & 0 & 0 & 0 & 1 & 0 & 0 & 0 & 1 & 0 & 0 & 0 & 1 & 0 & 0 & 0 & 0 & 0 & 0 & 0 \\ 
X_{10,9} & 0 & 0 & 0 & 0 & 0 & 1 & 1 & 1 & 1 & 0 & 0 & 0 & 0 & 0 & 0 & 0 & 0 & 0 & 0 & 0 & 0 & 0 & 0 & 1 & 0 & 0 & 0 \\ 
Y_{10,9} & 0 & 0 & 0 & 0 & 0 & 1 & 1 & 1 & 1 & 0 & 0 & 0 & 0 & 0 & 0 & 0 & 0 & 0 & 0 & 0 & 0 & 0 & 0 & 1 & 0 & 0 & 0 \\ 
X_{12,10} & 0 & 0 & 0 & 0 & 0 & 0 & 0 & 0 & 0 & 0 & 0 & 0 & 0 & 0 & 0 & 0 & 0 & 1 & 1 & 1 & 1 & 0 & 0 & 0 & 0 & 0 & 1 \\ 
Y_{12,10} & 0 & 0 & 0 & 0 & 0 & 0 & 0 & 0 & 0 & 0 & 0 & 0 & 0 & 0 & 0 & 0 & 0 & 1 & 1 & 1 & 1 & 1 & 0 & 0 & 0 & 0 & 1 \\ 
X_{11,12} & 0 & 0 & 0 & 0 & 0 & 0 & 0 & 0 & 0 & 0 & 0 & 0 & 0 & 1 & 1 & 1 & 1 & 0 & 0 & 0 & 0 & 0 & 0 & 0 & 0 & 1 & 0 \\ 
Y_{11,12} & 0 & 0 & 0 & 0 & 0 & 0 & 0 & 0 & 0 & 0 & 0 & 0 & 0 & 1 & 1 & 1 & 1 & 0 & 0 & 0 & 0 & 0 & 0 & 0 & 0 & 1 & 0
\end{array}
\right)
\eeq
}
%=================================================================

%======================================================================
\bibliographystyle{JHEP}
\bibliography{ref}
%======================================================================

\end{document}